\documentclass[twocolumn]{aastex631}

\usepackage{lipsum}
\usepackage{enumitem}
\usepackage{amsmath}
\usepackage{url}
\usepackage{CJK}
\usepackage{graphbox}
\usepackage[colorinlistoftodos]{todonotes}
\setuptodonotes{inline,color=green!40}
\usepackage[normalem]{ulem}
\usepackage{xcolor}

\shorttitle{Tidal Distortions in DF2 and DF4}
\shortauthors{Keim et al.}

\begin{document}
\begin{CJK*}{UTF8}{gbsn}

\title{Tidal Distortions in NGC1052-DF2 and NGC1052-DF4: Independent Evidence for a Lack of Dark Matter}

\correspondingauthor{Michael A. Keim}
\email{michael.keim@yale.edu}

\author[0000-0002-7743-2501]{Michael A. Keim}
\affiliation{Department of Astronomy, Yale University, PO Box 208101, New Haven, CT 06520-8101, USA}

\author[0000-0002-8282-9888]{Pieter van Dokkum}
\affiliation{Department of Astronomy, Yale University, PO Box 208101, New Haven, CT 06520-8101, USA}

\author[0000-0002-1841-2252]{Shany Danieli}\altaffiliation{NASA Hubble Fellow}
\affiliation{Department of Astrophysical Sciences, 4 Ivy Lane, Princeton University, Princeton, NJ 08544, USA}
\affiliation{Institute for Advanced Study, 1 Einstein Drive, Princeton, NJ 08540, USA}

\author[0000-0002-2406-7344]{Deborah Lokhorst}
\affiliation{Department of Astronomy \& Astrophysics, University of Toronto, 50 St. George St., Toronto, ON M5S 3H4, Canada}
\affiliation{Dunlap Institute for Astronomy and Astrophysics, University of Toronto, Toronto ON, M5S 3H4, Canada}
\affiliation{NRC Herzberg Astronomy \& Astrophysics Research Centre, 5071 West Saanich Road, Victoria, BC V9E2E7, Canada}

\author[0000-0001-9592-4190]{Jiaxuan Li (李嘉轩)}
\affiliation{Department of Astrophysical Sciences, 4 Ivy Lane, Princeton University, Princeton, NJ 08544, USA}
%\affiliation{Kavli Institute for Astronomy and Astrophysics, Peking University, 5 Yiheyuan Road, Haidian District, Beijing 100871, China}

\author[0000-0002-5120-1684]{Zili Shen}
\affiliation{Department of Astronomy, Yale University, PO Box 208101, New Haven, CT 06520-8101, USA}

\author[0000-0002-4542-921X]{Roberto Abraham}
\affiliation{Department of Astronomy \& Astrophysics, University of Toronto, 50 St. George St., Toronto, ON M5S 3H4, Canada}
\affiliation{Dunlap Institute for Astronomy and Astrophysics, University of Toronto, Toronto ON, M5S 3H4, Canada}

\author[0000-0002-4175-3047]{Seery Chen}
\affiliation{Department of Astronomy \& Astrophysics, University of Toronto, 50 St. George St., Toronto, ON M5S 3H4, Canada}
\affiliation{Dunlap Institute for Astronomy and Astrophysics, University of Toronto, Toronto ON, M5S 3H4, Canada}

\author[0000-0002-8931-4684]{Colleen Gilhuly}
\affiliation{Department of Astronomy \& Astrophysics, University of Toronto, 50 St. George St., Toronto, ON M5S 3H4, Canada}
\affiliation{Dunlap Institute for Astronomy and Astrophysics, University of Toronto, Toronto ON, M5S 3H4, Canada}

\author[0000-0002-7490-5991]{Qing Liu (刘青)}
\affiliation{Department of Astronomy \& Astrophysics, University of Toronto, 50 St. George St., Toronto, ON M5S 3H4, Canada}
\affiliation{Dunlap Institute for Astronomy and Astrophysics, University of Toronto, Toronto ON, M5S 3H4, Canada}

\author[0000-0001-9467-7298]{Allison Merritt}
\affiliation{Max-Planck-Institut f\"ur Astronomie, K\"onigstuhl 17, D-69117 Heidelberg, Germany}

\author[0000-0001-8367-6265]{Tim B. Miller}
\affiliation{Department of Astronomy, Yale University, PO Box 208101, New Haven, CT 06520-8101, USA}

\author[0000-0002-7075-9931]{Imad Pasha}
\affiliation{Department of Astronomy, Yale University, PO Box 208101, New Haven, CT 06520-8101, USA}

\author[0000-0002-5283-933X]{Ava Polzin}
\affiliation{Department of Astronomy, Yale University, PO Box 208101, New Haven, CT 06520-8101, USA}

\begin{abstract}
\renewcommand{\thempfootnote}{$\dagger$}
Two ultra diffuse galaxies in the same group, NGC1052-DF2 and NGC1052-DF4, have been found to have little or no dark matter and to host unusually luminous globular cluster populations. Such low mass diffuse objects in a group environment are easily disrupted and are expected to show evidence of tidal distortions. In this work we present deep new imaging of the NGC1052 group, obtained with the Dragonfly Telephoto Array, to test this hypothesis. We find that both galaxies show strong position angle twists and are significantly more elongated at their outskirts than in their interiors. The group's central massive elliptical NGC1052 is the most likely source of these tidal disturbances. The observed distortions imply that the galaxies have a low total mass or are very close to NGC1052. Considering constraints on the galaxies' relative distances, we infer that the dark matter halo masses of these galaxies cannot be much greater than their stellar masses. Calculating pericenters from the distortions, we find that the galaxies are on highly elliptical orbits, with a ratio of pericenter to present-day radius $R_{\rm peri}/R_0{\sim}0.1$ if the galaxies are dark matter-free and $R_{\rm peri}/R_0{\sim}0.01$ if they have a normal dark halo. Our findings provide strong evidence, independent of kinematic constraints, that both galaxies are dark matter deficient. Furthermore, the similarity of the tidal features in NGC1052-DF2 and NGC1052-DF4 strongly suggests that they arose at comparable distances from NGC1052. In Appendix A, we describe {\tt sbcontrast}, a robust method to determine the surface brightness limit of images.\footnote{Publicly available via `{\tt pip install sbcontrast}.'}
\end{abstract}
\keywords{Dark matter (353) --- Galaxy evolution (594) --- Galaxy structure (622) --- Low surface brightness galaxies (940) --- Tidal distortion (1697)}

%--------------------------------------------------------------
% Introduction
%--------------------------------------------------------------

\section{Introduction} \label{Sec:Introduction}

The galaxies \object{NGC1052-DF2} and \object{NGC1052-DF4} both have little to no dark matter based on the velocity dispersion of kinetic tracers \citep{2018Natur.555..629V,2018ApJ...863L..15W,2019ApJ...874L...5V,2019ApJ...874L..12D,2019A&A...625A..76E} and host a unique population of globular clusters nearly two magnitudes brighter than expected \citep{2018ApJ...856L..30V,2020MNRAS.496.3741M,2021ApJ...909..179S}. They are likely associated with the massive elliptical NGC1052 \citep{2021ApJ...914L..12S} and fall into the category of ultra diffuse galaxies; large, faint objects with the sizes of $L_*$ galaxies but the stellar masses of dwarf galaxies \citep{2015ApJ...798L..45V}.

Both their luminous globular clusters and low dark matter content set NGC1052-DF2 and NGC1052-DF4 apart from almost all other galaxies known. The near-universal globular cluster luminosity function has a peak at $M_V\approx-7.5$ mag~\citep{2012Ap&SS.341..195R}, whereas the NGC1052-DF2 and NGC1052-DF4 distributions peak at $M_V\approx-9$ mag \citep{2021ApJ...909..179S}. Galaxies with the stellar mass of NGC1052-DF2 and NGC1052-DF4 (approximately 2.0$\times$10$^8$ M$_\odot$ and 1.5$\times$10$^8$ M$_\odot$, respectively; \citealt{2018Natur.555..629V,2019ApJ...874L...5V}) ought to have a $>$300 times more massive dark matter halo \citep{2013ApJ...770...57B}. However, their inferred total masses are of the same order as their stellar masses alone. 

The observation of two independent galaxies each displaying the same remarkable properties suggests that the low dark matter content is not the consequence of mass measurement uncertainties. This is further supported by the survival of both galaxies' massive globular clusters, which would rapidly sink to the center of the galaxies were a cuspy halo present \citep{2019ApJ...877..133D}. Yet, an important question remains regarding the apparent lack of tidal distortions in the Dragonfly Telephoto Array (Dragonfly; \citealt{2014PASP..126...55A}) images which first identified NGC1052-DF4 and brought renewed interest in NGC1052-DF2 \citep{2018Natur.555..629V,2019ApJ...874L...5V}. A kpc-scale object with a dwarf galaxy's stellar content and no dark matter is fragile, and material on its outskirts should be easily unbound by the gravitational pull of other galaxies in a group environment. However, both galaxies have a smooth, spheroidal morphology in the initial Dragonfly images. As already noted by \citet{2018Natur.555..629V}, at face value this lack of distortions is difficult to reconcile with the idea that the galaxies are dark matter deficient.

One possible explanation for the inconsistency is that the galaxies are in the outskirts of the group, far from the central galaxy. Indeed, the recently measured line-of-sight distance between NGC1052-DF2 and NGC1052-DF4 of 2.1$\pm$0.5 Mpc \citep{2021ApJ...914L..12S} implies that at least one of the galaxies may be at a considerable distance from the center of the NGC1052 group. It is possible that they are on highly elliptical orbits, or that one or both of the galaxies was ejected from the NGC1052 group in the distant past.  This may be in line with proposed formation scenarios that include high velocity collisions (see, e.g., \citealt{2019MNRAS.488L..24S} and \citealt{2020ApJ...899...25S}) or backsplash orbits \citep{2021NatAs...5.1255B}. However, given the long ($\gtrsim$ 1 Gyr) dynamical time scales in the outskirts of dark matter-free galaxies one would still expect to see an imprint of the  tidal forces that acted on these feeble systems at closest approach.

A second explanation of the apparent lack of tidal features is that the initial Dragonfly images' depth, a 1$\sigma$ surface brightness limit of $\mu_g \approx 29$ mag arcsec$^{-2}$ on 12$\arcsec$ scales \citep{2018Natur.555..629V}, was insufficient to detect them. Recent studies have obtained deeper
imaging and have indeed found evidence for distortions.
\citet{2020ApJ...904..114M} find that NGC1052-DF4 becomes disrupted at $\approx$44$\arcsec$, where the galaxy's profile reaches a surface brightness of $\mu_g\approx29$ mag arcsec$^{-2}$. Similarly, NGC1052-DF2 appears to be elongated at large radii in Fig.\ 3 of \citet{2019A&A...624L...6M}. However, a recent analysis by \citet{2021ApJ...919...56M} does not interpret NGC1052-DF2 as being tidally distorted.

In this work we return to Dragonfly to explore tidal features associated with the dark matter deficient galaxies, producing one of the deepest images yet taken by the array. We start with a brief discussion of the utility of tidal structures to probe a galaxy's mass and orbit in Section~\ref{Sec:Tides}. In Section~\ref{Sec:Data}, we outline our observations and the data reduction procedures we use to isolate low surface brightness emission. In Section~\ref{Sec:Results} we describe and analyze our images of NGC1052-DF2 and NGC1052-DF4 and use their derived morphology to constrain their masses and orbits. We place our work in a wider context in Section~\ref{Sec:Discussion}, and give a final summary of our findings in Section~\ref{Sec:Conclusions}. Where relevant, we assume a flat, Lambda Cold Dark Matter ($\Lambda$CDM) cosmological model with a Hubble constant $H_{0}=$ 67.4 km s$^{-1}$ Mpc$^{-1}$ and density parameters $\Omega_{\textrm{M}}=$ 0.315 and $\Omega_{\Lambda}=$ 0.685 \citep{2020A&A...641A...6P}.

%--------------------------------------------------------------
% Tidal Disruptions
%--------------------------------------------------------------

\begin{figure*}
    \centering
    \includegraphics[width=0.47329439252\textwidth]{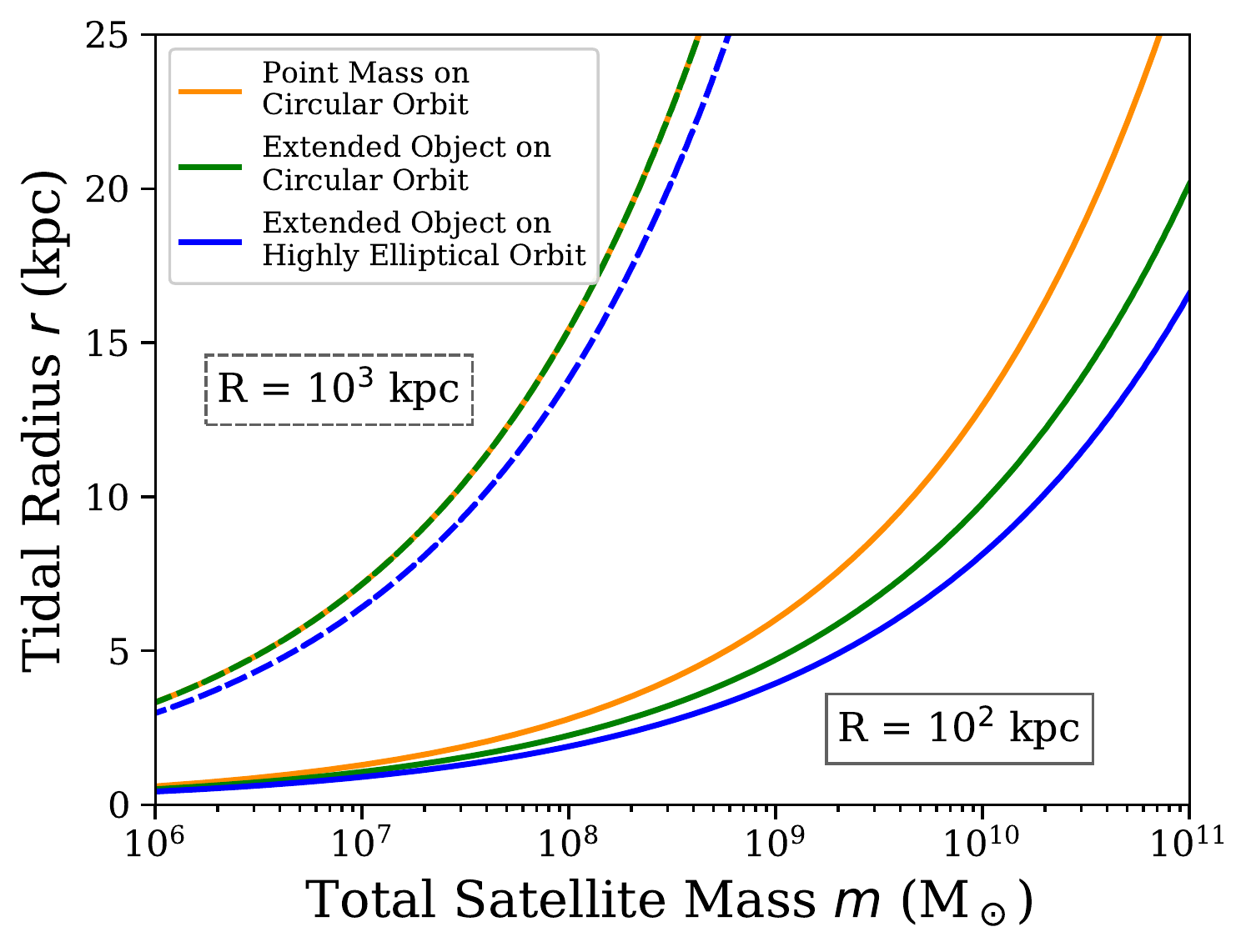}\quad\includegraphics[width=0.47\textwidth]{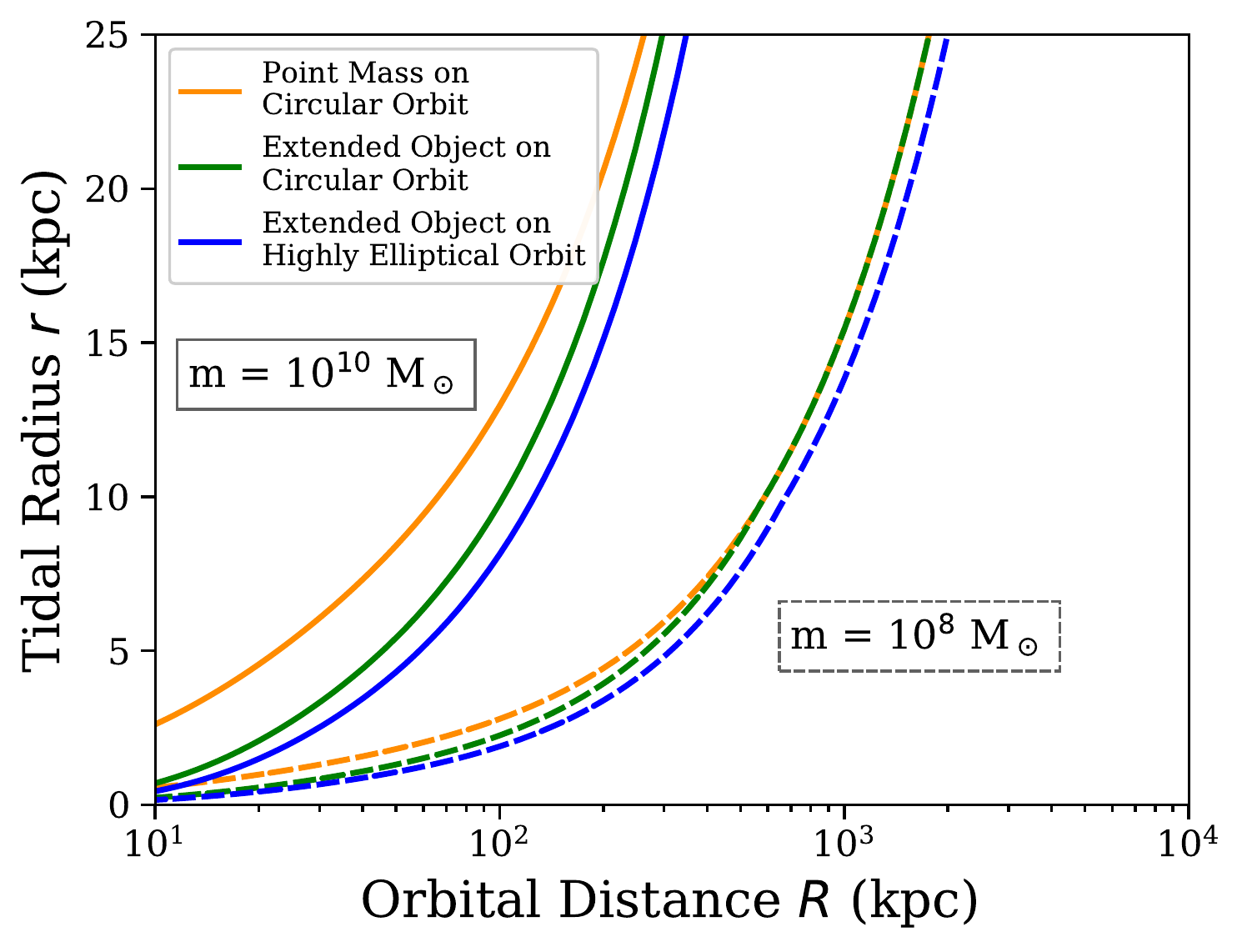}
    \caption{A demonstration of how the tidal radius depends on properties of the satellite and its orbit using Eq.~(\ref{Eq:rtid}). Material in the satellite beyond the indicated tidal radius will be stripped by NGC1052. \textit{Left panel}: The tidal radius as a function of total satellite mass $m$ at two different distances from NGC1052, 10$^2$ and 10$^3$ kpc. \textit{Right panel}: The tidal radius as a function of orbital distance $R$ for two different satellite masses, 10$^8$ and 10$^{10}$ M$_\odot$. We first show $r_{\rm tid}$ for a point mass on a circular orbit (\textit{orange lines}), then compare to an extended satellite with a Navarro-Frenk-White (NFW; \citealt{1997ApJ...490..493N}) profile using a concentration $c = 9.6 \times (m/10^{13}\textrm{ M}_\odot)^{-0.13}\times(1+z)^{-1}$ \citep{2005ApJ...635..931B} and a virial radius $r_{\rm vir} = \left( G m / 100 H^2(z) \right)^{1/3}$ (\textit{green lines}). Finally, we include the same extended satellite on a highly elliptical orbit with an angular speed $\Omega = \sqrt{2} V_{\rm circ}/R$, i.e. the minimum possible tidal radius for a bound orbit (\textit{blue lines}). NGC1052's mass profile $M(R)$ is taken from \citet{2019MNRAS.489.3665F}, an NFW profile with $c$ = 7.0, $M_{\rm vir}$ = 6.2$\times$10$^{12}$ M$_\odot$, and $R_{\rm vir}$ = 390 kpc, untruncated to account for other objects within the group, and a 4$\times$10$^{11}$ M$_\odot$ central baryonic component.\label{Fig:Tides}}
\end{figure*}

\section{Tidal Distortions} \label{Sec:Tides}

Satellite galaxies in a group environment are tidally disturbed by the gravitational pull of other galaxies in the group. At a point known as the tidal radius, material in the satellite becomes unbound by the perturbing galaxy and is stripped away. If we consider two galaxies orbiting at a distance $R$ and account for a centrifugal force, the radius $r_{\rm tid}$ where the tidal force from a perturber $M$ is greater than the self gravity of a satellite $m$ is given by
\begin{equation} \label{Eq:rtid}
    r_{\rm tid} = R \times \left( \frac{m(r_{\rm tid}) / M(R)}{2 + \frac{\Omega^2 {R^3}}{G M(R)} - \frac{\textrm{d} \ln{M}}{\textrm{d} \ln{R}}|_{R} } \right)^{1/3}
\end{equation}
where $m(r)$ and $M(R)$ are the satellite and perturber's mass profiles, respectively, and $\Omega = |\vec{V} \times \vec{R}| / R^2$ is the angular speed of the satellite \citep{1962AJ.....67..471K,2010gfe..book.....M}. Note that other definitions of the tidal radius exist in the literature, as reviewed by \citet{2018MNRAS.474.3043V}, with Eq.~(\ref{Eq:rtid}) making the fewest assumptions. In the limit that the galaxies are on a circular orbit sufficiently large so that $m(r)$ and $M(R)$ may be treated as point masses, Eq.~(\ref{Eq:rtid}) reduces to
\begin{equation} \label{Eq:rJ}
r_{\rm J} = R\times\left(\frac{m}{3 M}\right)^{1/3},
\end{equation}
i.e., the classical Jacobi radius \citep{2008gady.book.....B}.

In Fig.~\ref{Fig:Tides}, we demonstrate how the tidal radius of a satellite galaxy moving through the potential of NGC1052 (as taken from \citealt{2019MNRAS.489.3665F}) depends on the satellite's orbital distance, orbital velocity, total mass, and mass distribution.\footnote{Note that here we only consider the satellite's halo to present a generalized case, since dark matter dominates the total mass and outer density profile of most galaxies. However, this is not appropriate for dark matter deficient galaxies. Thus, in Section~\ref{Sec:Results} we also consider a fixed stellar component based on the observed S\'ersic profile.} The group's central elliptical NGC1052 represents the most likely perturber due to its high mass and relative proximity to NGC1052-DF2 and NGC1052-DF4 \citep{2001MNRAS.327.1004B,2020ApJ...895L...4D,2021ApJ...914L..12S}.\footnote{We will show later that the morphologies of the galaxies are consistent with this assumption. The galaxies have very similar tidal distortions, even though they are not close to each other on the sky. This similarity, shown in Section~\ref{Sec:Results}, implies a similar underlying tidal field -- as expected when NGC1052 is the perturber.}

Fig.~\ref{Fig:Tides} shows that less massive satellites are more fragile and will be tidally stripped at smaller radii within the satellite, leading to more dramatic distortions. A low mass satellite must be orbiting at a much farther distance to have the same tidal radius as a more massive satellite. This effect is amplified if we consider an object's spatial extent, since material throughout the satellite will have a lower self gravity, experiencing the inward pull of less enclosed mass at greater distances. It is also amplified if we consider angular speeds exceeding that of a circular orbit, as this leads to a greater inward acceleration toward the perturber making it easier for material to be gravitationally unbound. Since the dark matter deficient, ultra diffuse galaxies NGC1052-DF2 and NGC1052-DF4 each have an especially low mass and large spatial extent, they are easily tidally disrupted. The primary driver is their unusual dark matter deficiency, with their spatial extent acting only as a secondary effect; we should not expect the larger population of ultra diffuse galaxies with normal dark matter content to be as easily distorted. This propensity for tidal effects is an important consequence, though not necessarily the cause, of the galaxies' dark matter deficiency. 

One clear observational signature of the tidal radius are S-shaped tidal tails which indicate stellar material actively being stripped away from the satellite by a perturbing galaxy (\citealt{2001ApJ...548L.165O}; \citealt{2006MNRAS.365.1263M,2012ApJ...755L..13K,2017ApJ...851...27M}). However, internal morphological distortions related to tidal interactions may be observed even without the detection of tidal tails. Such features may be identified through the stretching or twisting of isophotes \citep{1973ApJ...182..671H,1982SAAS...12..115K,2002AJ....124..127J,2006MNRAS.369.1321D,2012ApJ...761L..31B,2017ApJ...851...27M}, although in this case inferring the precise location of the tidal radius becomes more complicated. These distortions may represent bound material within the tidal radius being heated by an ongoing interaction, or they may be tidal relics -- remnants of an interaction that occurred in the past, for instance at the pericentric passage of an elliptical orbit where tidal interactions are strongest. Indeed, a number of works have made the approximation that the onset of observed distortions is equal to the tidal radius at pericenter (e.g. \citealt{1957ApJ...125..451V,1973ApJ...179..423F,1995MNRAS.277.1354I,2004MNRAS.347..119B,2014A&A...566A..44B}). Assuming the dissipation of tidal features through mechanisms like phase mixing had a negligible effect, this allows for a constraint on the orbital pericenter.

By relating tidal features to the current tidal radius, one may also use distance measurements to constrain the satellite's mass. However, this relationship may vary considerably across elliptical orbits as the strength of ongoing interactions change, so such analysis is best informed by numerical study (e.g. \citealt{2002AJ....124..127J,2013MNRAS.433..878L,2016ApJ...819...20G}). Indeed, simulations have found that breaks in a satellite's surface brightness profile are generally reasonable indicators for the instantaneous tidal radius: for orbital locations beyond pericenter, breaks at $r_{\rm break}$ may be found that are closer in than the current tidal radius $r_{\rm tid}$ by a factor of a few. This variation is most extreme for highly elliptical orbits, and for an apocenter-to-pericenter ratio of 20 may reach up to $r_{\rm tid} \approx 3 \times r_{\rm break}$ (\citealt{2013MNRAS.433..878L}; see Fig. 4). Given the large $\approx$2 Mpc separation between NGC1052-DF2 and NGC1052-DF4 \citep{2021ApJ...914L..12S}, either one or both of the galaxies are far from NGC1052 and likely beyond their orbital pericenter. In this case, Fig.~\ref{Fig:Tides} indicates that even if the dark matter deficient galaxies are $>$1 Mpc away from NGC1052, we might still be able to observe breaks in NGC1052-DF2 and NGC1052-DF4's surface brightness profiles lying slightly inwards of the current tidal radius.

\begin{figure*}
    \centering
    \includegraphics[width=\textwidth]{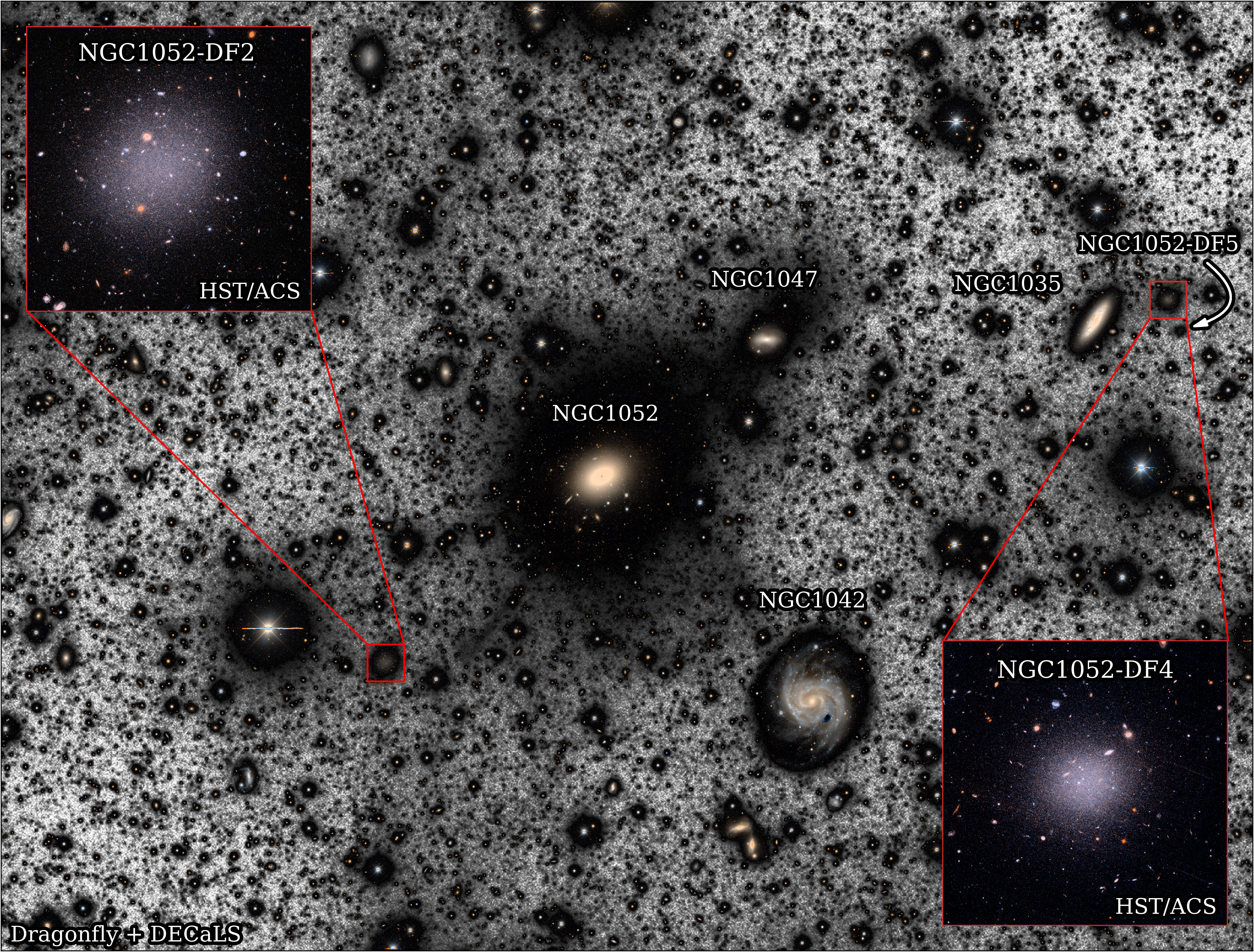}
    \caption{The NGC1052 field. A composite $g+r$ Dragonfly image (\textit{greyscale}) covers the central 1.0$\times$0.75 deg region cutout from the original 12 deg$^2$ field. High surface brightness objects in the same region are displayed using data from DECaLS (\textit{multi-color}). The 1.75$\arcmin{\times}$1.75$\arcmin$ areas surrounding NGC1052-DF2 and NGC1052-DF4 are shown using deep \textit{HST} images from \citet{2021ApJ...914L..12S} and \citet{2020ApJ...895L...4D} observed with the ACS over 40 and 12 orbits, respectively (\textit{zoom-in panels}). \label{Fig:Field}} % Displayed resolution is not sufficient to reflect that achieved by DECaLS and \textit{HST} data.
\end{figure*} % 

We caution readers that it is possible NGC1052-DF2 and NGC1052-DF4's trajectories may not be well described by past simulations which study tidal breaks, preventing a hard constraint on the current tidal radius. Moreover, the concept of tidal radii in general has flaws that have been extensively reviewed in the literature (e.g. \citealt{2001ApJ...559..716T,2003ApJ...598...49Z,2006MNRAS.366..429R,2008gady.book.....B,2010gfe..book.....M,2018MNRAS.474.3043V}), breaking down in specific regimes where underlying approximations are invalid (e.g. $r_{\rm tid} \approx R$) and being generally incorrect in assuming a perfectly spheroidal surface where material is unbound. Finally, we note that isophote twists may also result from the projection of a triaxial system with radially-varying axis ratios rather than tidal disruption, although the resulting position angle changes tend to be small \citep{1995AJ....110.2027V}.

%; otherwise the galaxies would only close on the sky by chance, and we should expect to find a far greater number of dark matter deficient ultra diffuse galaxies with bright globular clusters in future low surface brightness surveys.

%--------------------------------------------------------------
% Data
%--------------------------------------------------------------

\section{Data} \label{Sec:Data}

%--------------------------------------------------------------

\subsection{Observations and Reduction} \label{Sec:Observation}

Images of the NGC1052 field were obtained with the 48-lens Dragonfly Telephoto Array (\citealt{2014PASP..126...55A,2020ApJ...895L...4D}). Each of the 48 Canon 400 mm f/2.8 II telephoto lenses that are in the array is outfitted with a Santa Barbara Imaging Group (SBIG) CCD camera, with an instantaneous field of view of $2 \fdg 6 \times 1 \fdg 9$ and a pixel scale of $2\farcs85 \ \mathrm{pixel}^{-1}$. Redundancy is achieved by offsetting the lenses from one another by $\approx 10\%$ of the field of view and by large dithers between individual exposures. Individual observing sequences consisted of nine exposures of $600\,\mathrm{s}$ each, dithered in a quasi-random pattern in a $45'$ box. Owing to the dithering, the final image covers $12\,\mathrm{deg}^2$, with reduced depth near the edges of the field. Images were taken in both $g$ and $r$ bands simultaneously with half (24) of the lenses equipped with Sloan-$g$ filters and half with Sloan-$r$. The observations took place during 18 nights between October 2018 and February 2019.

The data reduction was performed using the customized Dragonfly Reduction Pipeline software, described in detail in \citet{2020ApJ...895L...4D} and in \citet{2018PhDTZhang}. Briefly, a set of algorithms assesses the quality of individual frames, retaining only high-quality data. The sky modeling and subtraction is performed in two stages, ensuring that low surface brightness features with spatial scales up to $0 \fdg 9 \times 0 \fdg 6$ are preserved. All the data processing steps were performed using batch processing protocols on the Canadian Advanced Network for Astronomical Research (CANFAR) cloud. The final $g$ and $r$-band co-added images consist of $2346$ and $2603$ frames respectively. This is equivalent to $17.2\,\mathrm{hrs}$ with the full 48-lens array ($8.14\,\mathrm{hrs}$ in $g$-band and $9.04\,\mathrm{hrs}$ in $r$-band). % \footnote{\url{https://jielaizhang.github.io/files/Zhang_Jielai_201811_PhD_Thesis_excludech4.pdf}} 

In addition to our new Dragonfly observation, this work also makes use of Dark Energy Camera (DECam) Legacy Survey (DECaLS; \citealt{2019AJ....157..168D}) data to model high surface brightness objects (see Section~\ref{Sec:MRF}) as well as deep \textit{Hubble Space Telescope (HST)} Advanced Camera for Surveys (ACS) imaging to study the morphology of the galaxies' inner regions (see Section~\ref{Sec:HST}). DECaLS $g$- and $r$-band data was accessed from Data Release 8 (DR8). \textit{HST} observations, as detailed in \citet{2020ApJ...895L...4D} and \citet{2021ApJ...914L..12S}, were taken during Cycles 24 (program 14644), 26 (program 15695), and 27 (program 15851), imaging NGC1052-DF4 over 8 orbits in the F814W filter and 4 orbits in the F606W filter and NGC1052-DF2 over 20 orbits in each filter. 

Fig.~\ref{Fig:Field} presents a composite view of the data utilized in our study, with low surface brightness emission from Dragonfly given in black and white, high surface brightness objects from DECaLS displayed in color \citep{2004PASP..116..133L}, and a zoomed-in view of the two dark matter deficient galaxies from \textit{HST} shown in inset panels. The field is a beautiful collection of spiral, elliptical, and dwarf galaxies. Such groups of galaxies can have strong gravitational effects on their satellites, and an example is seen in Fig.~\ref{Fig:Field}:  NGC1047 is strongly disrupted by NGC1052.

%--------------------------------------------------------------

\begin{figure*}
    \centering
    \includegraphics[width=\textwidth]{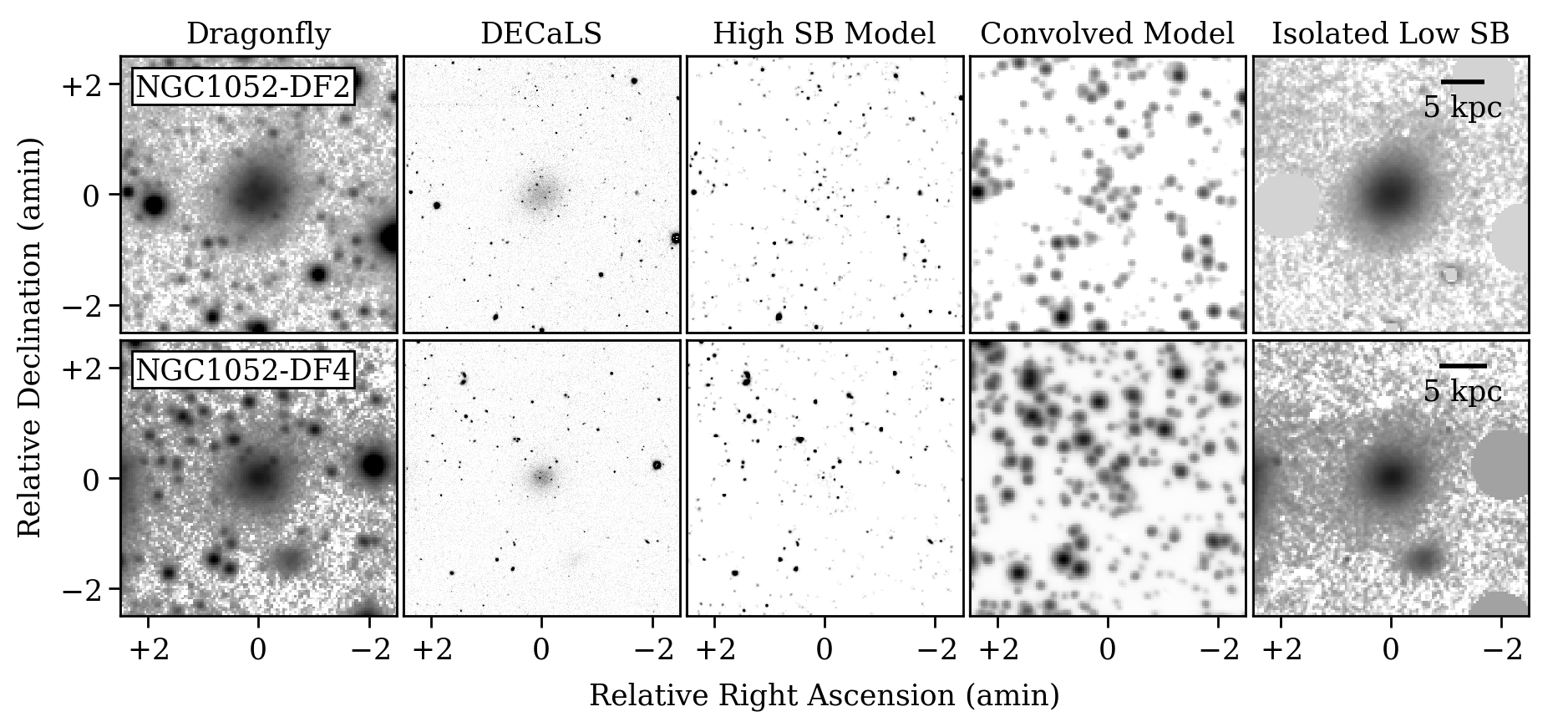}
    \caption{A visualization of MRF as we apply it to the NGC1052 field in the $r$-band for NGC1052-DF2 (top) and NGC1052-DF4 (bottom). \textit{Left panel}: The initial Dragonfly image with both high and low surface brightness emission. \textit{Second panel}: The initial, higher resolution DECaLS image. \textit{Third panel}: A flux model of compact objects detected in the high resolution image. \textit{Fourth panel}: The same flux model convolved with a characteristic kernel derived from bright stars in order to match Dragonfly resolution. \textit{Right panel}: The final, masked residual Dragonfly image after subtracting the convolved flux model. We adopt normalizations that best display the MRF technique. Right ascension and declination are given relative to $\alpha_{\rm DF2} = 02^{\mathrm h}41^{\mathrm m} 46\fs8$, $\delta_{\rm DF2} = -08^{\circ} 24^{\prime} 09\farcs7$ and $\alpha_{\rm DF4} = 02^{\mathrm h}39^{\mathrm m} 15\fs1$, $\delta_{\rm DF4} = -08^{\circ} 06^{\prime} 57\farcs6$ and scale bars are based on distances from \citet{2020ApJ...895L...4D} and \citet{2021ApJ...914L..12S}. \label{Fig:MRF}} % $\alpha_{\rm DF2} = 02^{\rm h}41^{\rm m} 47^{\rm s}$, $\delta_{\rm DF2} = -08^{\circ} 24^{\prime} 12^{\prime\prime}$ and $\alpha_{\rm DF4} = 02^{\rm h}39^{\rm m} 15^{\rm s}$, $\delta_{\rm DF4} = -08^{\circ} 06^{\prime} 58^{\prime\prime}$
\end{figure*}

\subsection{Multi-resolution Filtering} \label{Sec:MRF}

The presence of compact, high surface brightness sources like foreground stars, globular clusters, and galaxies complicates our search for low surface brightness tidal features, obscuring tidal tails and hindering fits to the ultra diffuse galaxies' emission. Thus, before continuing our analysis we must remove high surface brightness emission. However, while Dragonfly's sensitivity is optimal for imaging low surface brightness structures, its spatial resolution -- a PSF FWHM of 5$\arcsec$ -- is not sufficient to resolve many compact sources. This may lead to the misidentification of groups of stars and galaxies as low surface brightness emission. We therefore adopt the multi-resolution filtering (MRF) procedure of \citet{2020PASP..132g4503V}, using higher resolution DECaLS data to create a model of high surface brightness emission which is then matched to Dragonfly's resolution and subtracted. A visualization of key MRF steps as they are applied to NGC1052-DF2 and NGC1052-DF4 is given in Fig.~\ref{Fig:MRF}. The remainder of this section summarizes our application of this procedure.

First, to mitigate sampling effects we bin DECaLS $g$- and $r$-band images into 2$\times$2 pixel regions and convolve with a $\sigma$ = 1 pixel Gaussian. To account for differences in the instrumental response between the DECam and Dragonfly filter systems we apply a color correction
\begin{equation}
    I_{\nu_{\rm DF}}(\alpha, \delta) = I_{\nu_{\rm DCLS}}(\alpha, \delta) \times \left( \frac{I_{g_{\rm DCLS}}(\alpha, \delta)}{I_{r_{\rm DCLS}}(\alpha, \delta)} \right)^{\gamma(\nu)},
\end{equation}
that is, the $\nu_{\rm DCLS}$ band at a right ascension $\alpha$ and declination $\delta$ is multiplied by the ratio of DECaLS $g$- to $r$-band flux raised to a filter-dependent power of either $\gamma(g) = 0.05$ or $\gamma(r) = 0.01$ determined by comparing the colors of stars in DECaLS and Dragonfly.

Next, we generate a high surface brightness flux model by identifying compact sources in the DECaLS data with \citet{2016JOSS....1...58B}'s implementation of Source Extractor \citep{1996A&AS..117..393B}. Low surface brightness emission with $\mu > 25.5$ mag arcsec$^{-2}$ is removed from the high surface brightness flux model in order to make sure such low surface brightness phenomena are preserved in the Dragonfly data. Stars brighter than a magnitude of 17 are also removed from the high surface brightness flux model. These bright stars are saturated in the high-resolution image leading to inaccurate flux values in the model, and their emission may have a large spatial extent, complicating our convolution used to match Dragonfly's resolution.

We then create the kernel that is needed to convolve the flux model to Dragonfly's resolution. This is done by identifying isolated, non-saturated bright stars and taking the ratio of the Fourier transform of the low resolution image to the high resolution model. We select the median kernel generated from 24 stars, the brightest stars under an upper limit (a fraction $f_r = 1.5$\,\%, $f_g = 10$\,\% of the brightest 10 objects) and sufficiently circular to eliminate galaxies (an axis ratio of $b/a = 0.7$). We convolve our flux model with the median kernel to get an image that matches Dragonfly's resolution. We then subtract this, along with a constant sky background, from the Dragonfly image.

Next, remaining bright stars are removed by subtracting a model PSF. The interior $<30\arcsec$ of this PSF model is generated by stacking images of stars. The exterior of the model is generated from an aureole function, which is approximated by a composite of power-laws, matched to the flux of the interior PSF at the boundary. At the location of each star, the model PSF is normalized to the star's total flux as given in the Pan-STARRS PS1 catalog \citep{2016arXiv161205560C} and subtracted. Finally, remaining artifacts brighter than a magnitude of 16 are masked. Further details are provided in \citet{2020PASP..132g4503V,2022ApJ...925..219L}.

We perform an additional pre-processing step to ensure the accuracy of the flux of compact objects in the high surface brightness model. We subtract an initial model of galaxy emission for NGC1052-DF2, NGC1052-DF4, and NGC1052-DF5 from the high resolution DECaLS image before running MRF. This is done in two steps. First, we use aggressive background removal within Source Extractor (\texttt{BACK\_SIZE = 8}). Second, we model the remaining galaxy emission using Photutils (\citealt{larry_bradley_2020_4044744}'s Python implementation of the isophote fitting algorithm of \citealt{1987MNRAS.226..747J}). A description of this fitting method (as applied to Dragonfly data) is given in Section~\ref{Sec:Inspection}.

\subsection{Limiting Depth} \label{Sec:Deep}

We use the {\tt sbcontrast} method to estimate the surface brightness depth of the data. This method calculates the contrast between a path of a particular spatial scale and its immediate surroundings, providing a well-defined measure of surface brightness significance. An earlier version was initially released as part of the MRF package \citep{2020PASP..132g4503V}. We now provide a detailed explanation in Appendix~\ref{Sec:sbcontrast}, and make available an updated version via `{\tt pip install sbcontrast}.' Since the remainder of our work devotes itself only to the composite $g+r$ averaged image and studies tidal features which arise in galaxies at radii of $\approx40\arcsec-60\arcsec$ (diameters of $80\arcsec-120\arcsec$), we consider the 3$\sigma$ limit of the $g+r$ image on 60$\arcsec$ scales. The depth of the Dragonfly data on other scales are shown in Appendix~\ref{Sec:sbcontrast}. The $3\sigma$ depth is $\mu_{lim}(3\sigma,60\arcsec) = 29.9$ mag arcsec$^{-2}$. We note that this empirical limit is not nearly as faint as what one would obtain from an extrapolation of the pixel-to-pixel variance, as is sometimes done; that (incorrect) limit would be $\mu_{lim}(3\sigma,60\arcsec) = 31.6$ mag arcsec$^{-2}$. %  (e.g. \citealt{2020A&A...644A..42R})

%--------------------------------------------------------------
% Results
%--------------------------------------------------------------

\section{Results} \label{Sec:Results}

\subsection{Image Analysis} \label{Sec:Inspection}

\begin{figure}
    \centering
    \includegraphics[width=0.4525\textwidth]{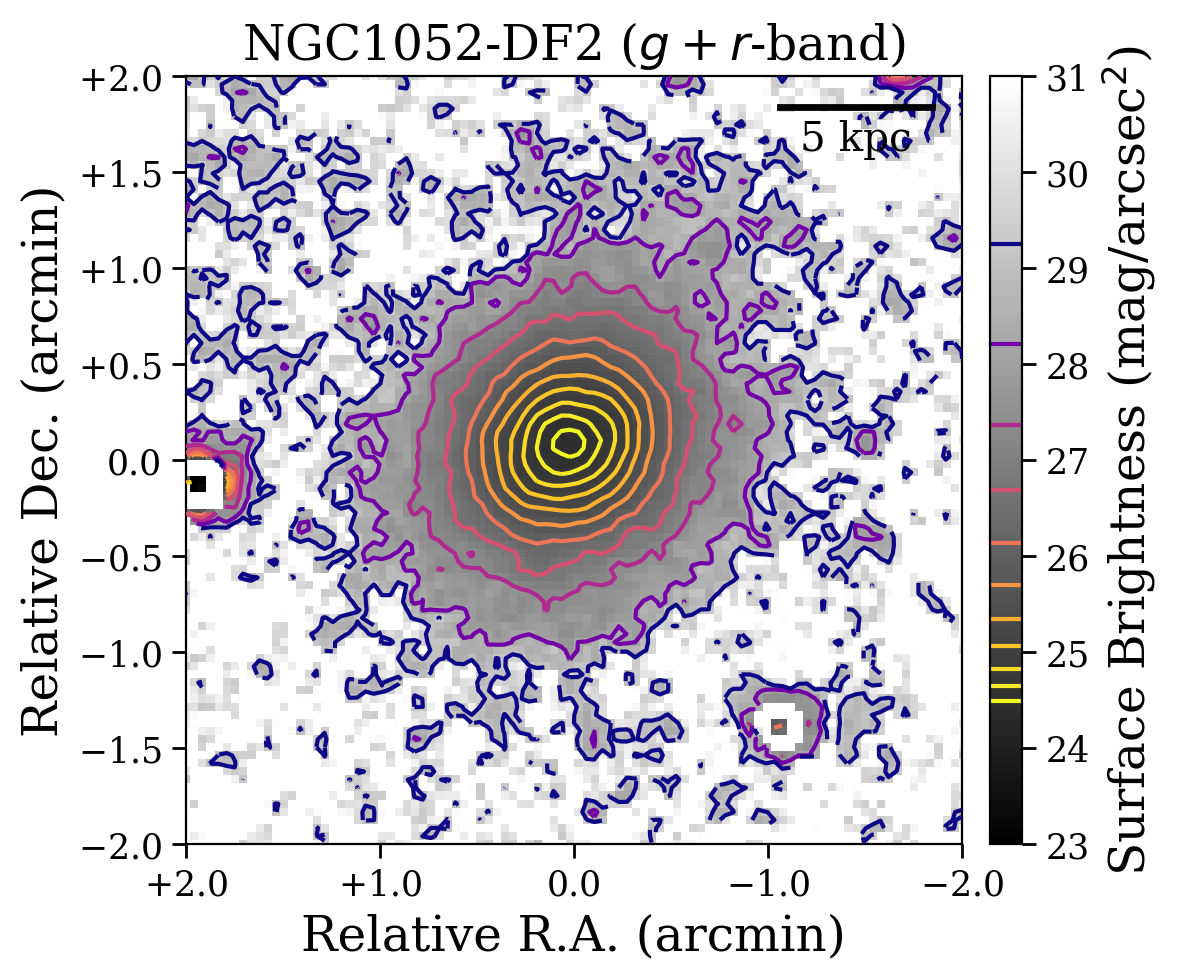} \\
    \includegraphics[width=0.4525\textwidth]{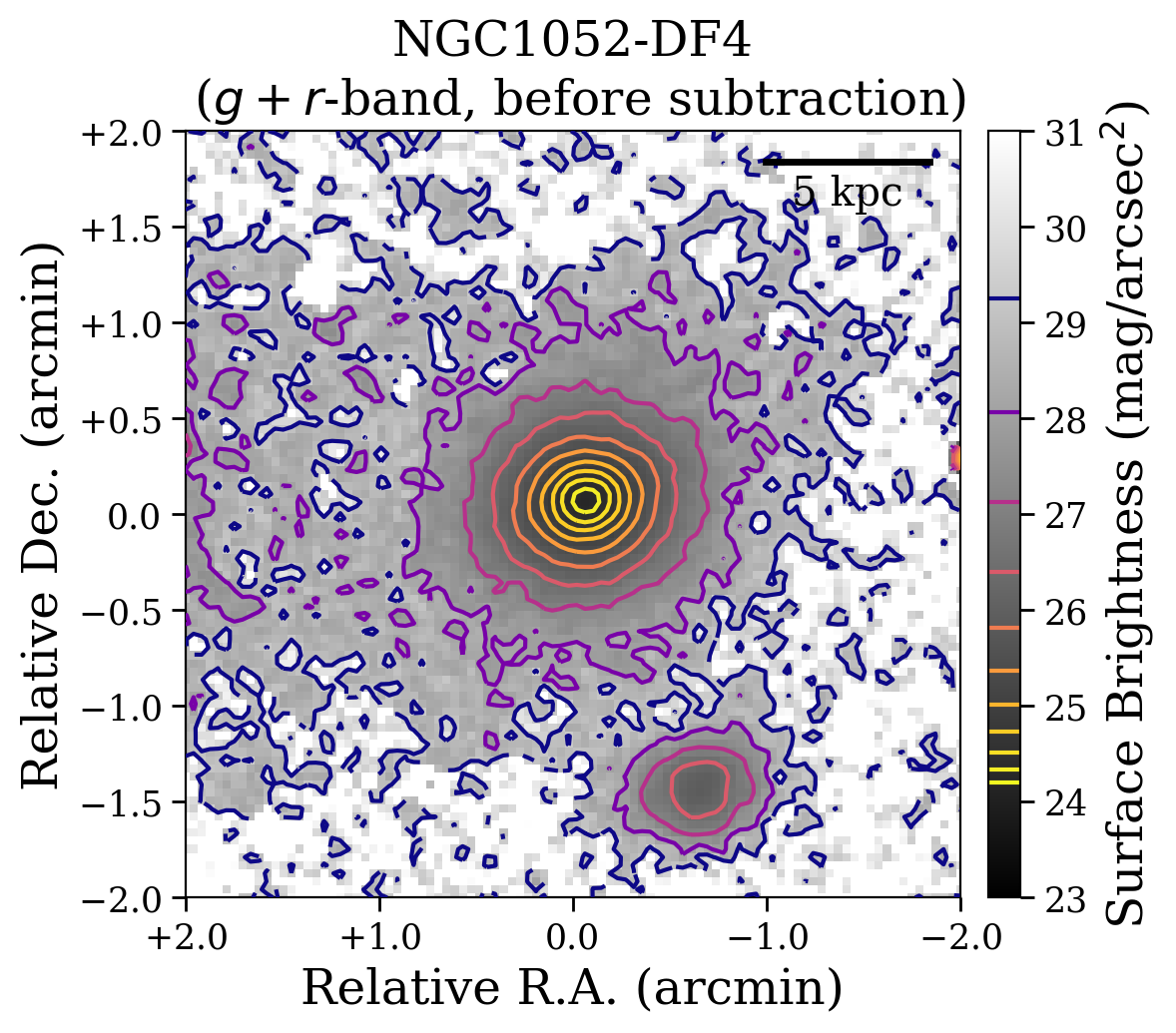} \\
    \includegraphics[width=0.4525\textwidth]{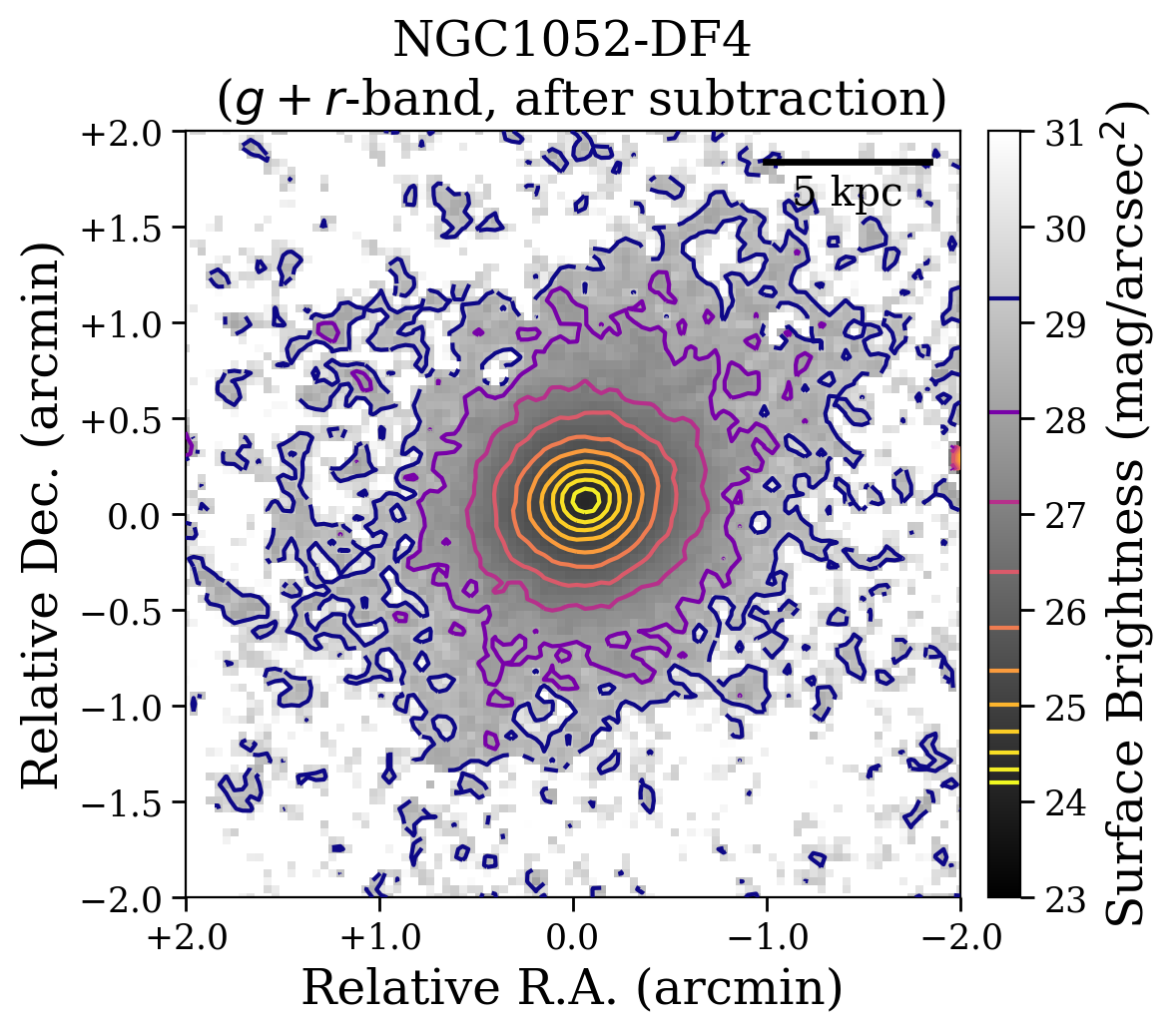}
    \caption{Low surface brightness maps of NGC1052-DF2 (\textit{top panel}) and NGC1052-DF4 before (\textit{middle panel}) and after (\textit{bottom panel}) subtracting models of NGC1052-DF5 and NGC1035. \label{Fig:Contours}}
\end{figure}

In Fig.~\ref{Fig:Contours} we present the surface brightness map of both dark matter deficient galaxies after applying MRF, averaging the $g$ and $r$ photometric bands to increase the signal-to-noise (S/N) ratio. We also apply a local background correction, subtracting the mean flux of the regions listed in Appendix~\ref{Sec:Background}. While the inner contours of NGC1052-DF2 and NGC1052-DF4 are round as in \citet{2018Natur.555..629V,2019ApJ...874L...5V}, the outer contours of both galaxies appear to become elliptical. This morphology is in agreement with the results of \citet[see Fig. 3]{2019A&A...624L...6M} and \citet{2020ApJ...904..114M}.

In the case of NGC1052-DF4, our morphological analysis is impeded by the presence of NGC1052-DF5 and NGC1035, whose light overlaps with NGC1052-DF4 in sky projection. Thus, in order to better understand NGC1052-DF4's outskirts and investigate the presence of external tidal features, we first model and subtract NGC1052-DF5 and NGC1035.

To generate our models, we make use of Photutils, \citet{larry_bradley_2020_4044744}'s Python implementation of the isophote fitting algorithm of \citet{1987MNRAS.226..747J}. This procedure fits the center, position angle, and ellipticity\footnote{Note that in this work we refer to ellipticity rather than eccentricity, with the former generally taking on smaller values to represent more elongated ellipses; e.g., an ellipticity of 0.2 corresponds to an eccentricity of 0.6.} of individual ellipses with fixed semimajor axis lengths to image data in order to identify the shape of isophotes of the same length. The procedure is repeated for ellipses with semimajor axes sampled geometrically (spaced by a factor of 1.1 by default, i.e. 10\%) between predefined minima and maxima. These were selected based on Dragonfly's PSF, visual inspection of residual images, and the ability of the algorithm to actually fit morphology, as regions with insufficient signal are modeled in a non-iterative mode (although such ellipses were still sometimes included to ensure reasonable residuals that did not leave behind any emission clearly belonging to the galaxy in question).

In order to simultaneously model the outskirts of NGC1052-DF4, NGC1052-DF5, and NGC1035 we adopt the following approach. First, we generate an initial model of NGC1052-DF4, subtract it from the original image, and fit a model of NGC1052-DF5 to the residual image. This NGC1052-DF5 model is then subtracted from our original image, allowing us to generate an improved model of NGC1052-DF4. We then subtract this improved NGC1052-DF4 model from our original image and fit a model of NGC1035 to the new residual image. After generating and subtracting a new model of NGC1052-DF4 from an image with both neighboring galaxies removed, we fit a final model of NGC1052-DF5. We find that this procedure has converged at this point, with further iterations leading to little model improvement. We show our resulting final models of NGC1052-DF5 and NGC1035, as well as the complete subtracted model including that from MRF, in Appendix~\ref{Sec:Residuals}. In Fig.~\ref{Fig:Contours}, we show the final image of NGC1052-DF4 with both neighboring galaxies removed.

\begin{figure*}
    \centering
    \includegraphics[width=\textwidth]{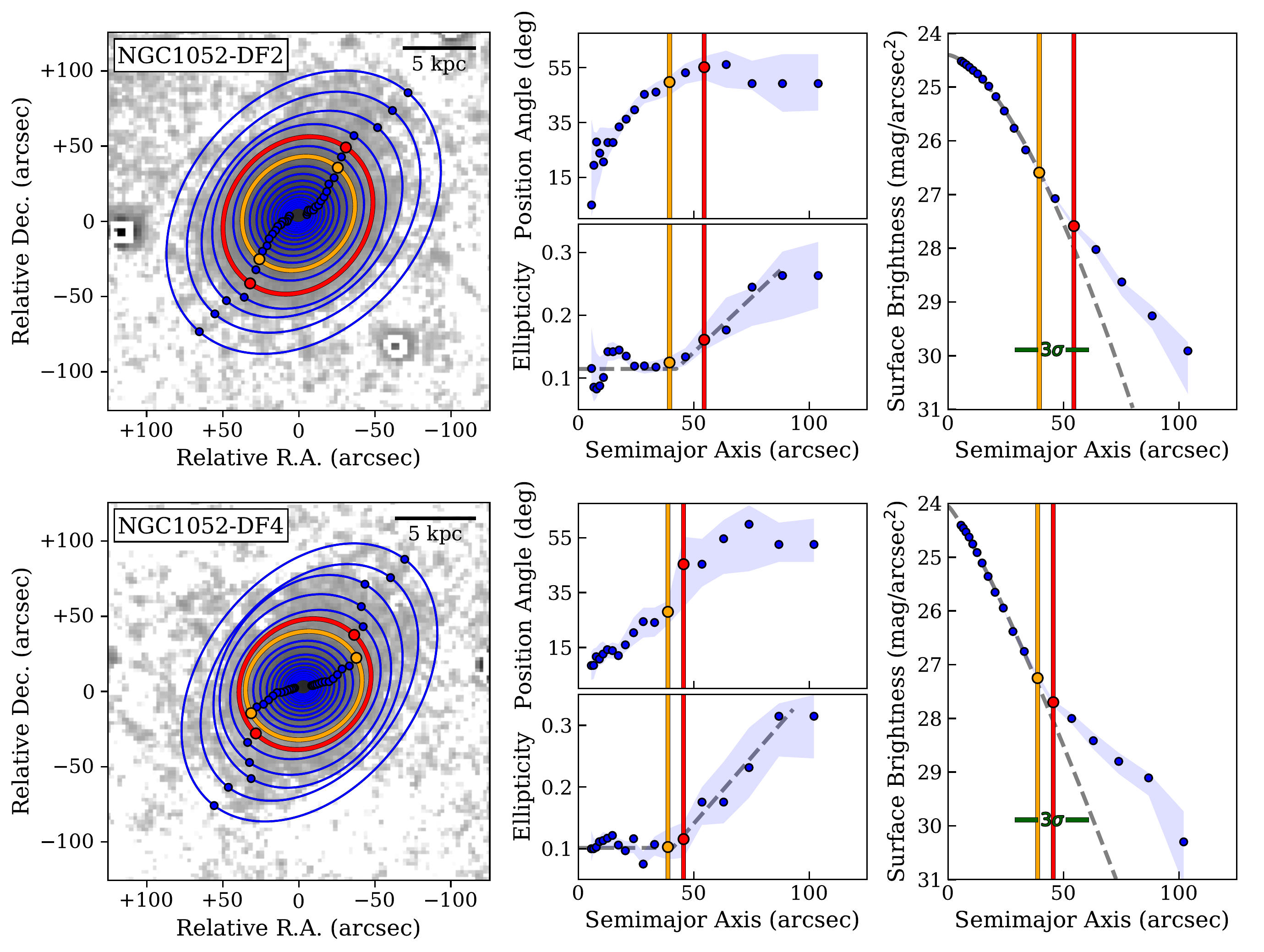}
    \caption{The morphology of NGC1052-DF2 (\textit{top}) and NGC1052-DF4 (\textit{bottom}) as derived from an isophotal analysis. \textit{Left panel}: best-fit elliptical isophotes of equal brightness, with vertices (\textit{circular markers}) given to illustrate trends in the position angle. \textit{Middle panel}: Position angle (\textit{top}) and ellipticity (\textit{bottom}) as a function of the semimajor axis of the fitted isophotes. The onset of distortions is determined through a fit (\textit{grey dashed line}) to Eq.~(\ref{Eq:Distort}), with the isophote nearest to $x_0$ indicated in all panels (\textit{orange circles and lines}). \textit{Right panel}: The surface brightness of each isophote, along with a S\'ersic fit (\textit{grey dashed line}). The tidal break is given as the innermost isophote that exceeds the S\'ersic fit by at least 0.2 mag arcsec$^{-2}$ and is indicated in all panels (\textit{red circles and lines}). As an estimate of uncertainty, we re-fit our model in 30 different locations, listed in Appendix~\ref{Sec:Error}, and report the 68\% confidence interval of values reached at a given semimajor axis length (\textit{blue shading}). We also indicate the $3\sigma$ surface brightness limit at 60$\arcsec$ scales as calculated in Section~\ref{Sec:Deep} (\textit{dark green bar}). Ellipses with diameters shorter than twice the PSF's FWHM are not shown and were excluded from our analysis. \label{Fig:Morphology}}
\end{figure*}

\subsection{Morphological Models} \label{Sec:Fit}

The contours of the two galaxies look remarkably similar, with both showing a gradual increase in the ellipticity and a gradual change in the position angle of the isophotes. We do not detect distinct tidal tails, suggesting
that we may have not directly observed the tidal radius of an ongoing interaction where unbound material is being stripped away. Still, the elongated contours in Fig.~\ref{Fig:Contours} may indicate material still bound to the galaxies being actively distorted or relic distortions which arose at pericentric passage.

To quantify these aspects we examine the shape of the two galaxies using the elliptical isophote fitting method described above in Section~\ref{Sec:Inspection}. To improve the precision of our morphological model, we further restrict our convergence criteria (so that either the terms representing the imperfection of an elliptical isophote must be below 1\% of the rms or the number of fitting iterations for an individual isophote reaches a maximum of 1,000) and increase the geometric step size to 17.5\% for a better signal to fit the galaxies' outskirts while maintaining sufficient radial coverage. The results are shown in Fig.~\ref{Fig:Morphology}; the fit models and residuals are given in Appendix~\ref{Sec:Residuals}. To provide an estimate of the uncertainty for isophote position angle, ellipticity, and surface brightness we inserted our originally fit model into 30 different locations near NGC1052 as listed in Appendix~\ref{Sec:Error} and re-fit the galaxy in this new environment using the same set of semimajor axis lengths. The shading in Fig.~\ref{Fig:Morphology} represents the fifth-largest and fifth-smallest instances of each quantity from these 30 fits (i.e, the approximate 68\% confidence interval).

As with the contours in Fig.~\ref{Fig:Contours}, in Fig.~\ref{Fig:Morphology} we see that both galaxies' isophotes become elliptical in their outer regions. There is a visually identifiable location where the galaxies begin a roughly linear increase in their ellipticity. Moreover, the isophotes become steadily more twisted as well, with the position angles of the galaxies' outermost regions significantly rotated compared to their interior. The morphological profiles appear to be similar, though NGC1052-DF4 reaches a higher ellipticity at comparable radii and becomes slightly more twisted compared to its orientation at inner radii. These similar, significant stretches and twists for both galaxies are unlikely to come from identical projection effects, and imply that the galaxies are experiencing, or have experienced, a similar underlying tidal field. There also is a point, again similar for both NGC1052-DF2 and NGC1052-DF4, where the galaxies' surface brightness profiles deviate from S\'ersic fits. We note that the profiles of the two galaxies do differ, with a S\'ersic index $n_{\rm DF2} = 0.64$ for NGC1052-DF2 compared to $n_{\rm DF4} = 0.85$ for NGC1052-DF4 and an effective (half light) radius $R_{\rm e,DF2} = 24\farcs 8$ compared to $R_{\rm e,DF4} = 19\farcs 8$. To mitigate the effect of the PSF, we only included isophotes with radii greater than 5$\arcsec$, i.e. diameters over twice the FWHM, in our model. We find that adopting the 2D S\'ersic model from \citet{2018Natur.555..629V,2019ApJ...874L...5V}, which accounted for the PSF using GALFIT's convolution technique \citep{2002AJ....124..266P}, has little effect on our findings. We note that Dragonfly has well controlled outer PSF wings due to anti-reflection coatings and highly baffled telephoto lenses \citep{2014PASP..126...55A} which would not affect the observed tidal features, especially given the galaxies' faint centers. Indeed, while scattered light from the galaxies' centers may contribute to the observed emission at their outskirts, in Appendix~\ref{Sec:PSF} we show that this PSF up-bending is negligible compared to the observed tidal break and lies below the limiting depth of our image. This reflects \citet{2020MNRAS.495.4570M}'s finding that Dragonfly's PSF has a negligible effect on the outskirts of TNG100 galaxies.

To provide a quantitative estimate for the location where tidal distortions arise, we fit the ellipticity profile $\varepsilon(r)$ of the two galaxies with the function
\begin{equation} \label{Eq:Distort}
    \varepsilon(r) =
        \left\{ \begin{array}{ll}
            \varepsilon_0 + a \times (r-r_{\rm distort}) & \text{if } r > r_{\rm distort}\\
            \varepsilon_0 & \text{if }  r \leq r_{\rm distort}
        \end{array} \right.
\end{equation}
using a horizontal line to account for the undisturbed region and an inclined line to represent the distorted outskirts, labeling their intersection $r_{\rm distort}$. We exclude the outermost isophote from this fit since it was modeled in the non-iterative mode. The onset of distortions occurs at $\approx$40$\arcsec$ in both galaxies, though the 68\% interval across the 30 re-fit models (used to estimate the errors in Fig.~\ref{Fig:Morphology}) differ slightly with $\theta_{\rm distort,DF2} = 39\arcsec^{+7}_{-10}$ and $\theta_{\rm distort,DF4} = 39\arcsec^{+10}_{-6}$. Taking into account distance measurement uncertainty \citep{2020ApJ...895L...4D,2021ApJ...914L..12S}, this corresponds to physical sizes of $r_{\rm distort, DF2} = 4^{+1}_{-1}$ kpc and $r_{\rm distort, DF4} = 4^{+1}_{-1}$ kpc. While this is technically within the detectable limits of previous works, the inflection point in the ellipticity profile was only identifiable due to the increasing $\varepsilon(r)$ beyond $r_{\rm distort}$. This would have been imperceptible in studies which were not able to detect structure out to $\gtrsim$75$\arcsec$, such as the initial Dragonfly image used to study NGC1052-DF2 and NGC1052-DF4.

As explained in Section~\ref{Sec:Tides}, this distortion radius may represent material, either bound or unbound, being disrupted by an ongoing interaction, or a disturbance that occurred at pericentric passage which has not yet relaxed back to an equilibrium state (although this may still be captured by the results of simulation unless the orbit is highly elliptical or unbound). Thus, $r_{\rm distort}$ may represent the approximate tidal radius at pericenter. In order to extract an upper bound of the instantaneous tidal radius, we must investigate the galaxies' surface brightness profiles. 

To pinpoint the region where the galaxies' surface brightness profiles have an excess of light, we fit a S\'ersic profile to the isophotes' surface brightness and identify $r_{\rm break}$ as the innermost isophote that exceeds the fit by 0.2 mag\,arcsec$^{-2}$ (as in \citealt{2020ApJ...904..114M}). We determine that the break in the galaxies' surface brightness, approximately half of the tidal radius upper bound, occurs at $\theta_{\rm break,DF2} = 55\arcsec$ in NGC1052-DF2 and $\theta_{\rm break,DF4} = 46\arcsec$ in NGC1052-DF4. Due to the low uncertainties in the surface brightness profiles at this radius, all re-fit models in the 68\% confidence interval occur at exactly this radius. The main uncertainty is the distance; using the distance uncertainties \citep{2020ApJ...895L...4D,2021ApJ...914L..12S} we find $r_{\rm break, DF2} = 5.9\pm0.3$ kpc and $r_{\rm break, DF4} = 4.4\pm0.4$ kpc.

\subsection{Comparison to \textit{HST} Star Counts} \label{Sec:HST}

Before moving to an interpretation of our results we test the central finding of significant radial change in the isophotes using star counts in \textit{HST} images. Dragonfly's large pixel size, and the necessary subtraction of compact sources using the MRF algorithm, make it difficult to study the galaxies' innermost regions. To circumvent this limitation we utilize the \textit{HST} observations from \citet{ 2021ApJ...914L..12S} and \citet{2020ApJ...895L...4D}, whose depth and high resolution facilitated the identification of individual red giants in both NGC1052-DF2 and NGC1052-DF4. This allows us to characterize the morphologies of the galaxies' central regions directly from the distribution of their stars. 

\begin{figure}
    \centering
    \includegraphics[width=0.5\textwidth]{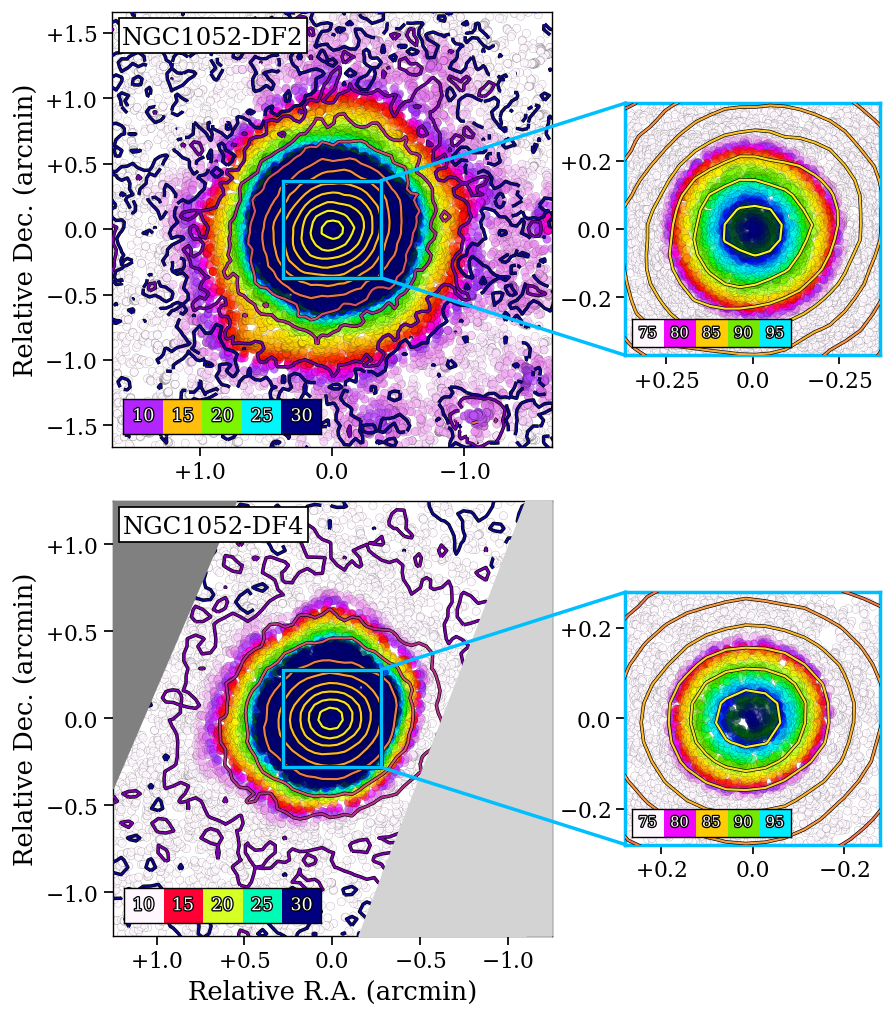}
    \caption{Red giants in NGC1052-DF2 (\textit{top panel}) and NGC1052-DF4 (\textit{bottom panel}) which were individually resolved by deep \textit{HST} observation, colored according to local stellar density. Contours from Fig.~\ref{Fig:Contours} are overlaid for comparison to Dragonfly, and zoom-in panels are included to aid in examination of central regions. Color scales are given as percentages of the peak density. Light grey shading indicates regions where star counts were not taken, and dark grey shading indicates the \textit{HST} image boundary. We caution that density estimates near this boundary may be unreliable due to high noise levels. \label{Fig:HST}}
\end{figure}

In Fig.~\ref{Fig:HST} we plot each galaxies' resolved red giant population, colored according to local stellar density. The stars were  identified with DOLPHOT \citep{2000PASP..112.1383D} and filtered to ensure reliable photometry using the quality cuts described by \citet{2021ApJ...914L..12S} and \citet{2020ApJ...895L...4D}, with a final total of 19,001 stars in NGC1052-DF2 and 9,287 for NGC1052-DF4 (note that the images were taken over 40 and 12 orbits respectively, with the former achieving a higher depth). Stellar density was determined using a kernel density estimate, summing over Gaussian-convolved stars to find the density at a given point. Applying the same isophote fitting technique we used in modeling Dragonfly data to the density map, we find that both galaxies have a central ellipticity of $\varepsilon \lesssim 0.1$. Thus, we confirm that the innermost regions of both galaxies are indeed relatively spheroidal, as seen in the inner contours of Fig.~\ref{Fig:Contours} and the findings of previous works \citep{2018Natur.555..629V,2019ApJ...874L...5V,2019A&A...624L...6M}.

\subsection{The Current Tidal Radius} \label{Sec:Current}

As explored in Section~\ref{Sec:Tides}, low mass satellites are easily disrupted by tidal fields, while high mass satellites must be close to a perturbing galaxy to be tidally stripped. Thus, the identification of tidal features may be used to characterize NGC1052-DF2 and NGC1052-DF4's orbital position and mass. To make such calculations, we adopt two approaches. In this section, we consider an ongoing interaction and explore the present day tidal radius $r_{\rm tid}$ as estimated from the profile break location $r_{\rm break}$. In Section~\ref{Sec:Pericenter} we consider a scenario where the distortions were created at the time of the most recent pericenter passage. We then interpret these results in the context of the galaxies' masses and relative distances.

%\subsubsection{Distance Implications} \label{Sec:Distance}

A first item of interest is the distance between the galaxies along the radial direction from earth, i.e. the line-of-sight. The similar distortions observed in Section~\ref{Sec:Fit} implies that the galaxies are experiencing a similar underlying tidal field, as expected if both are being disturbed by NGC1052. In this case, for a fixed satellite dark matter halo mass there is a maximum distance that the galaxies may be from NGC1052 that would be consistent with the observed tidal distortions. That is, the largest possible tidal radius inferable from our measurements determines the farthest orbital distance NGC1052-DF2 and NGC1052-DF4 could be from NGC1052. This can then be converted to a maximum distance along the line-of-sight, by also considering the galaxies' projected separation on the sky.\footnote{We may also use $r_{\rm distort}$ to find the minimum possible line-of-sight separation from NGC1052 a satellite could be and exhibit the observed tidal features, although we find this is only relevant in the case that NGC1052-DF2 is almost completely dark matter free. See Section~\ref{Sec:Pericenter} for continued discussion.} 

\begin{figure}
    \centering
    \includegraphics[width=0.47\textwidth]{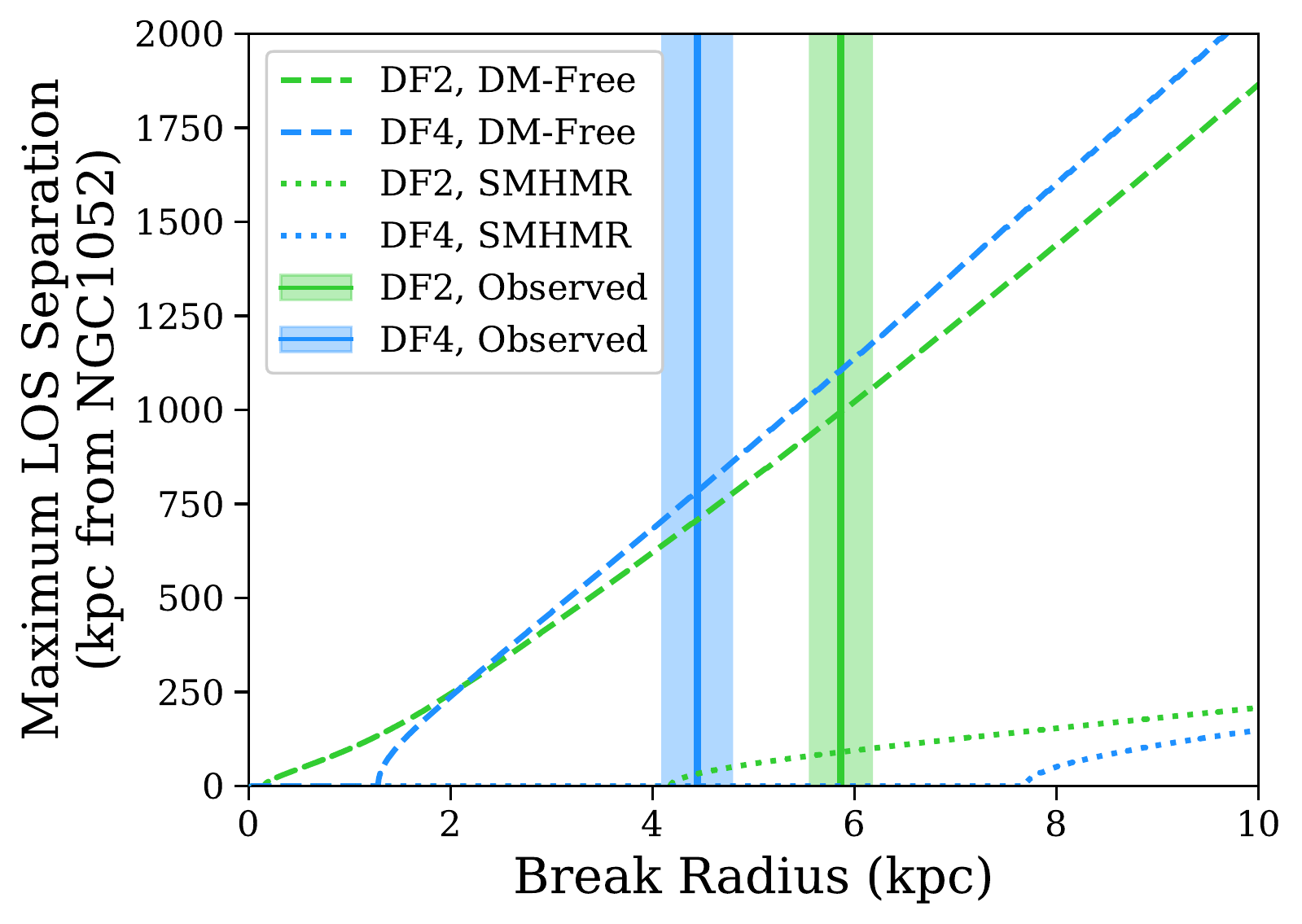}
    \caption{Deriving maximum line-of-sight separations from NGC1052 consistent with our observation. For a range of $r_{\rm break}$ values we infer what the maximum line-of-sight from the satellite (\textit{DF2 in green, DF4 in blue}) to NGC1052 would be using an integrated S\'ersic profile as a stellar mass distribution and either no dark matter (\textit{dashed lines}) or a dark matter halo with an SMHMR from \citet{2013ApJ...770...57B} and the same NFW parameters as in Fig.~\ref{Fig:Tides} (\textit{dotted lines}). The observed $r_{\rm break}$ (\textit{solid lines}) is included with shading to represent distance measurement uncertainties and (negligible) random uncertainties. Note that this does not capture definitional uncertainty for the tidal radius. \label{Fig:rbreak}}
\end{figure}

Since we are interested in the maximum line-of-sight distance between the galaxies and NGC1052, we begin by considering the maximum possible 3D, absolute orbital distance a satellite can be from NGC1052,  $R_{\rm max}$, as given by solving for $R = R_{\rm max}$ in Eq.~(\ref{Eq:rtid}) where $M(R)$ is NGC1052's mass profile. We take $m(r_{\rm tid})$ as the sum of $m_\star(r_{\rm tid})$, the satellite's enclosed stellar mass, and $m_h(r_{\rm tid})$, the satellite's enclosed dark matter, each at the tidal radius. We find $m_\star(r)$ by integrating over each galaxies' S\'ersic profile and assuming a stellar mass-to-light ratio as in \citet{2018Natur.555..629V,2019ApJ...874L...5V} that is constant across the galaxy, which is supported by the lack of radial variation in the galaxies' color (although, since the majority of light is contained within the tidal features, $m_\star(r_{\rm tid})$ may also be well approximated by point masses). We ignore the galaxies' gaseous content since this has been measured to be  negligible compared to the stellar mass \citep[lower by 2--3 orders of magnitude]{2019MNRAS.482L..99C,2019ApJ...871L..31S}. We take $m_h(r)$ for a given total mass $m_{\rm vir}$ from an NFW profile with a concentration $c = 9.6 \times (m_{\rm vir}/10^{13}\textrm{ M}_\odot)^{-0.13}\times(1+z)^{-1}$ (appropriate for satellite galaxies following \citealt{2005ApJ...635..931B}) and a virial radius $r_{\rm vir} = \left( G m_{\rm vir} / 100 H^2(z) \right)^{1/3}$. $M(R)$ is taken from \citet{2019MNRAS.489.3665F}, an NFW profile with $c$ = 7.0, $M_{\rm vir}$ = 6.2$\times$10$^{12}$ M$_\odot$, $R_{\rm vir}$ = 390 kpc, and a 4$\times$10$^{11}$ M$_\odot$ central baryonic component. We allow $M(R)$ to continue increasing beyond $R_{\rm vir}$ to account for other objects within the group contributing to NGC1052's tidal field such as NGC1042 and NGC1047 \citep{2019RNAAS...3...29V}. Since we are interested in the maximum value of $R$, we take $\Omega = \sqrt{2}V_{\rm circ}/R$ (note that even highly elliptical bound orbits will only approach this angular speed at pericenter, so that our resulting $R_{\rm max}$ is an overestimate) and $r_{\rm tid} = 3 \times r_{\rm break}$ to account for decreased interaction strength far from pericenter (the approximate maximum value of $r_{\rm tid}/r_{\rm break}$ reached for a simulation of a highly elliptical orbit with a apocenter-to-pericenter of 20; \citealt{2013MNRAS.433..878L}). To solve Eq.~(\ref{Eq:rtid}) and find the maximum line-of-sight distance, we create a radial grid well sampled in log-space representing a wide range of possible $R_{\rm max}$ values. We calculate Eq.~(\ref{Eq:rtid}) from $M(R_i)$ at each gridpoint and where $R_{i}$ is closest to the corresponding $R_{\rm max}$. Finally, we use the satellite's sky position and trigonometric relations to find the corresponding separation along the radial direction from earth.

In Fig.~\ref{Fig:rbreak} we repeat this procedure to find the maximum offset from NGC1052 along the line-of-sight for both an $m_{\rm vir}$ from the stellar mass-halo mass relation (SMHMR, taken from \citealt{2013ApJ...770...57B}) and for a galaxy where the dark matter content is negligible. We find that, if NGC1052-DF4 had a `normal' dark matter content, it would need to be \textit{closer to NGC1052 than it appears even in projection on the sky} to have the observed $r_{\rm break}$ (or any $r_{\rm break}\lesssim$7.5 kpc for that matter). This strongly supports the conclusion that NGC1052-DF4 is dark matter deficient. Similarly, NGC1052-DF2 would need to be well within NGC1052's virial radius at a line-of-sight separation of $\lesssim$100 kpc (a difference caused by NGC1052-DF4's greater sky projected distance). This would imply a small DF2-DF4 line-of-sight separation which is $4\sigma$ lower than the central value of the tip of the red giant branch estimate of \citet{2021ApJ...914L..12S}. However, if the galaxies indeed are dark matter deficient, NGC1052-DF2 and NGC1052-DF4 may be offset from NGC1052 by up to 850 and 1050 kpc, respectively, which together is within the error bounds of \citet{2021ApJ...914L..12S}. 

\begin{figure}
    \centering
    \includegraphics[width=0.47\textwidth]{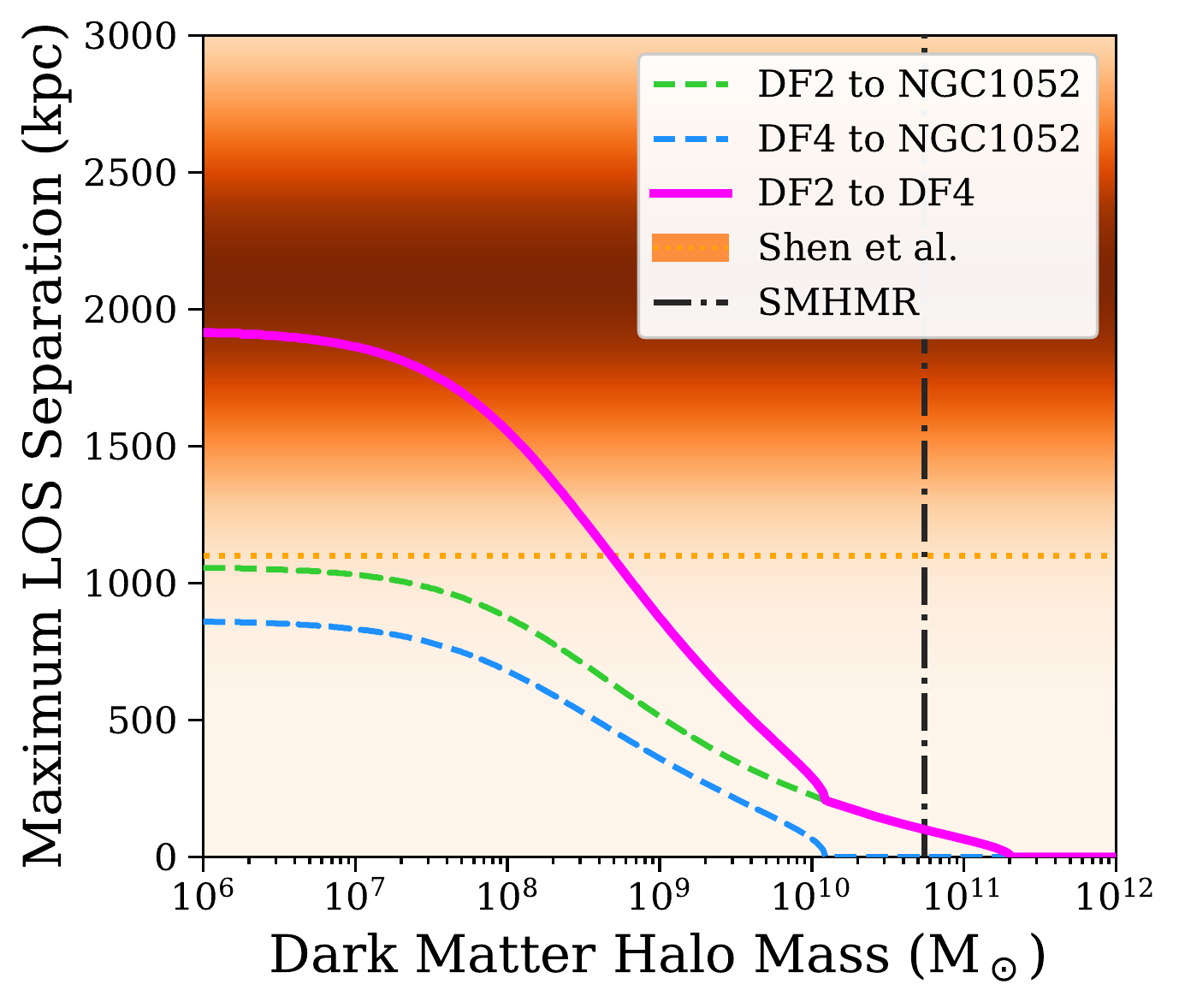}
    \caption{Deriving a dark matter mass upper limit from maximum line-of-sight separations. Assuming a fixed stellar mass profile, for a given dark matter mass we calculate the largest possible line-of-sight distances from NGC1052 to NGC1052-DF2 (\textit{green dashed line}) and NGC1052-DF4 (\textit{light blue dashed line}) that would still be consistent with our observed tidal radius. We add these together (\textit{magenta solid line}) and compare to \citet{2021ApJ...914L..12S}'s DF2-to-DF4 measurement (\textit{orange shading with dotted line at $-2\sigma$}). We emphasize that the x-axis does not include stellar mass, which is left fixed and is dominant at low halo masses. \label{Fig:DM}}
\end{figure}

Thus, by combining our observed tidal distortions with \citet{2021ApJ...914L..12S}'s distance constraint, we may obtain an independent constraint on the galaxies' masses. \citet{2021ApJ...914L..12S} found that -- while their absolute distances have a large uncertainty of $\approx 1.5$\,Mpc -- NGC1052-DF2 and NGC1052-DF4 have a well constrained {\em separation} of 2.1$\pm$0.5 Mpc along the line-of-sight, implying that one or both galaxies is at considerable distance away from NGC1052. However, as we found above, if the galaxies have a highly massive dark matter halo, they cannot be significantly offset from NGC1052 along the line-of-sight. Considering these facts along with the galaxies' stellar content, we may draw further implications about their dark matter. 

For the halo mass of NGC1052-DF2 and NGC1052-DF4 to be consistent with \citet{2021ApJ...914L..12S}, the maximum line-of-sight distances of the two galaxies added together should be above the lower bound of the DF2-DF4 distance constraint. However, if this dark matter halo mass is significantly larger than the true value, then even if the galaxies are on opposite sides of NGC1052, at the maximum distance along the line-of-sight consistent with our observations, they could not be as far apart as measured by \citet{2021ApJ...914L..12S}.

In Fig.~\ref{Fig:DM}, we take our observed $r_{\rm break}$ and consider a range of $m_{\rm vir}$ to find which dark matter halo masses are within $2\sigma$ agreement with \citet{2021ApJ...914L..12S}. We use the same procedure as described above, now adjusting $m_{\rm vir}$ of NGC1052-DF2 and NGC1052-DF4, while keeping their stellar mass, $r_{\rm break}$, and NGC1052's mass profile fixed. In theory, one could derive separate conditions for both NGC1052-DF2 and NGC1052-DF4 by assuming one had no dark matter. However, for the sake of simplicity and realism we assume that they each have approximately the same halo mass and estimate a single constraint for both galaxies. Indeed, we expect this to be the case given their similarities, in particular their stellar masses.

We find that if the galaxies have a dark matter halo above $\approx$5$\times$10$^8$ M$_\odot$, (a total mass of $\approx$7$\times$10$^8$ M$_\odot$), even if the galaxies are on opposite sides of NGC1052 at their maximum possible distance along the line-of-sight they cannot be within 2$\sigma$ of the central DF2-DF4 separation measurement of \citet{2021ApJ...914L..12S}. Taking our results at face value and ignoring systematic errors, this suggests that the galaxies are not just dark matter deficient (compared to the SMHMR) but nearly dark matter free.

\subsection{The Tidal Radius at Pericenter} \label{Sec:Pericenter}

In the previous section we assumed that the observed tidal features could be related to the instantaneous tidal radius using the results of numerical simulation (accounting for the fact that, for highly elliptical orbits with apocenter-to-pericenter ratios $\geq 20$, beyond pericenter breaks may be observed a factor of $\approx$3 inwards of $r_{\rm tid}$; \citealt{2013MNRAS.433..878L}). Given the limitations discussed in Section~\ref{Sec:Tides}, in this section we instead explore the assumption that the onset of distortions $r_{\rm distort}$ is approximately the tidal radius at pericentric passage $r_{\rm tid}(R_{\rm peri})$, and that subsequent interaction as well as the dissipation of these features through mechanisms like phase mixing had a negligible effect. In this case, we may use Eq.~(\ref{Eq:rtid}) to provide a range of possible values for the galaxies' closest approach to NGC1052.

\begin{figure}
    \centering
    \includegraphics[width=0.47\textwidth]{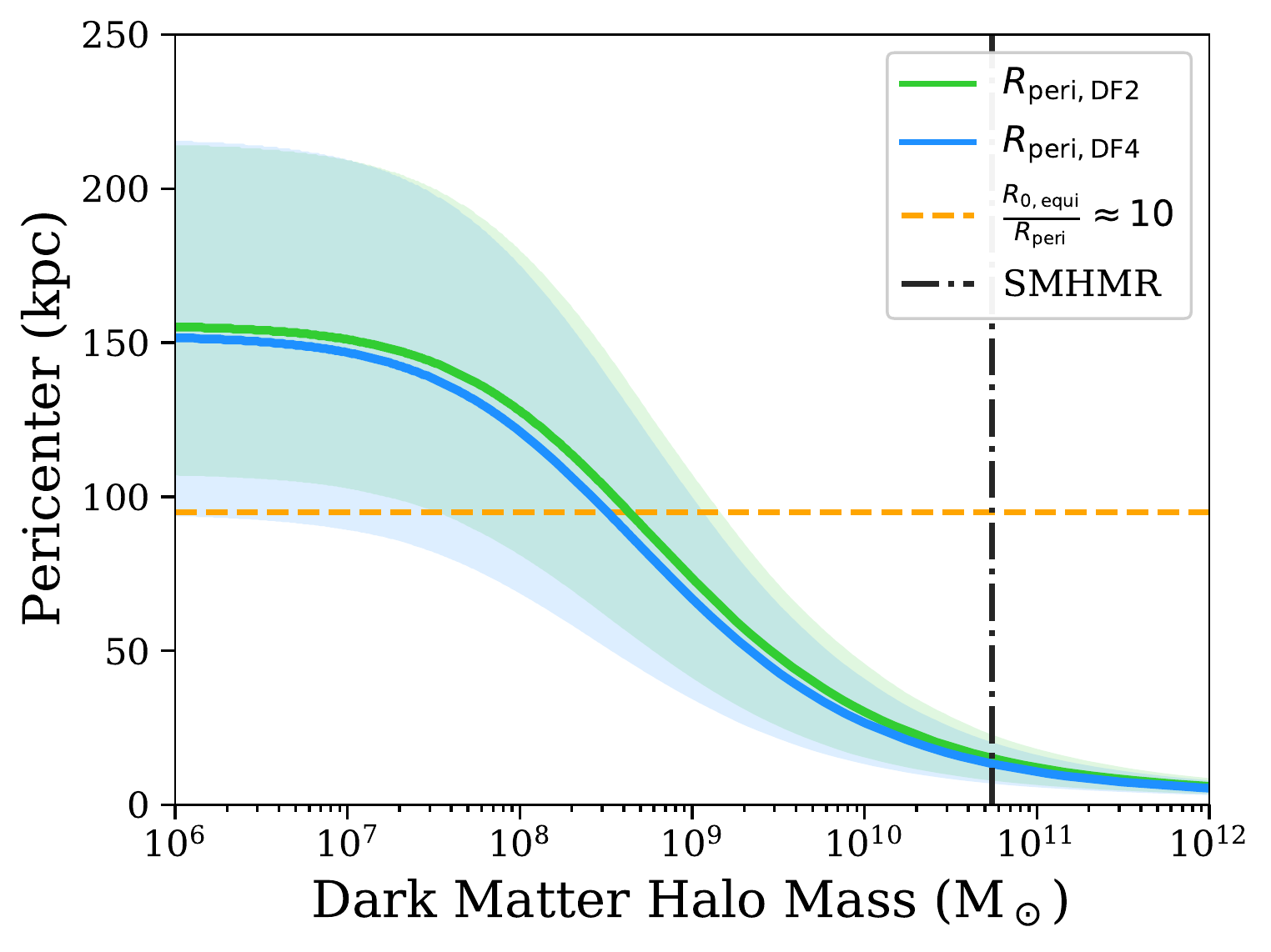}
    \caption{Deriving constraints on the orbital location of pericentric passage. For a fixed stellar mass profile, we calculate $R_{\rm peri}$ at given dark matter masses for NGC1052-DF2 (\textit{green}) and NGC1052-DF4 (\textit{light blue}), with shading to account for a range of orbital ellipticities and the uncertainty in $r_{\rm distort}$. We include a comparison to the SMHMR (\citealt{2013ApJ...770...57B}; \textit{black, dash-dotted line}) and a pericenter which is one-tenth the present day orbital position (assuming DF2 and DF4 are equidistant to NGC1052), which is approximately the 68\% confidence limit for generic orbits (\citealt{1999ApJ...515...50V}; \textit{orange, dashed line}). We emphasize that the x-axis does not include stellar mass, which is left fixed and is dominant at low halo masses. \label{Fig:Peri}}
\end{figure}

If the orbit is bound, the angular speed at pericenter may be no smaller than $\Omega_{\rm min} = V_{\rm circ}/R_{\rm peri}$ and no greater than $\Omega_{\rm max} = \sqrt{2}V_{\rm circ}/R_{\rm peri}$, since $\vec{V}$ is perpendicular to $\vec{R}$ at pericenter. Using Eq.~(\ref{Eq:rtid}) and \citet{2019MNRAS.489.3665F}'s NFW profile, in Fig.~\ref{Fig:Peri} we draw 1,000 realizations each of $\Omega$ and $r_{\rm distort}$ and report the 68\% confidence interval for pericenters. $\Omega$ is drawn from a uniform distribution between $\Omega_{\rm min}$ and $\Omega_{\rm max}$, while $r_{\rm distort}$ is drawn from a Gaussian distribution based on the mean and uncertainties from Section~\ref{Sec:Fit}. We include a comparison to $R_{\rm peri} \approx 100$ kpc, as this is approximately one tenth of their current orbital position assuming they are equidistant to NGC1052, whereas \citet{1999ApJ...515...50V} found that only $\lesssim$15\% of generic orbits should be more elliptical than $R_{\rm apo}/R_{\rm peri} = 10$.

Several implications may be drawn from these ranges. First, such 10$^2$ kpc-scale pericenter estimates are over an order magnitude less than \citet{2021ApJ...914L..12S}'s 2.1$\pm$0.5 Mpc DF2-DF4 distance constraint. Thus, if $r_{\rm distort}$ indeed represents $r_{\rm tid}$ at pericentric passage, at least one of the galaxies has traveled great distances of order 10$^3$ kpc since pericenter and was once much closer to NGC1052 than it is today. Moreover, if the galaxies have the dark matter content that is expected from the  stellar mass-halo mass relation \citep{2013ApJ...770...57B}, they would need to have had extremely close pericentric passages to show the observed distortions. It is extremely unlikely that both galaxies independently have an orbit with a present-day radius to pericenter ratio $R_{\rm peri}/R_0 \sim 0.01$.

%--------------------------------------------------------------
% Discussion
%--------------------------------------------------------------

\section{Discussion} \label{Sec:Discussion}

\subsection{Constraints from the Tidal Analysis}

In this work we found clear evidence of tidal features in NGC1052-DF2 and NGC1052-DF4, as would be expected for low mass dark matter deficient galaxies in a group environment. Both exhibited morphological distortions, becoming twisted and elongated in their outskirts, as well as breaks in their surface brightness profiles.\footnote{Note that such twists are not likely to be associated with spiral density waves for NGC1052-DF2 and NGC1052-DF4, namely as the galaxies do not appear to be disks. Moreover, such low mass galaxies have little differential rotation to drive spiral structure.} We considered two approaches for our analysis of these tidal features.

In Section~\ref{Sec:Current}, we assumed that $r_{\rm break}$ could be related to the instantaneous tidal radius $r_{\rm tid}$ as computed from the galaxies’ current orbital position. We found that, if the galaxies had a `normal' amount of dark matter, they must both be quite close to NGC1052, in conflict with the observed distance between DF2 and DF4 of 2.1 Mpc. In the case of NGC1052-DF4 the inferred distance to NGC1052 is even closer than the galaxy appears in sky projection. However, if the galaxies are dark matter free, they can each be over 1 Mpc away and still have the observed distortions due to NGC1052's tidal field. Comparing these distances to the $2.1\pm0.5$ DF2-DF4 distance constraint by \citet{2021ApJ...914L..12S}, we found that even if the galaxies are on opposite sides of NGC1052, they can only be in 2$\sigma$ consistency with \citet{2021ApJ...914L..12S} if they have a dark matter halo mass of $\lesssim$5$\times$10$^{8}$ M$_\odot$.

In Section~\ref{Sec:Pericenter}, we instead assumed that $r_{\rm distort}$ is equal to $r_{\rm tid}$ at pericenter. We found that, no matter what dark matter content the galaxies have, their pericenters are far less than the current DF2-DF4 separation. Thus, the galaxies are likely on extreme orbits. Assuming the galaxies are approximately equidistant from NGC1052, this implies a minimum apocenter-to-pericenter ratio of 5 even in the dark matter free case. However, if the galaxies had a normal amount of dark matter, they would need to have passed within $\approx$20 kpc of NGC1052 at pericenter, implying an apocenter-to-pericenter ratio of $\gtrsim$50. We stress that this applies to {\em both} galaxies, an orbital coincidence that is extremely unlikely.

Both lines of argument provide strong evidence that NGC1052-DF2 and NGC1052-DF4 are indeed dark matter deficient. Even so, we note that providing an exact limit on the dark matter content requires us either assuming a true pericenter or conducting simulations more applicable to the NGC1052 system. Indeed, an issue with Section~\ref{Sec:Current} is that the relationship between $r_{\rm break}$ and $r_{\rm tid}$ can only be constrained by simulation, and it is unclear that previous simulations examining this relationship are applicable to the NGC1052 system which includes Mpc-scale current orbital distances and galaxies that somehow came to lack dark matter. An additional issue with Section~\ref{Sec:Current} is that our constraints come from the galaxies being opposite sides of, and approximately equidistant from, NGC1052 so that we could add up their line-of-sight offsets and compare to \citet{2021ApJ...914L..12S}. However, if we take current distance measurements at face value despite their large error bars (the only distance measurement with errors under 1 Mpc is the relative separation between NGC1052-DF2 and NGC1052-DF4), NGC1052-DF2 is much farther away (1.9 Mpc) from NGC1052 than NGC1052-DF4. Thus, NGC1052-DF2's distortions would barely be in 2$\sigma$ agreement with \citet{2021ApJ...914L..12S} even in the dark matter free case. Moreover, an advantage of the analysis in Section~\ref{Sec:Pericenter} is that it directly probes the point where the strongest interactions actually occur, i.e. at pericenter. We therefore consider the scenario explored in Section~\ref{Sec:Pericenter} to be the more robust of the two.

\subsection{Formation Scenario}

A key question remains regarding how NGC1052-DF2 and NGC1052-DF4 came to lack dark matter and host such luminous globular clusters. Unfortunately, we cannot well constrain their formation scenario from studying tidal features alone. Though tidal stripping has been proposed as an explanation for NGC1052-DF2 and NGC1052-DF4's missing dark matter \citep{2018MNRAS.480L.106O,2021MNRAS.501..693M,2021MNRAS.502.1785J,2021MNRAS.503.1233O} with simulation using self-interacting dark matter producing exactly the 0.2-0.3 ellipticities measured in Section~\ref{Sec:Fit} (\citealt{2020PhRvL.125k1105Y}, following correspondence with the authors), the observation of tidal distortions is expected purely based on the galaxies' dark matter deficiency, independent of whether such tidal interactions originally caused their lack of dark matter. Such a tidal stripping scenario would require that the galaxies have a high metallicity for their mass, since they would have originated from a far more massive galaxy, however the galaxies are relatively blue with a lower metallicity than a tidal origin would imply \citep{2018Natur.555..629V,2019ApJ...874L...5V}. Moreover, while stripping via the tidal field of the same perturber, NGC1052, is able to explain how two objects could both be nearly dark matter free, a complete description of the galaxies' formation should also include the origin of the galaxies' bright globular clusters.

Efforts to explain both unusual properties include \citet{2020ApJ...899...25S}'s study of the dark matter deficient remnants of high velocity collisions, with extreme pressure inducing the formation of bright globular clusters as further explored by \citet{2021ApJ...917L..15L}, as well as \citet{2021MNRAS.506.4841T}'s merger hypothesis and \citet{2022MNRAS.510.3356T}'s exploration of globular cluster feedback as an explanation for dark matter deficiency. Indeed, the galaxies may have arisen from pure baryonic material, either fragmented on infall or flung out by quasar winds \citep{1998MNRAS.298..577N}. An important point of our study is that we expect to observe tidal distortions in all  scenarios where the galaxies have little or no dark matter.

Even so, the similarity between the outer regions of NGC1052-DF2 and NGC1052-DF4 does provide new information to help understand the galaxies' formation process. These galaxies were already known to have similar kinematics, globular cluster populations, stellar masses, sizes, surface brightnesses, and stellar populations; now we add tidal distortions to this remarkable list. Qualitatively, this similarity suggests that the two galaxies have experienced a similar tidal field, which in turn implies that they likely are roughly equidistant from the giant elliptical galaxy NGC1052, or at least were at similar 3D distances at pericentric passage. Moreover, the small pericenters of Section~\ref{Sec:Pericenter} as compared to large present day orbital distances in particular support both collisional formation processes, since such high velocity encounters would be most likely to occur near the center of the group and result in objects with highly elliptical or unbound orbits, and tidal stripping, since elliptical orbits may include strong tidal effects.

% More exotic explanations involve alternatives to cold dark matter~\citep{2020ApJ...898..132C} as. 

% One possibility is that we live in a flawed simulation and the NGC1052 group was once stuck near a periodic boundary \citep{2020MNRAS.491.1278S}. This would explain two outlier galaxies in the same group, yet is concerning given the boundary's proximity and the potential manifestation of more bugs.

\subsection{Comparison to Previous Works} \label{Sec:Comparison}

Three recent works have undertaken similar efforts to investigate the NGC1052 field for evidence of tidal features \citep{2019A&A...624L...6M,2020ApJ...904..114M,2021ApJ...919...56M}. Our observational results largely agree with these previous studies, though in some cases our interpretation of these results differ. 

\citet{2019A&A...624L...6M}'s examination of the NGC1052 field in their Fig.~1 is similar to our own Fig.~\ref{Fig:Field}, and we confirm the tidal features to the south-east and south-west of NGC1052. Like \citet{2019A&A...624L...6M}, we do not observe external streams, tails, or bridges associated with either NGC1052-DF2 or NGC1052-DF4. Our NGC1052-DF2 initial image and generated model also appear similar to their Fig.\ 3. However, our NGC1052-DF4 results differ in that the outskirts are significantly more elliptical. 

\begin{figure}
    \centering
    \includegraphics[width=0.225\textwidth]{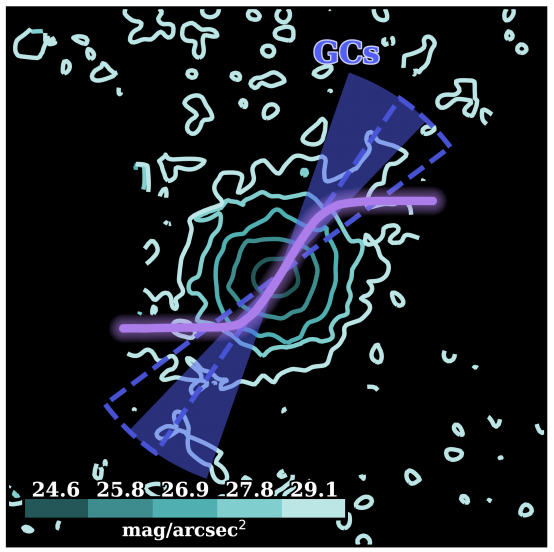}\:\includegraphics[width=0.225\textwidth]{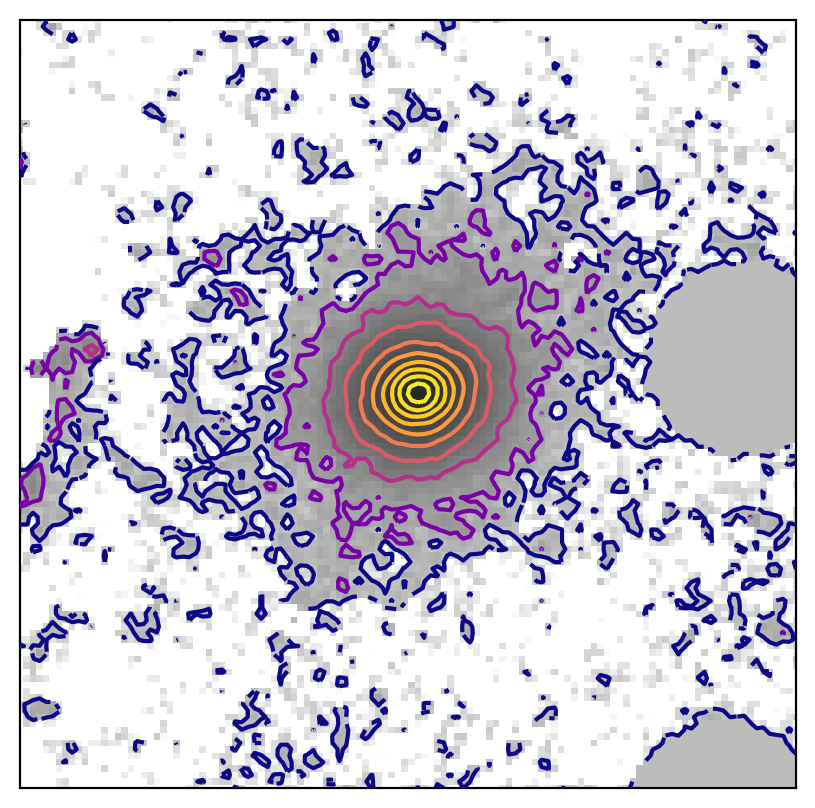} \\
    \caption{A comparison between the S-shape observed by \citet{2020ApJ...904..114M} (\textit{reproduced in left panel}) and our own image of NGC1052-DF4 (\textit{right panel}). \label{Fig:Comp}}
\end{figure}

Our analysis of NGC1052-DF4 is in better agreement with the results of \citet{2020ApJ...904..114M}, who generate a morphological model that is quite similar to our own. Our outermost isophotes have slightly lower surface brightness, which likely has to do with our subtraction of NGC1052-DF5, but are otherwise in agreement. However, a significant difference between our image and that of \citet{2020ApJ...904..114M} is the presence of increased flux to the south-east and north-west which they identify as S-shaped tidal tails and evidence of an ongoing interaction with the galaxy NGC1035 (see Fig.~\ref{Fig:Comp} for a direct comparison to our own work). This is apparent in their raw data, and persists after subtracting a model for the light of NGC1035. While we observe a significant excess of light in the direction of NGC1035 prior to subtracting a NGC1035 model (due to its overlap with NGC1052-DF2 in projection), we do not observe this excess after removing NGC1035, and we do not find evidence for tidal tails in either case.

Finally, \citet{2021ApJ...919...56M} find that NGC1052-DF2 has no radial variation in its ellipticity or position angle and suggest that this -- and an exponential surface brightness profile -- implies the presence of a low-inclination disk. With respect to the NGC1052-DF2's ellipticity, the difference between our work and theirs appears not be in the outskirts of the galaxy, where both works find $\varepsilon \gtrsim 0.2$, but in the interior of the galaxy where \citet{2021ApJ...919...56M} estimates an ellipticity of $\varepsilon \approx 0.5$; a structure they identify as `bulge-like,' though we note that no actual photometric bulge is present. We find a much lower ellipticity in the central regions down to 5$\arcsec$, and confirm this result with star counts in the deep \textit{HST} image of NGC1052-DF2 (see Fig.~\ref{Fig:HST}). This difference may be due to the difficulty of fitting the morphology of a region with a relatively constant surface brightness. With respect to NGC1052-DF2's surface brightness profile, our fit model is nearly identical to that of \citet{2021ApJ...919...56M}. The difference lies in the interpretation, where we fit a S\'ersic model rather than a disk and attribute excess light in the outer parts to tidal distortions.

Our interpretation of both galaxies' profiles is similar to that of \citet{2020ApJ...904..114M} for NGC1052-DF4, namely tidal distortions. \citet{2020ApJ...904..114M,2021ApJ...919...56M} propose that the similarities between NGC1052-DF2 and NGC1052-DF4 are coincidental and have two independent causes: a face-on massive disk for NGC1052-DF2 and a tidal interaction\footnote{With NGC1035, not NGC1052, though we note the best estimates available place NGC1035 at 15.6$\pm$2.2 Mpc, NGC1052 at 20.2$\pm$1.2 Mpc, and NGC1052-DF4 at 20.0$\pm$1.6 Mpc \citep{2017ApJ...843...16K,2001MNRAS.327.1004B,2020ApJ...895L...4D}.} for NGC1052-DF4 which stripped its dark matter. In our interpretation, the similarity between the outskirts of NGC1052-DF2 and NGC1052-DF4 has a common cause: tidal effects from NGC1052, which are strong because of the low masses of the galaxies.

\subsection{Limitations and Future Efforts} \label{Sec:Future_Efforts}

The discovery of tidal features suggests NGC1052-DF2 and NGC1052-DF4 are indeed dark matter deficient. However, the accuracy of our inferred constraints is hampered by systematic uncertainties inherent to tidal theory. A more appropriate method to derive robust limits from the observed distortions would be through improved numerical simulation, exploring the region of parameter space consistent with observations of the NGC1052 system, including the galaxies' large present day orbital distance, in order to constrain the range of NGC1052-DF2 and NGC1052-DF4 masses across all possible orbits.

Within the analytical picture presented, a significant obstacle has been the relation between observed distortions and the true tidal radius where material is unbound. If the tidal radius is indeed within the galaxies' stellar profile, deeper observation may reveal tidal streams of unbound stars being stripped by NGC1052, thereby providing a more robust upper limit on the instantaneous tidal radius at the satellite's current orbital position. Alternatively, if such streams are extremely faint it may be possible to map stripped material through associated globular clusters. However, our results imply that this radius could be over 2$\arcmin$ from the galaxies' centers. Thus, it is uncertain whether such tidal tails exist, and with sufficient baryonic content, to be observed.

The next clear observational step to further our work would be an improved NGC1052 distance constraint. Just as \citet{2021ApJ...914L..12S} was able to cancel out systematics in the DF2-DF4 distance estimate, a similar measurement could be obtained for NGC1052 to achieve $<$1 Mpc uncertainty. This would allow us to test whether or not the two galaxies are indeed at similar distances from NGC1052.

%--------------------------------------------------------------
% Conclusions
%--------------------------------------------------------------

\section{Summary and Conclusions} \label{Sec:Conclusions}

In this work we found evidence for tidal distortions associated with NGC1052-DF2 and NGC1052-DF4, two galaxies that had been found to lack dark matter based on their kinematics. Both galaxies show strong position angle twists and become elongated in their outskirts, as is visually apparent in Fig.~\ref{Fig:Contours} and quantified by an isophotal analysis in Fig.~\ref{Fig:Morphology}. This is an important result in itself, since we expect such low mass objects to be easily tidally disrupted; unless the galaxies are several hundred kiloparsecs away from any other massive galaxy, they should not remain spheroidal in their outskirts if they are truly dark matter deficient. This is true no matter how the galaxies initially came to lack dark matter.

First, we related these distortions to the tidal radius and derived the maximum separation the galaxies could be from the massive elliptical NGC1052 along the line-of-sight and still show the observed disturbances. We found that, if NGC1052-DF4 had a `normal' amount of dark matter, its sky-projected distance from NGC1052 is too great to be consistent with our results. Similarly, NGC1052-DF2's radial distance would need to be within 100 kpc of NGC1052's to be consistent with its expected halo mass. However, the galaxies could each be up to $\approx$1 Mpc from NGC1052, if they are truly dark matter deficient. We compared these distances together with \citet{2021ApJ...914L..12S}'s 2.1$\pm$0.5 Mpc DF2-DF4 measurement and found that the galaxies could have a maximum dark matter content of $\approx5\times$10$^8$ M$_\odot$ (a total mass of $\approx7\times$10$^8$ M$_\odot$) to be within 2$\sigma$ consistency. Second, we related the distortions to the tidal radius at pericenter and calculated the range of orbital distances at pericentric passage consistent with our observations. We found that the galaxies are likely on highly elliptical orbits, with a maximum pericenter of $\approx$200 kpc in the dark matter free case, and down to $\approx$20 kpc if the galaxies to had a `normal' dark matter content. However, it is unlikely for both galaxies to independently follow such extreme orbits with pericenter to present-day radii of $R_{\rm peri}/R_{\rm 0}\sim10^{-2}$.

Our results are completely independent of previous kinematic constraints and provide strong evidence that NGC1052-DF2 and NGC1052-DF4 are dark matter deficient. Looking forward, future numerical studies and improved distance measurements may infer more robust mass and orbit constraints from our observations. Such constraints may better inform efforts to understand how these galaxies came to be, and what they imply for our understanding of galaxy formation and the nature of dark matter.

%--------------------------------------------------------------
% Acknowledgments
%--------------------------------------------------------------

\section*{Acknowledgments} \label{Sec:Acknowledgments}
We thank the anonymous referee for a detailed and insightful report that improved the manuscript, and Frank van den Bosch for highly useful conversations exploring the limits of tidal analysis and for his idea to use our results to constrain the distance of pericentric passage. S. Danieli is supported by NASA through Hubble Fellowship grant HST-HF2-51454.001-A awarded by the Space Telescope Science Institute, which is operated by the Association of Universities for Research in Astronomy, Incorporated, under NASA contract NAS5-26555. The authors thank the excellent and dedicated staff at the New Mexico Skies Observatory. Support from NSF grants AST1312376 and AST1613582, NSERC, the Dunlap Institute (funded by the David Dunlap Family), and STScI grants HST-GO-14644, HST-GO-15695, and HST-GO-15851 is gratefully acknowledged.

The Legacy Surveys consist of three individual and complementary projects: the Dark Energy Camera Legacy Survey (DECaLS; Proposal ID \#2014B-0404; PIs: David Schlegel and Arjun Dey), the Beijing-Arizona Sky Survey (BASS; NOAO Prop. ID \#2015A-0801; PIs: Zhou Xu and Xiaohui Fan), and the Mayall z-band Legacy Survey (MzLS; Prop. ID \#2016A-0453; PI: Arjun Dey). DECaLS, BASS and MzLS together include data obtained, respectively, at the Blanco telescope, Cerro Tololo Inter-American Observatory, NSF's NOIRLab; the Bok telescope, Steward Observatory, University of Arizona; and the Mayall telescope, Kitt Peak National Observatory, NOIRLab. The Legacy Surveys project is honored to be permitted to conduct astronomical research on Iolkam Du'ag (Kitt Peak), a mountain with particular significance to the Tohono O'odham Nation.

\textit{HST} data presented in this paper can be accessed from the Mikulski Archive for Space Telescopes (MAST) at the Space Telescope Science Institute via \dataset[10.17909/t9-rd26-dv46]{https://doi.org/10.17909/t9-rd26-dv46}.

\facilities{Dragonfly, HST (ACS), and DECam (DECaLS).}
\software{MRF \citep{2020PASP..132g4503V}, Astropy \citep{2013A&A...558A..33A, 2018AJ....156..123A}, Photutils \citep{larry_bradley_2020_4044744}, Source Extractor \citep{1996A&AS..117..393B}, and SEP \citep{2016JOSS....1...58B}.}

%--------------------------------------------------------------
% Appendix
%--------------------------------------------------------------

\appendix

\section{A Method to Quantify the Surface Brightness Depth of Images}
\label{Sec:sbcontrast}

Whereas the point-source depth of astronomical images is well-defined (see, e.g., \citealt{2003AJ....125.1107L}), determining the surface brightness depth is notoriously difficult. The measured depth is dependent on the methodology, on the size scale for which the depth is determined, on the reduction steps that have been applied to the image, and often even on where, exactly, in the image the measurement is done. Generally the surface brightness limits are deeper on larger scales, as the relative noise fluctuations are reduced when averaging over more pixels.

In practice, authors commonly measure the surface brightness depth by quantifying the variance between measured fluxes in boxes of a fixed size that are placed in empty regions (e.g. \citealt{2005ApJ...631L..41M,2014ApJ...787L..37M,2019ApJ...883L..32V,2020ApJ...904..114M}). However, the specifics of the methodologies such as masking, aperture size, and placement are not standardized, and large scale gradients affect the variation on small scales. Furthermore, as the relation between depth and scale generally does not follow the Poisson expectation, it is not straightforward to interpret a measurement on, e.g., $10\arcsec$ scales in the context of a galaxy with a size of $60\arcsec$.

\subsection{The {\tt sbcontrast} Method} \label{Sec:sbmethod}

Here we describe {\tt sbcontrast}, a robust method to quantify the surface brightness depth of an image on arbitrary scales. An earlier implementation of the algorithm was released as part of the MRF package \citep{2020PASP..132g4503V}, though the method was not described in the published version of that paper.\footnote{A brief description appeared in an appendix of the first, pre-print arXiv version of the paper, prior to acceptance and publication; see \url{https://arxiv.org/pdf/1910.12867v1.pdf}.} The method defines the surface brightness limit as the {\em contrast} between a region of a particular spatial scale and its immediate surroundings. This definition is appropriate for the detection of low surface brightness objects of a particular size against a locally-smooth background which may have variations on much larger scales. It avoids the two conflicting issues that plague surface brightness limit estimates: first, there is no need to extrapolate from smaller scales to the scale of interest and second, the measurement is not affected by large scale gradients in the image.

The algorithm requires three inputs: the input image, a user-supplied spatial scale $s$ (which can be a single value or a list of scales), and an object mask. Object masks are standard products in packages such as SExtractor \citep{1996A&AS..117..393B} and MRF \citep{2020PASP..132g4503V}. The algorithm is robust against imperfect masking, or even a zero-filled mask if most pixels in the image are part of empty regions. The pixel scale $p$ (in arcseconds) and photometric zeropoint ZP can be supplied or read from the header. Optional parameters that offer fine control of the inclusion or rejection of pixels are $f_{\rm min}$ (default 0.8) and $n_{\rm min}$ (default 6), explained below.

The first step in the algorithm is to create a binned contrast map with pixels that have a size of $s\times s$. The user-supplied value of $s$ is rounded if needed so that the binning factors are integers. The value in each binned pixel is not the mean or the sum of the original pixels but the biweight location \citep{1990AJ....100...32B} of the unmasked pixels. The biweight is robust against outliers and is equal to the mean if there are no deviant pixels. If the number of original pixels that were masked exceeds $(1-f_{\rm min})s^2p^{-2}$ then the binned pixel is flagged as unusable. For example, if $p=0\farcs 5$ and the desired spatial scale $s=10\arcsec$, up to $400$ original pixels go into each binned pixel. For $f_{\rm min}=0.8$, at most 80 of these 400 pixels may be masked in the original image. If the number of masked pixels is $\leq 80$ then the biweight location of the remaining pixels in the original image is the value of the pixel in the contrast map. If it exceeds 80 then the pixel in the contrast map is flagged and not used in the subsequent analysis.

The next step is to calculate the local background of each binned pixel. This background is the biweight location of the surrounding usable pixels in the contrast map. This background is then subtracted from the pixel, allowing the calculation of a contrast on the desired scale that is independent of variation on larger scales. The maximum number of usable pixels is 8; if the number of background pixels that are flagged as unusable in the previous step exceeds $8-n_{\rm min}$ then the pixel is flagged as unusable too and it is disregarded in the rest of the analysis.

The final step is to calculate the variation in the contrast map and to convert this variation to a surface brightness limit. The variation $\sigma_{\rm ADU}(s)$ is the biweight scale of all usable pixels in the contrast map. The biweight scale is identical to the rms for a Gaussian distribution but robust against outliers. It is the repeated use of robust measures of location and scale that make the method insensitive to the presence of poorly- or un-masked bright objects in the image.
Expressed in surface brightness, we have
\begin{equation} \label{Eq:sblimit}
    \sigma_{\mu}(s) = ZP - 2.5\log_{10}{\sigma_{\rm ADU}(s)}+5\log_{10}{p}-2.5\log_{10}c, 
\end{equation}
where $\sigma_{\rm ADU}(s)$ is the $1\sigma$ surface brightness variation on scale $s$ and $c \approx 0.94$ is a correction factor that accounts for the small artificial increase in the variation due to the noise in the background estimation for each pixel. Since the local background for each bin is computed from 8 neighboring bins, the estimate will be imperfect introducing an error which increases the variance. This increase is $\sqrt{1^2 + (1/\sqrt{n})^2}$ , with $n$ being the number of pixels that went into the background estimate; for $n=8$ this becomes $\sqrt{1^2+1/8} = \sqrt{1.125}$. Thus, to correct for the increase we multiply the measured $\sigma_{ADU}$ by $c = 1/\sqrt{1.125} \approx 0.94$. Converting the variation to a $3\sigma$ or $5\sigma$ limiting surface brightness is trivial:
\begin{equation}\label{Eq:Nsig}
\mu_{\rm lim}(N\sigma, s) = \sigma_{\mu}(s) - 2.5\log_{10} N,
\end{equation}
with $N=3$ or $N=5$ the $N\sigma$ limit that is sought.

The {\tt sbcontrast} code is publicly accessible via `{\tt pip install sbcontrast}.' It can be run directly from the command line on fits files (e.g. `{\tt sbcontrast image.fits -masks masks.fits}' -- see {\tt sbcontrast -h} for options), or within Python scripts on {\tt numpy} arrays via function import (i.e. `{\tt from sbcontrast import sblimit}'). Refer to the {\tt README} file for up-to-date information. We ask that researchers utilizing {\tt sbcontrast} please cite this work. Additional user guides supplementing this appendix may also, in the future, become available via the github or the authors' personal websites.

\subsection{Demonstration Using Artificial Data}

\begin{figure}
    \centering
    \includegraphics[align=t,width=0.225\textwidth]{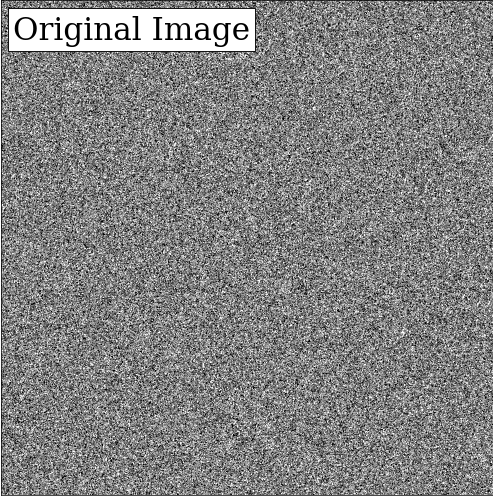}\:\:\:\:\includegraphics[align=t,width=0.225\textwidth]{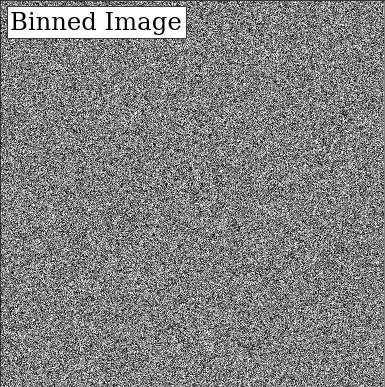}\:\:\:\:\includegraphics[align=t,width=0.225\textwidth]{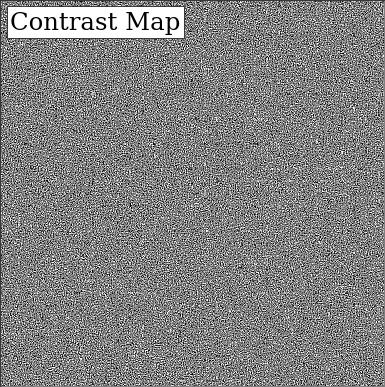}\:\:\:\:\includegraphics[align=t,width=0.275\textwidth]{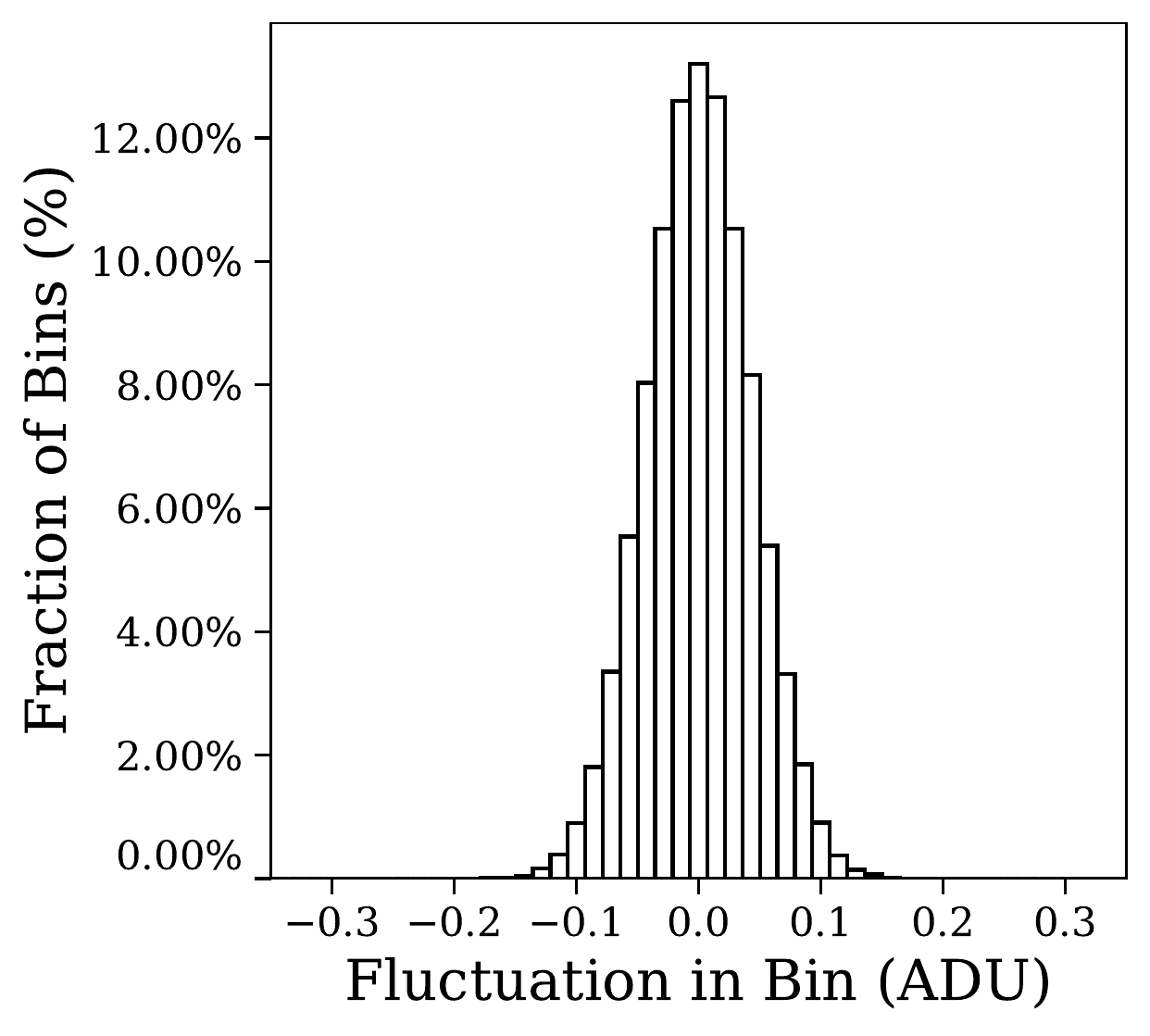}\\
    \includegraphics[align=t,width=0.225\textwidth]{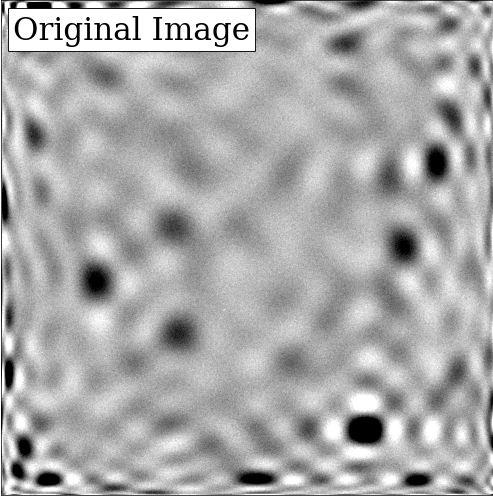}\:\:\:\:\includegraphics[align=t,width=0.225\textwidth]{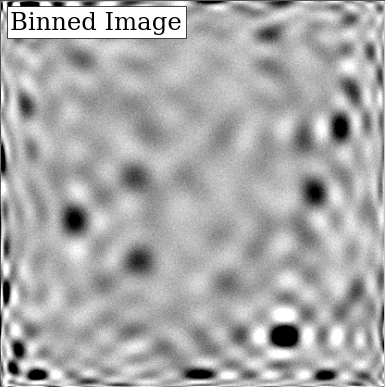}\:\:\:\:\includegraphics[align=t,width=0.225\textwidth]{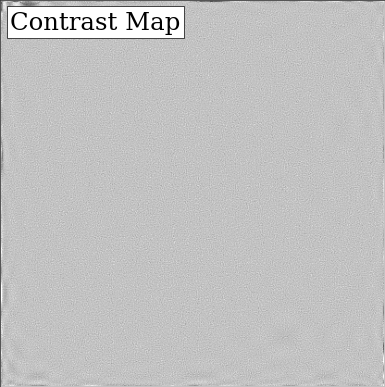}\:\:\:\:\includegraphics[align=t,width=0.275\textwidth]{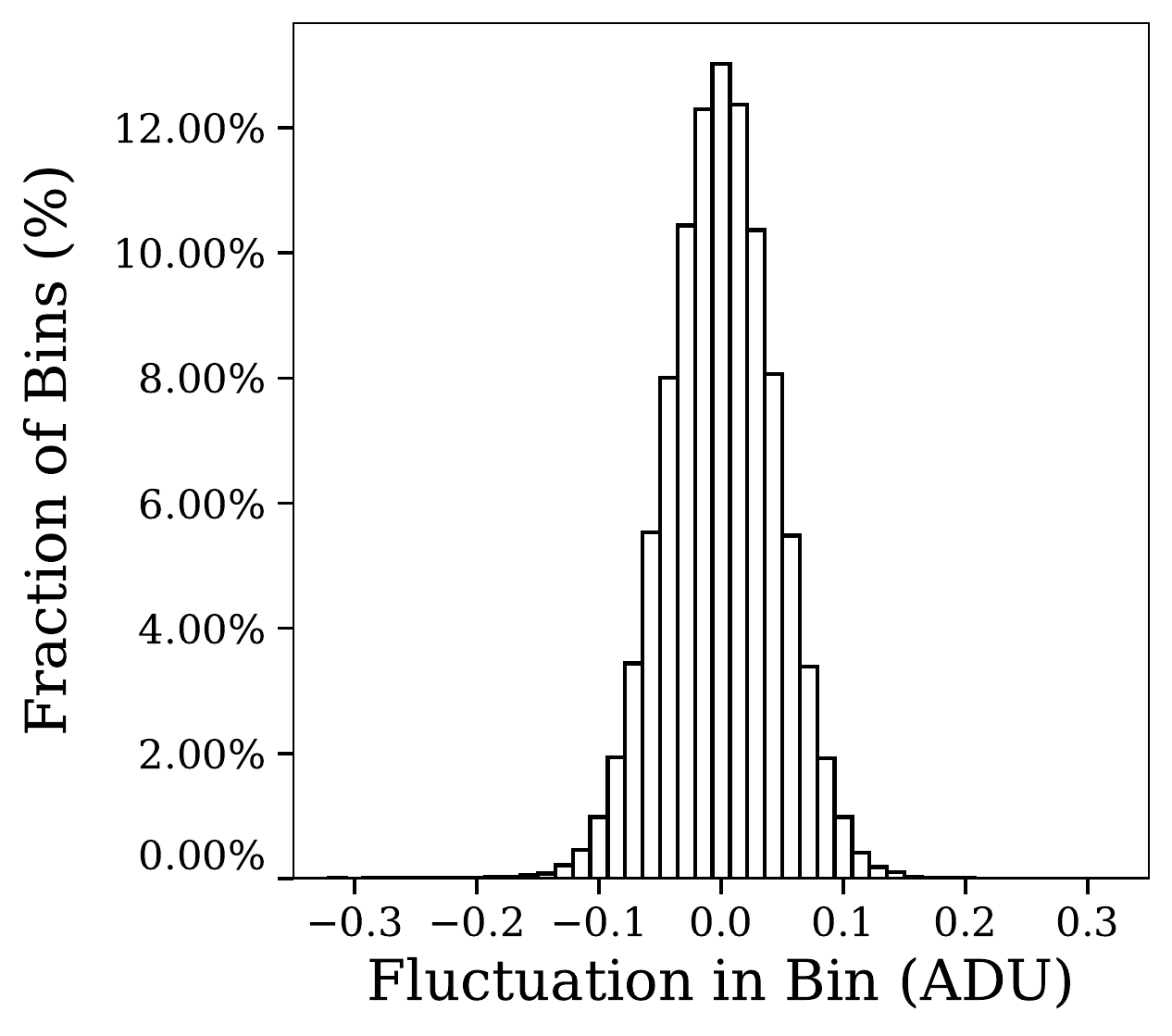}\\
    \includegraphics[align=t,width=0.225\textwidth]{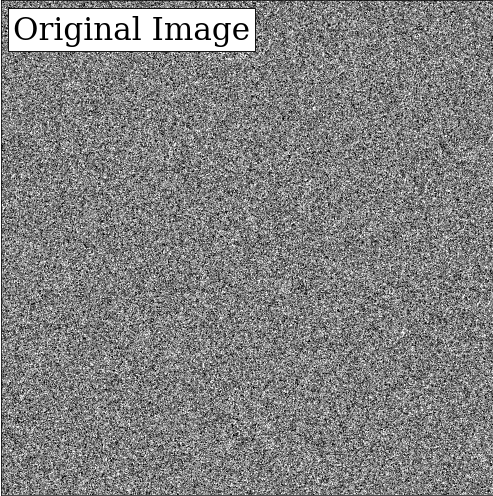}\:\:\:\:\includegraphics[align=t,width=0.225\textwidth]{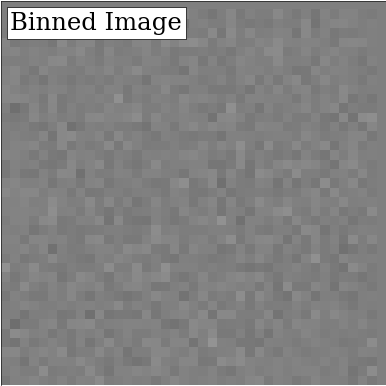}\:\:\:\:\includegraphics[align=t,width=0.225\textwidth]{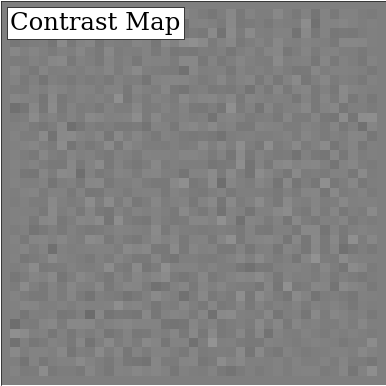}\:\:\:\:\includegraphics[align=t,width=0.275\textwidth]{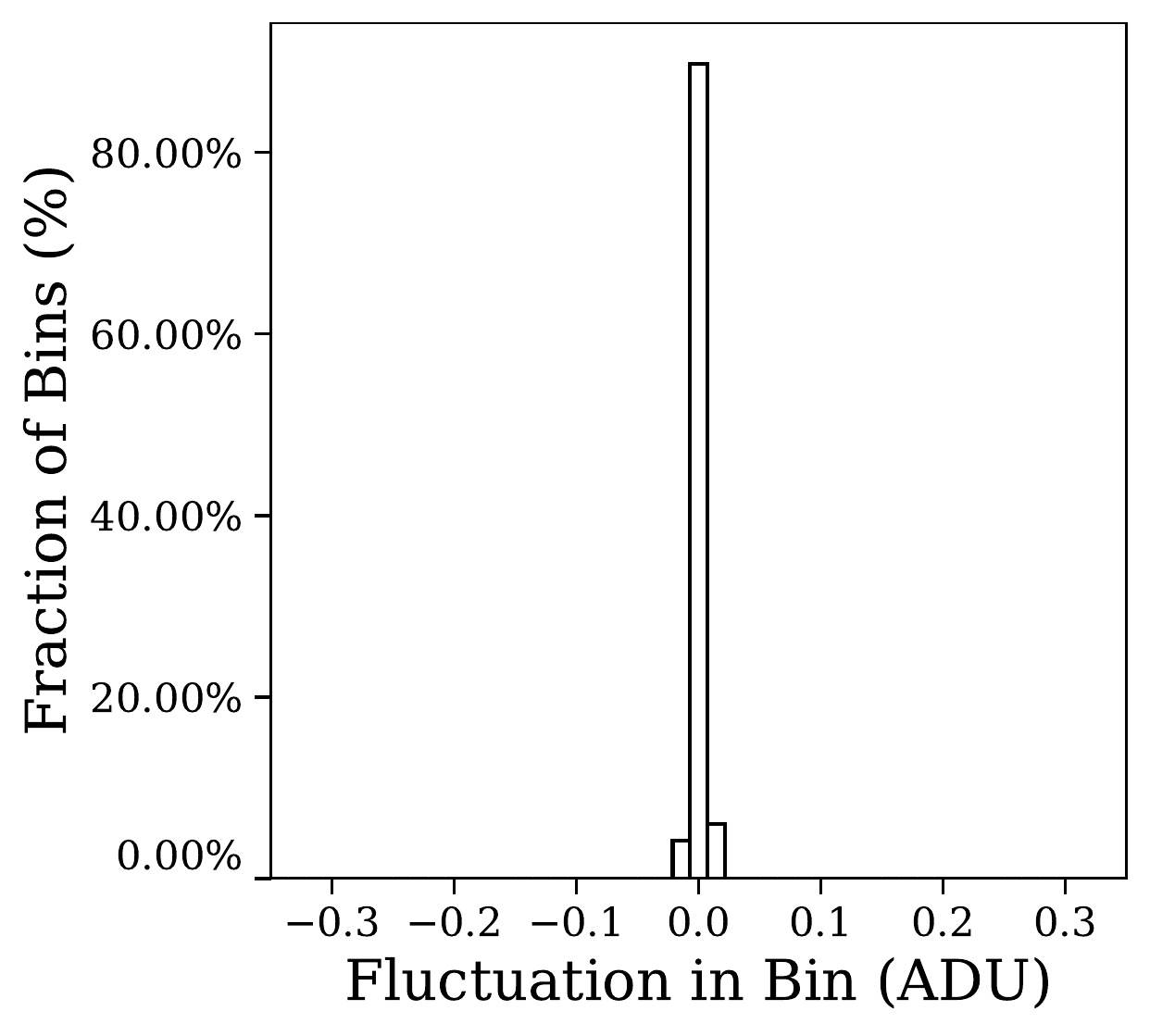}\\
    \includegraphics[align=t,width=0.225\textwidth]{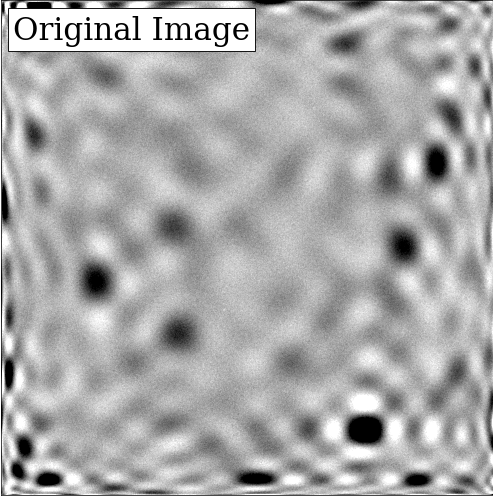}\:\:\:\:\includegraphics[align=t,width=0.225\textwidth]{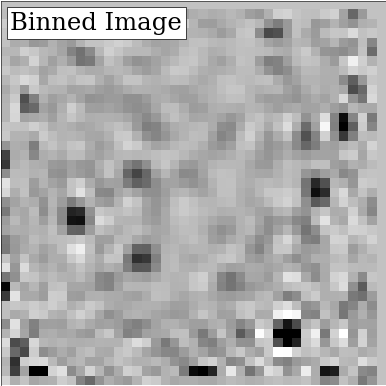}\:\:\:\:\includegraphics[align=t,width=0.225\textwidth]{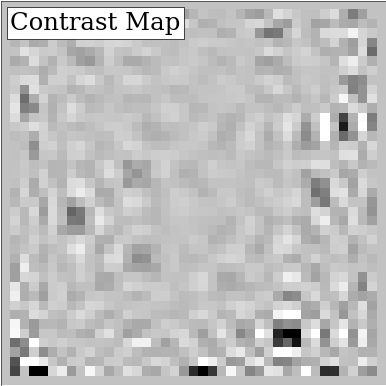}\:\:\:\:\includegraphics[align=t,width=0.275\textwidth]{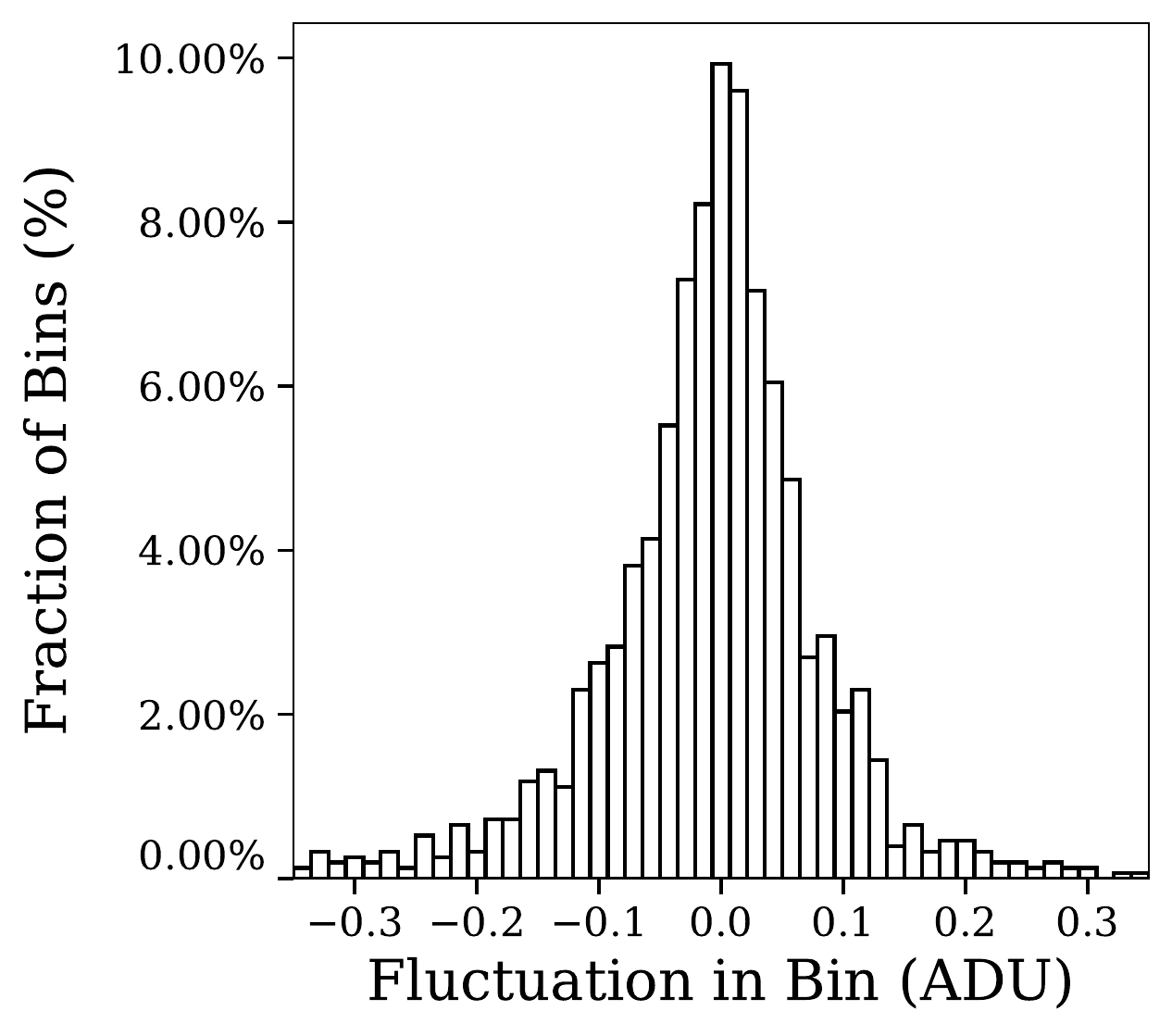}\\
    \caption{A visualization of {\tt sbcontrast} as we apply it for scales of $s=6\arcsec$ (\textit{top}) and $s=60\arcsec$ (\textit{bottom}) to an image of random noise and the same image with large scale variations inserted. \textit{Left column:} The original images. \textit{Second column:} The images binned into $s \times s$ boxes, with the value of each bin given as a biweight estimate of location. \textit{Third column:} The contrast map generated from local background subtraction. \textit{Right column:} The distribution of fluctuations in bins from the contrast map. \label{Fig:sbexample}}
\end{figure}

\begin{figure}
    \centering
    \includegraphics[width=0.625\textwidth]{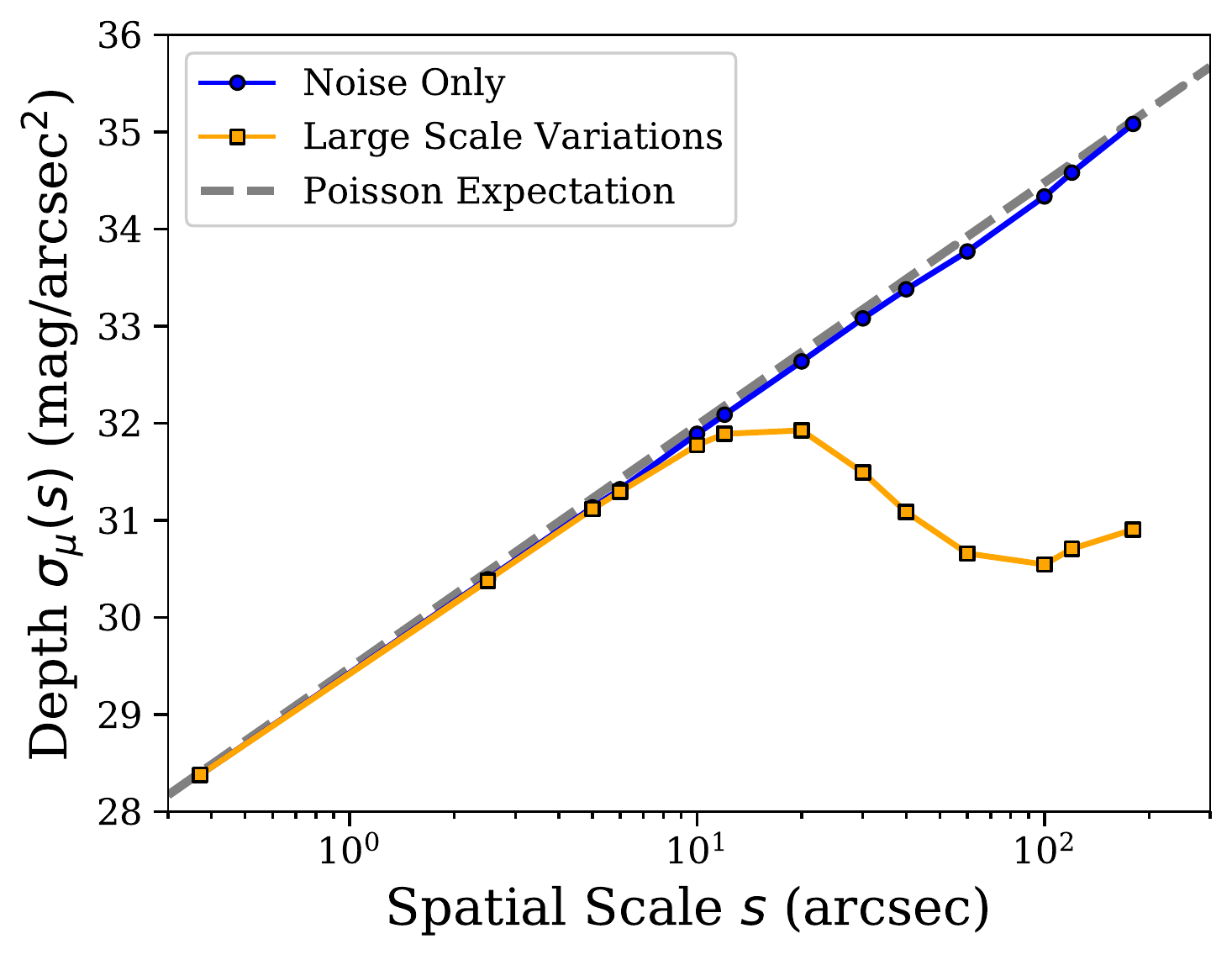}
    \caption{The calculated depth as a function of spatial scale for random noise (\textit{blue lines and circles}) and random noise with large scale variations (\textit{orange lines and squares}). These empirical results are compared to the expectation from Poisson statistics (\textit{dashed grey line}). The {\tt sbcontrast} results for the noise-only image match the Poisson expectation, validating the method. The depth of the image with background variations varies with spatial scale in a complex way. \label{Fig:examplescales}}
\end{figure}

Here we demonstrate the application and utility of the method using artificial data. First we consider an image of random Gaussian-distributed noise, whose calculated limiting surface brightness should follow the Poisson expectation. We then add a 2D polynomial to this image, simulating the effects of imperfect sky subtraction, flat fielding errors, or cirrus. In Fig.~\ref{Fig:sbexample}, we show each step as described above, as applied to both images. For large $s$, the distribution of fluctuations for the image with large scale variations is significantly broadened as compared to the image with just random noise, even though the pixel-to-pixel rms noise of the two images are nearly equivalent.

In Fig.~\ref{Fig:examplescales} we calculate the limiting depth for both images at a range of scales. The results are compared to the expectation from Poisson statistics (see Appendix A of \citealt{2020A&A...644A..42R}). We find that {\tt sbcontrast} matches the expected limit for the image consisting of pure Gaussian noise. This validates the method: in the idealized case of an image with no large scale background issues and uncorrelated noise the method precisely matches the Poisson expectation.\footnote{Note that for actual images, on the single pixel scale, i.e. $s = p$, {\tt sbcontrast} may diverge from the pixel-to-pixel rms due to correlations introduced by re-sampling.} As expected, the limiting depth in the image with background variations is much shallower than expected from Poisson statistics. Indeed, the relation between depth and scale can not only flatten but even invert at large scales: it may be possible to identify a low surface brightness galaxy with an effective radius $r_{\rm eff}=5\arcsec$ against a background with gradients on the scale of $60\arcsec$, whereas a galaxy with $r_{\rm eff}=60\arcsec$ with the same surface brightness cannot be reliably identified in those circumstances. The {\tt sbcontrast} method implicitly takes this effect into account, and provides the real-world limit on the appropriate scale for the science question.

\subsection{Limitations of the Method}

The method has several limitations. First, it assumes that actual objects in the image are masked. This is straightforward to accomplish for bright objects such as stars and galaxies, as discussed above, but more difficult for large, low surface brightness features such as Galactic cirrus. In a field with extensive cirrus emission the method will return depths that are too pessimistic on very large scales, as the cirrus itself will dominate the fluctuations. The method is therefore not well suited for determining the limiting depth for detecting cirrus, unless the cirrus itself is carefully masked or occupies only a small fraction of the field. In practice a neighboring empty field that was observed with the same instrument to the same depth may provide the most robust surface brightness limit on the spatial scale of the cirrus.

Second, the method does not address the problem of the actual {\em detection} of low surface brightness objects. The depth is defined as the contrast between an area with its local surroundings, but standard detection algorithms such as Source Extractor \citep{1996A&AS..117..393B} will often not be able to identify objects that span many pixels whose per-pixel S/N is much smaller than 1. The contrast map itself can be viewed as a crude detection map, as $\geq 5\sigma$ pixels in that map are good candidates for large, low surface brightness objects. However, as a detection method it is a crude approach, as it is most sensitive to square objects that are exactly centered on a binned pixel.

Finally, we emphasize that the method should not be used on scales that exceed the scale of background subtraction in the data reduction. Nearly all imaging pipelines include a step where the background is fit by a 2D polynomial, to remove large scale flat fielding errors and/or to perform sky subtraction. These polynomials are generally of low order but they tend to be applied on individual chips rather than on the full CCD mosaic (see \citealt{2019PASJ...71..114A} for efforts to do full-field sky subtraction). Individual chips typically span only a few arcminutes, which means that all structure on scales greater than $\sim 60\arcsec$ is artificially removed.

\begin{figure}
    \centering
    \includegraphics[width=1.00\textwidth]{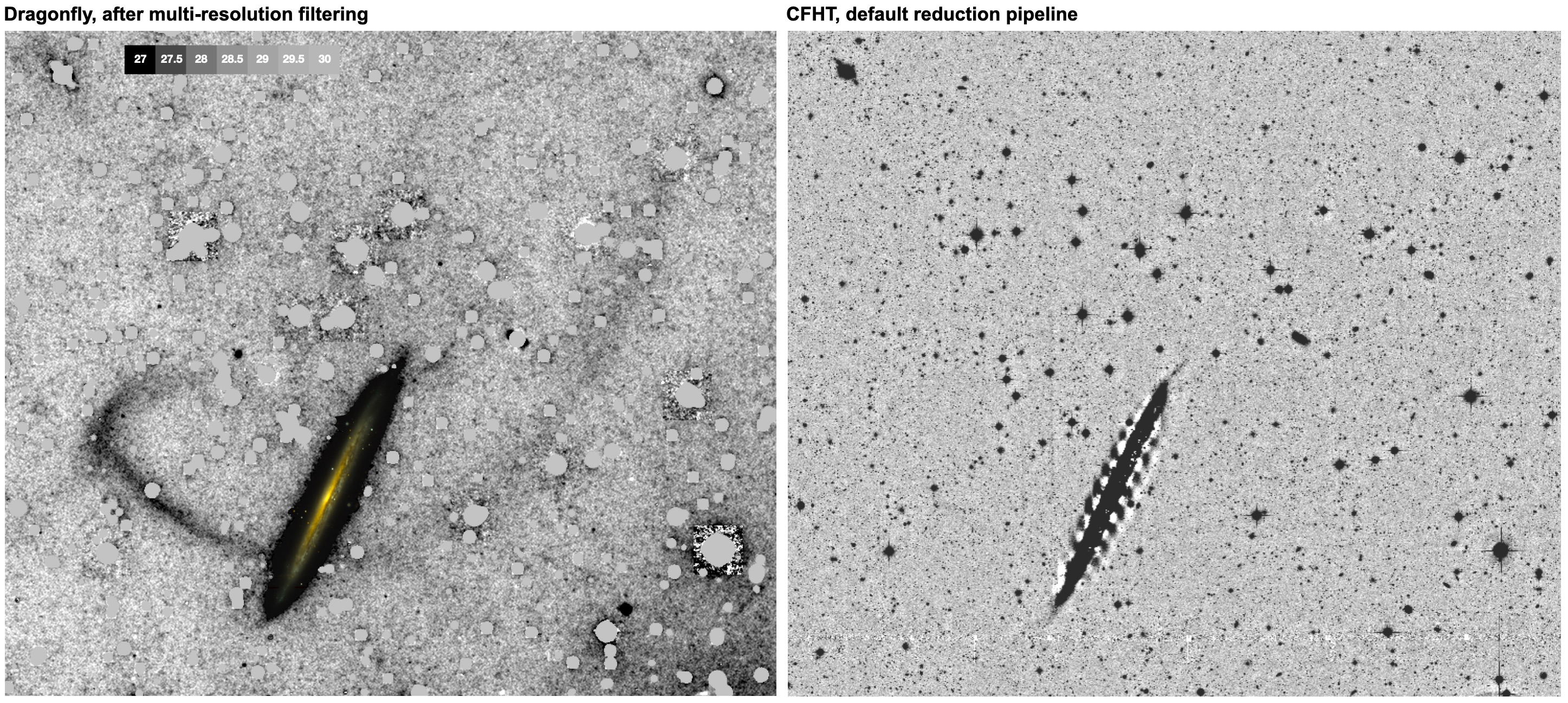}
    \caption{Illustration of the effects of sky subtraction on the surface brightness depth. {\em Left:} Dragonfly $g+r$ image of the galaxy NGC5907, adapted from Fig.\ 1 of \cite{2019ApJ...883L..32V}. The scale is $50\arcmin \times 39\arcmin$. The well-known giant stream has a surface brightness of $27-29$\,mag\,arcsec$^{-2}$. {\em Right:} Publicly available $g$ band image of the same area of sky obtained with CFHT. The {\tt sbcontrast} method indicates a $1\sigma$ surface brightness limit of 30.7\,mag\,arcsec$^{-2}$ for this image on the scale of the stream. However, the stream is not detected, as it was subtracted in the data reduction process. \label{Fig:NGC5907}}
\end{figure}

An extreme example is shown in Fig.~\ref{Fig:NGC5907}. Here we compare a deep Dragonfly image of the edge-on spiral galaxy NGC5907 (from \citealt{2019ApJ...883L..32V}) to a deep $g$ band image obtained with CFHT. The public CFHT data release used a pipeline that focused on the shapes of faint background galaxies, with aggressive sky subtraction in the reduction. As a result, the outer regions of the galaxy are removed and replaced by ringing of the polynomial. The well-known tidal feature of the galaxy \citep{1998ApJ...504L..23S,2019ApJ...883L..32V} is also entirely removed. The {\tt sbcontrast} method returns a $1\sigma$ depth of 30.7 mag arcsec$^{-2}$ on scales of $60\arcsec$ for this image (the width of the tidal feature). The surface brightness of the tidal feature is $\approx 28$\,mag\,arcsec$^{-2}$ \citep{2019ApJ...883L..32V}, well above this limit, and one might conclude that there are no tidal features associated with NGC5907. The reason for this mismatch is that the {\tt sbcontrast} limit should not be calculated on this scale, as all structure was removed in the reduction. This is not really a limitation of the method, but a reflection of the fact that the steps that were applied in the data reduction should be understood when calculating surface brightness limits on very large scales (using any method).\footnote{We note that the image in Fig.\ 13 is not representative for the capabilities of CFHT: some of the most impressive low surface brightness imaging results have been obtained with this telescope, using careful reduction techniques that preserve large scale structures \citep{2015MNRAS.446..120D,2016A&A...593A.126B,2018MNRAS.475L..40D}.}

\subsection{Depth of the Dragonfly NGC1052 Image}

\begin{figure}
    \centering
    \includegraphics[width=0.3\textwidth]{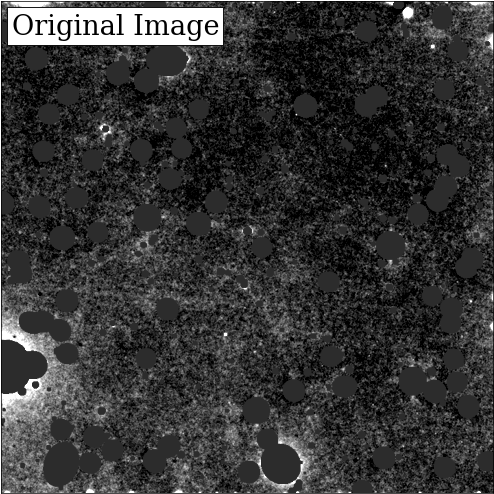}\:\includegraphics[width=0.3\textwidth]{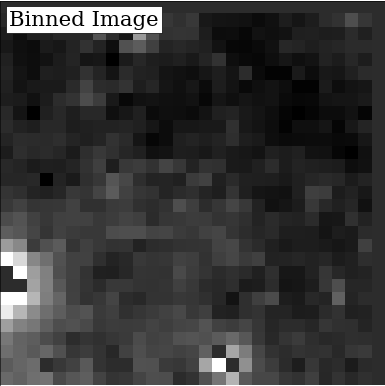}\:\includegraphics[width=0.3\textwidth]{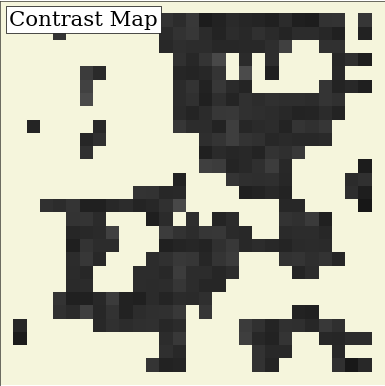}\\
    \includegraphics[align=c,width=0.4625\textwidth]{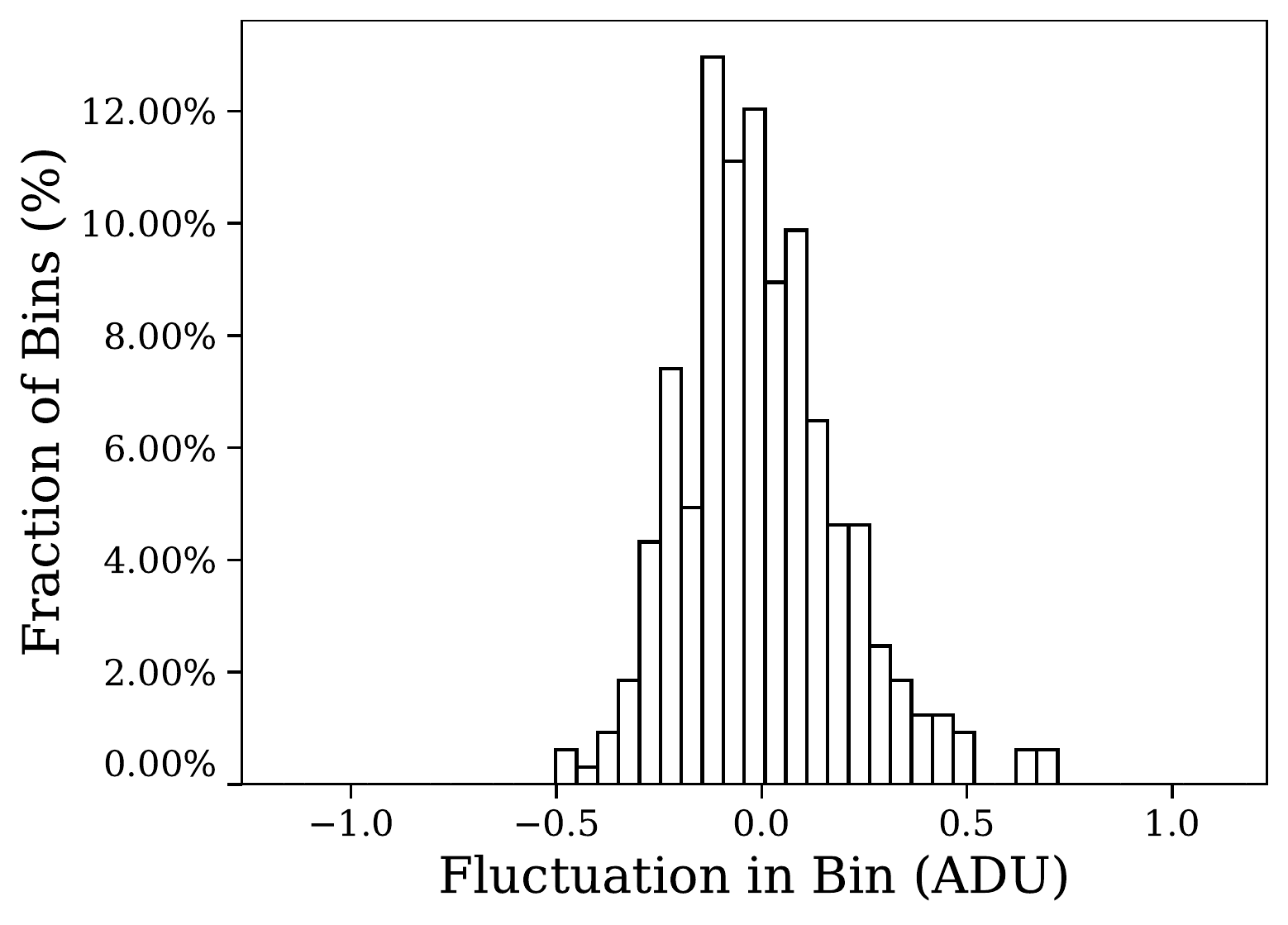}\:\includegraphics[align=c,width=0.4375\textwidth]{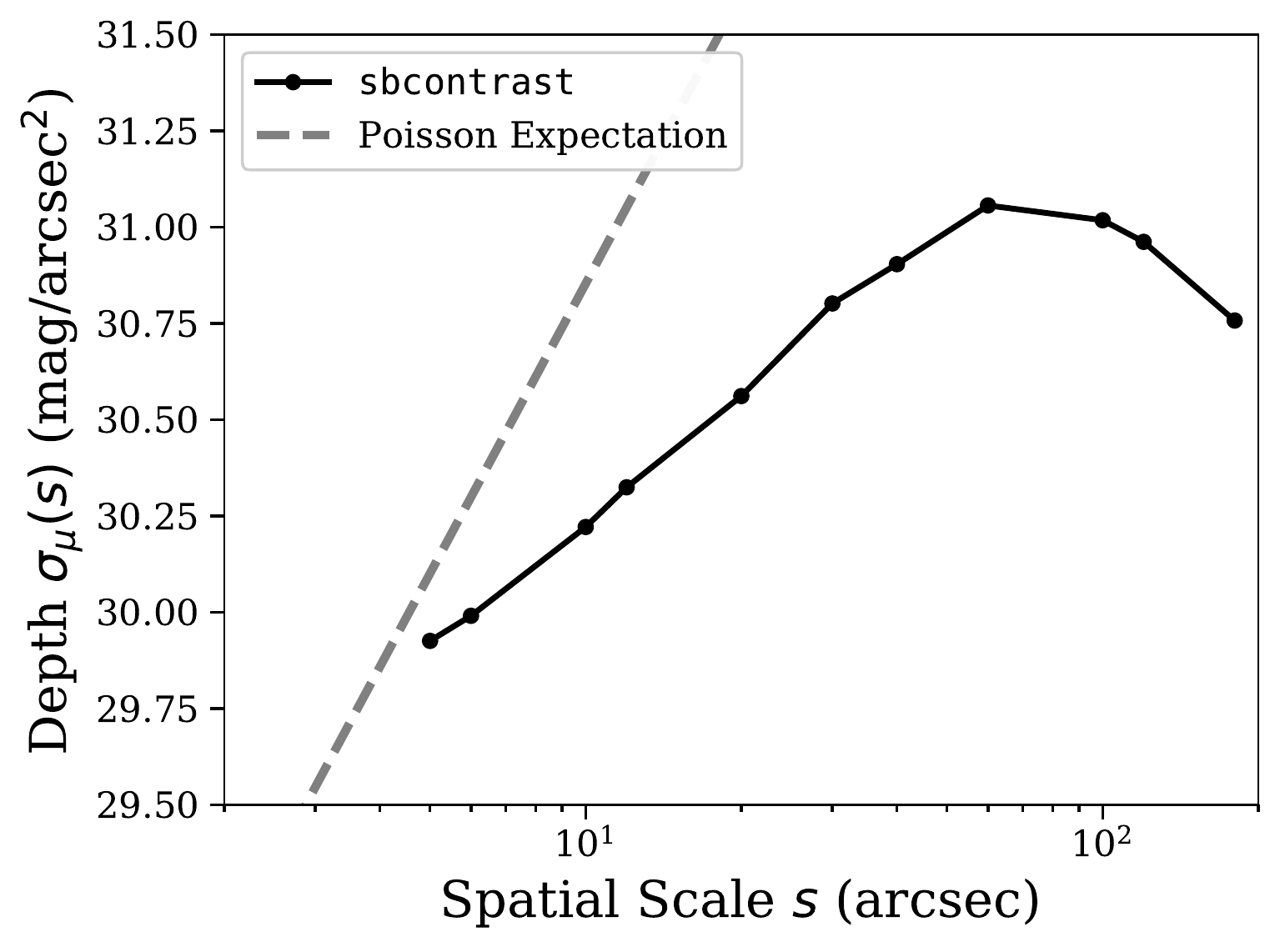}\\
    \caption{A visualization of {\tt sbcontrast} as we apply it to the NGC1052 field. \textit{Top left panel:} The original image. \textit{Top middle panel:} The biweight locations of the image binned into $60\arcsec$ boxes. \textit{Top right panel:} The functions for each bin after subtracting local background and masking poorly determined pixels (\textit{beige}). \textit{Bottom left panel:} The distribution of fluctuations in bins from the $60\arcsec$ contrast map. \textit{Bottom right panel:} The surface brightness limit as calculated at a range of spatial scales. We also provide the same limit based purely on extrapolating the variance at the single pixel scale (\textit{dashed line}). \label{Fig:sblimit}}
\end{figure}

Fig.~\ref{Fig:sblimit} details our application of {\tt sbcontrast} to the NGC1052 field, giving the original image (showing a cutout of the mosaic $\approx$0.5 deg north of NGC1052), the binned image with sizes of 60$\arcsec$ as referenced in the main body of our work, the contrast map as calculated from the binned image, a histogram of the variations in each bin of the contrast map, and the final limit at a range of spatial scales, including values other than the 60$\arcsec$ limit references in the main text (at $1\sigma$, which may be converted to the 3$\sigma$ limit using Eq.~\ref{Eq:Nsig}). For comparison, we again give the expected behavior for Poisson statistics and find that the actual depth increases more slowly, echoing results that had previously been obtained by placing random apertures in images (e.g. \citealt{2003AJ....125.1107L}).

\section{Local Background Correction} \label{Sec:Background}
The regions surrounding NGC1052-DF2 and NGC1052-DF4 had a significant amount of excess light originating from neighboring galaxies and bright stars excluded from our high surface brightness model. To better analyze emission from the dark matter deficient galaxies themselves, we therefore made a local correction to each galaxy separately by subtracting the mean flux of pixels in the regions listed in Table~\ref{Table:Regions}.

\begin{deluxetable}{ccc}
\tablecaption{Background Regions\label{Table:Regions}}
\tablewidth{0pt}
\tablehead{
\colhead{R.A.} & \colhead{Dec.} & \colhead{Radius} \\
\colhead{(deg)} & \colhead{(deg)} & \colhead{(arcsec)}
}
\startdata
\multicolumn3c{NGC1052-DF2} \\
\hline
40.47333 & -8.42436 & 20.3 \\
40.43832 & -8.36521 & 17.8 \\
40.41294 & -8.39201 & 17.8 \\
\hline
\multicolumn3c{NGC1052-DF4} \\
\hline
39.82145 & -8.14999 & 21.4 \\
39.78004 & -8.12458 & 21.0 \\
39.78511 & -8.07410 & 27.2 \\
\enddata
\end{deluxetable}

\section{Isophote Fitting and Model Subtraction} \label{Sec:Residuals}
\begin{figure}
    \centering
    \includegraphics[width=0.32\textwidth]{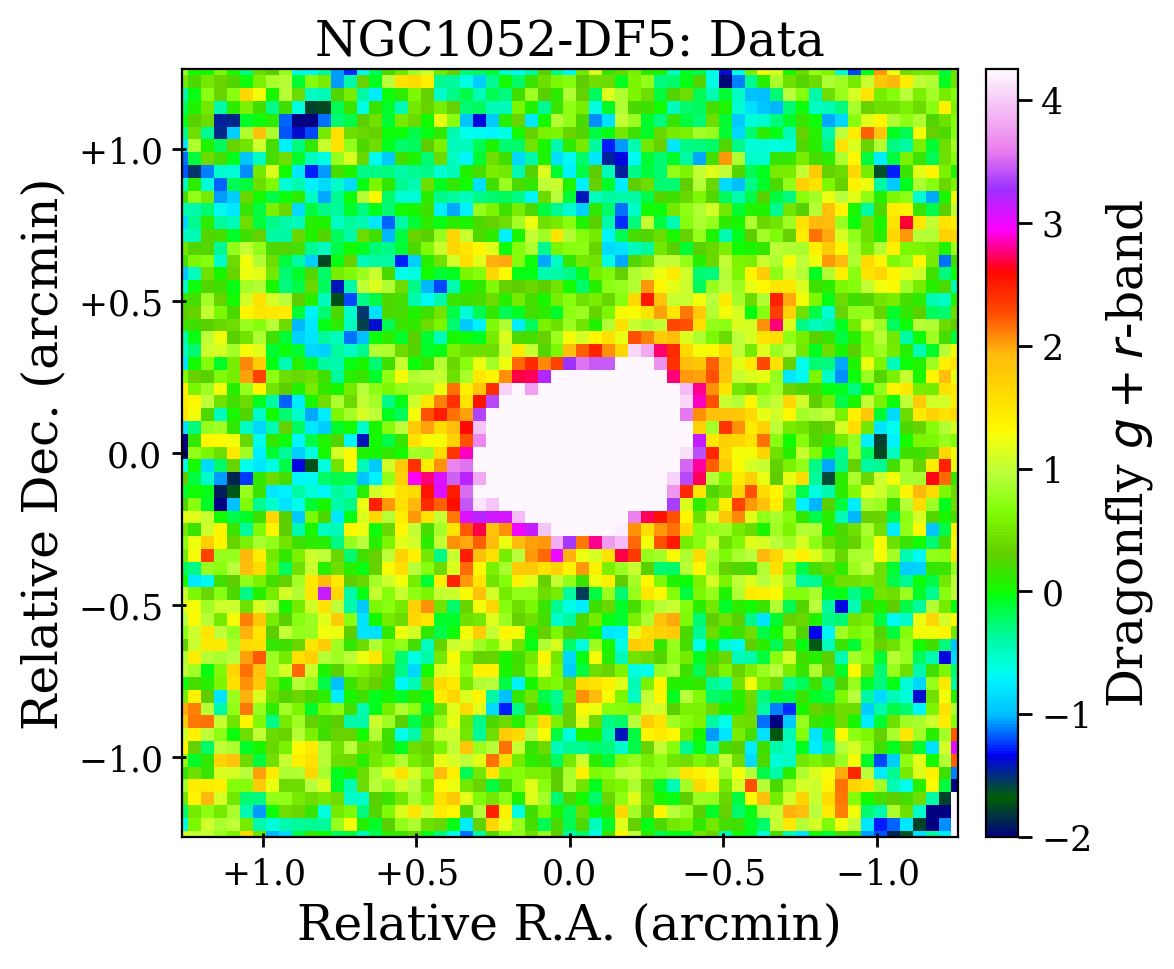}   \includegraphics[width=0.32\textwidth]{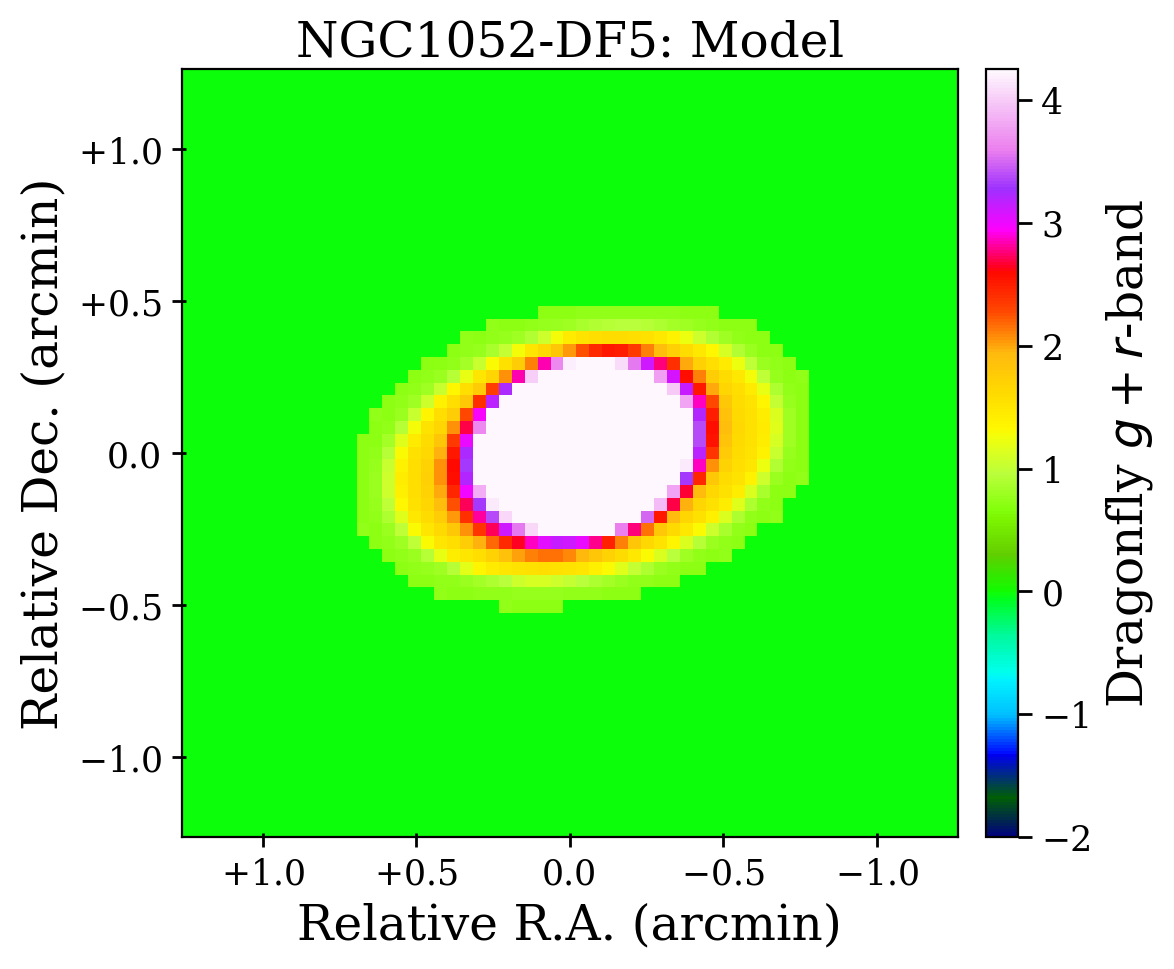}   \includegraphics[width=0.32\textwidth]{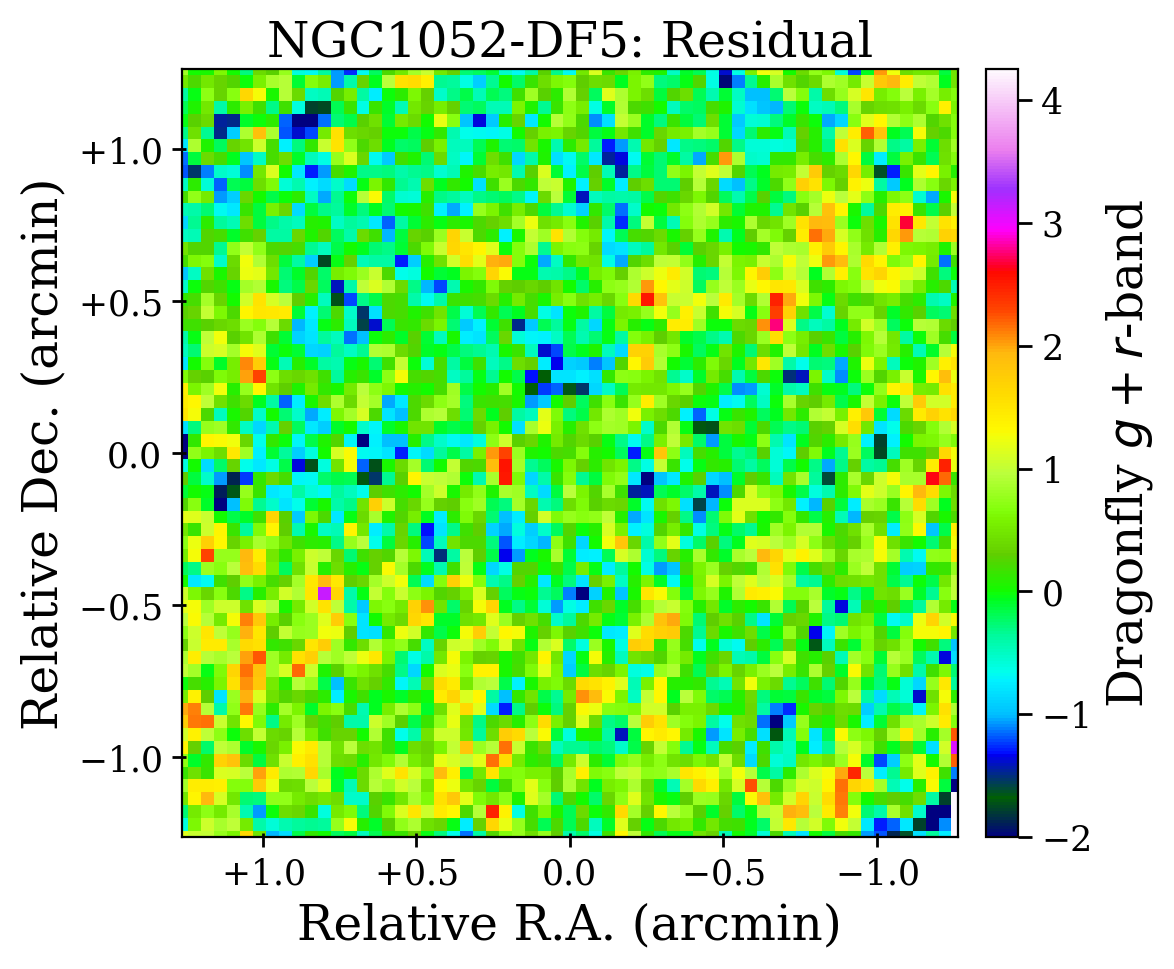}   \\
    \includegraphics[width=0.32\textwidth]{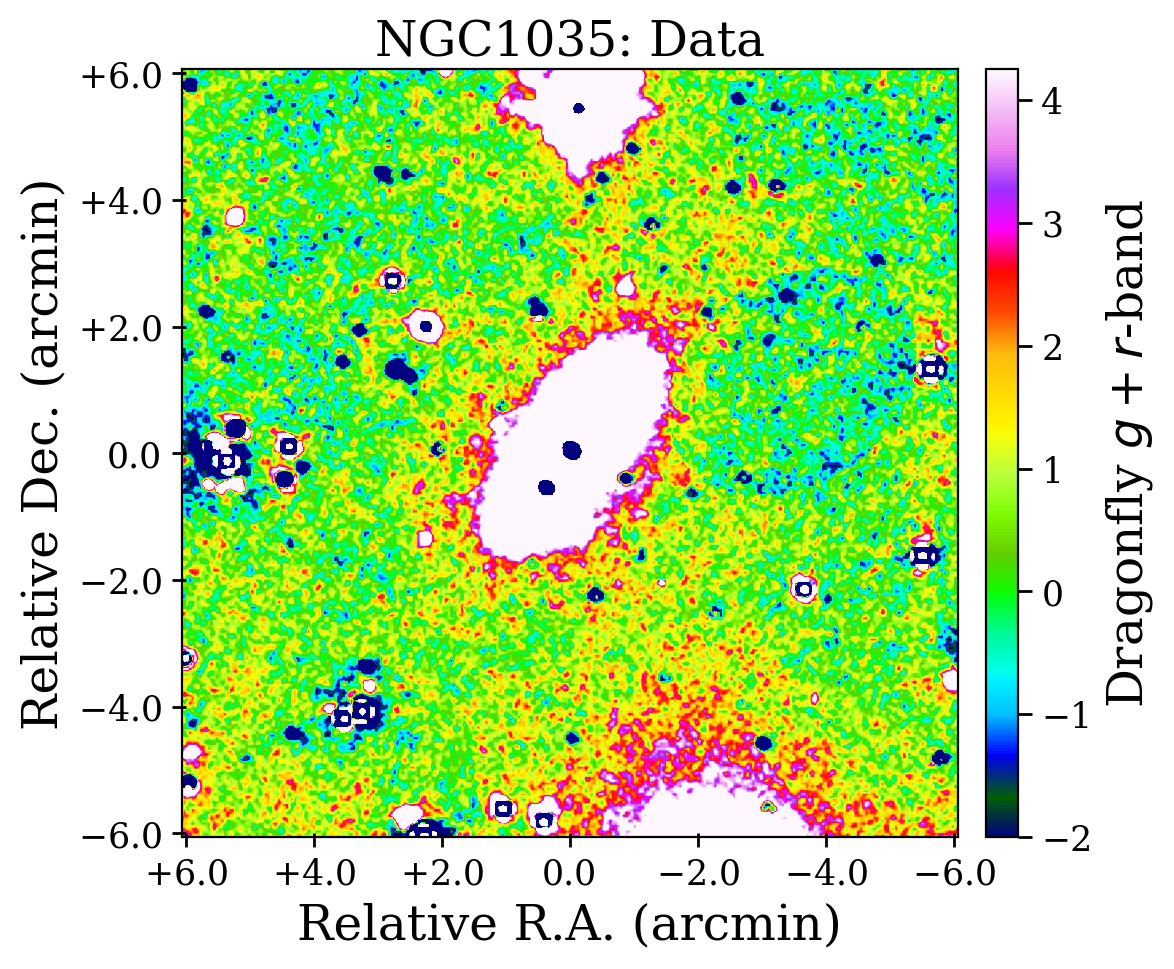}  \includegraphics[width=0.32\textwidth]{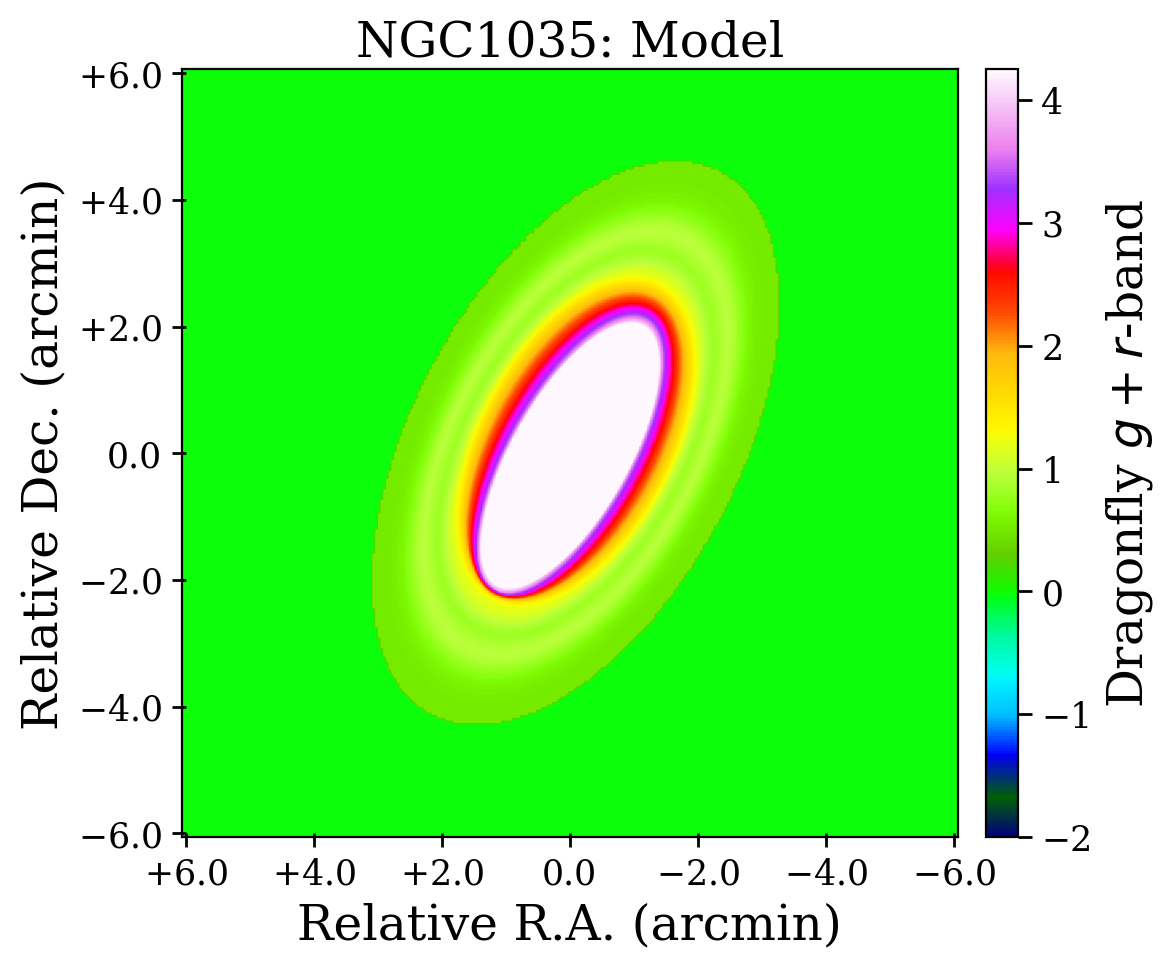}  \includegraphics[width=0.32\textwidth]{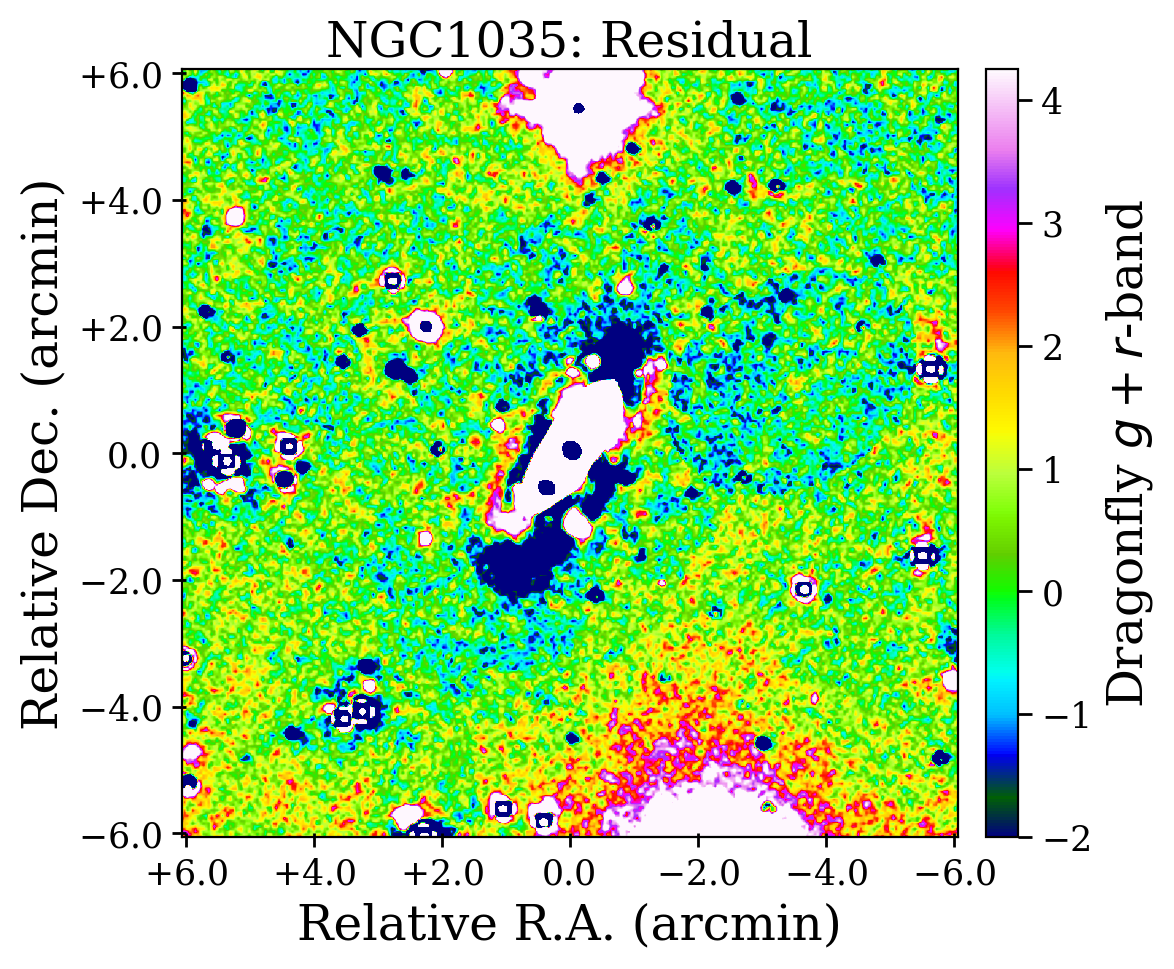}  \\ 
    \includegraphics[width=0.32\textwidth]{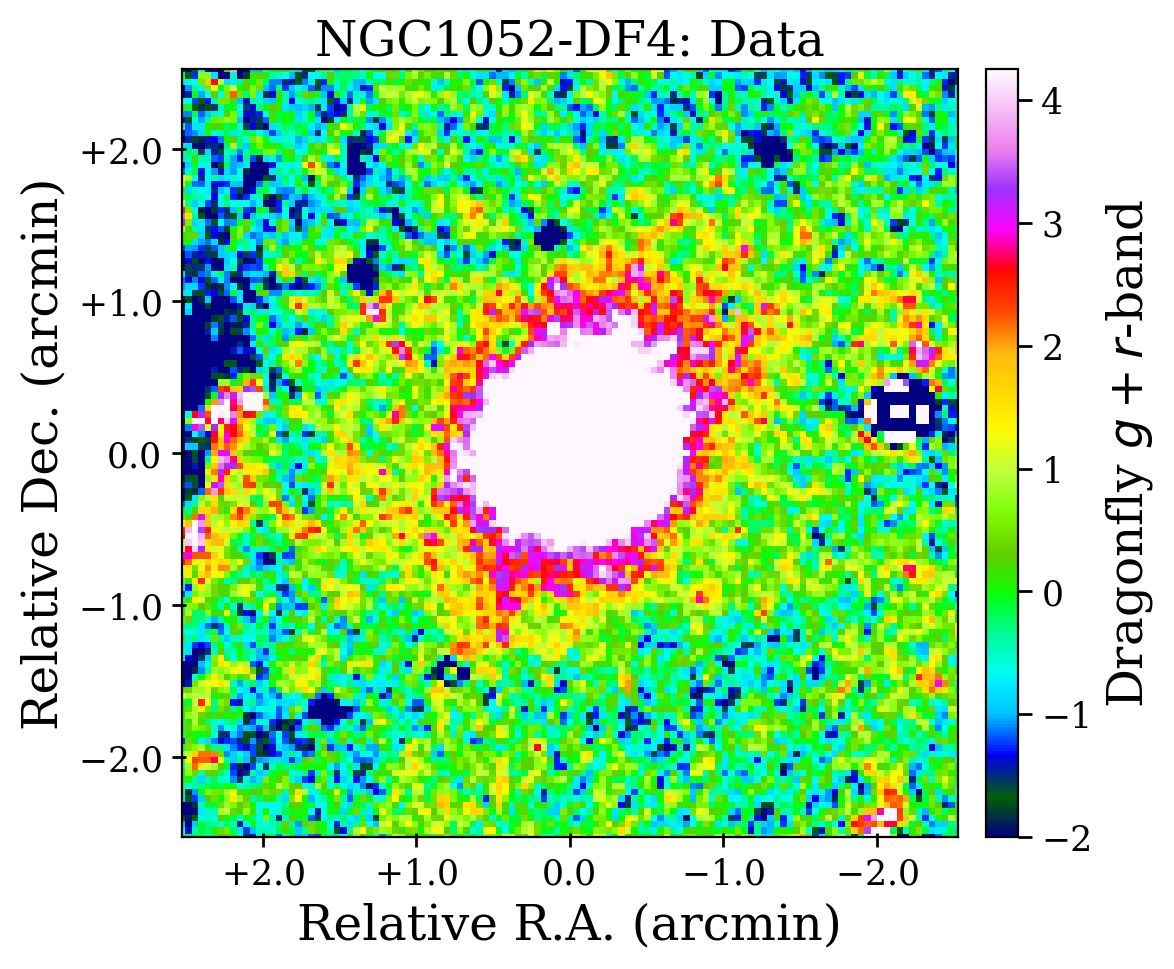} \includegraphics[width=0.32\textwidth]{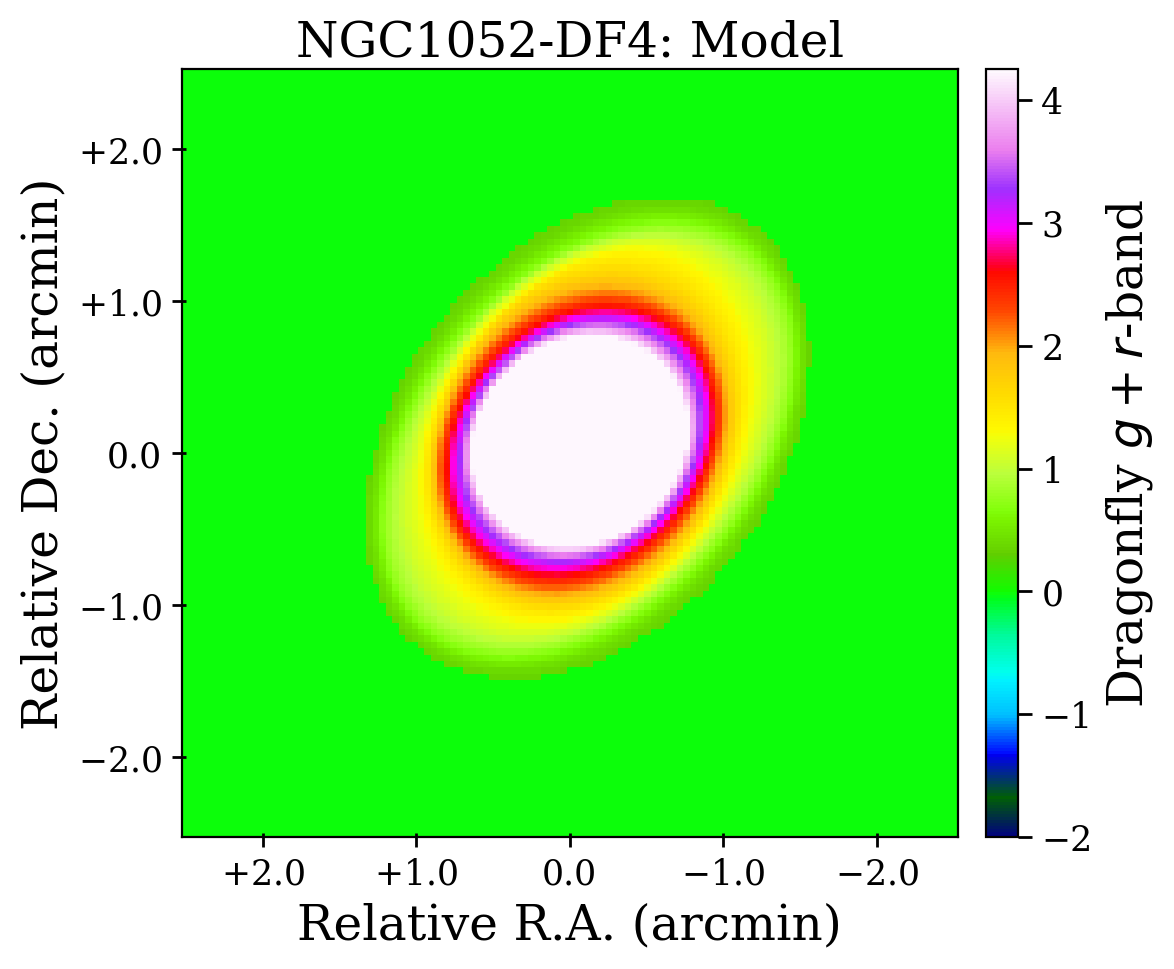} \includegraphics[width=0.32\textwidth]{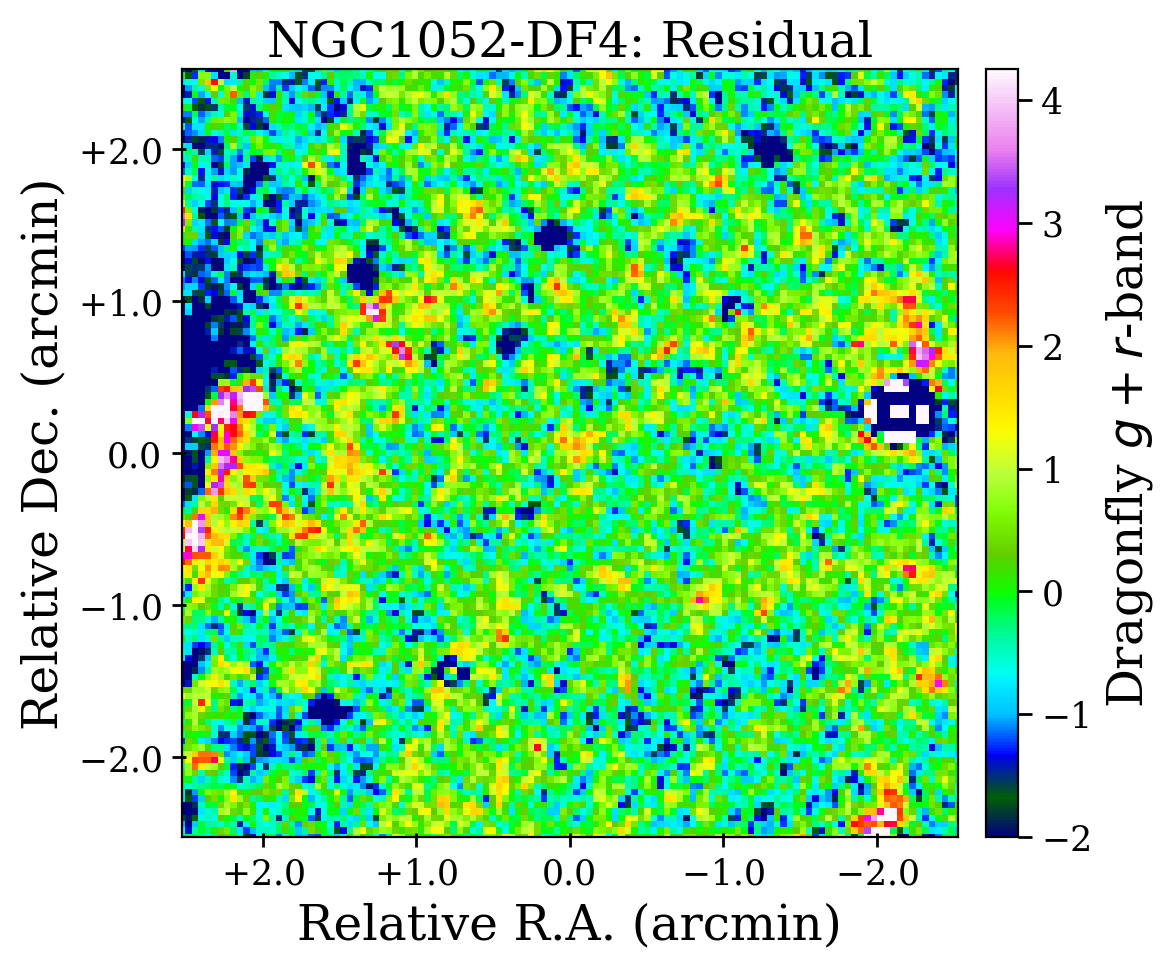} \\ 
    \includegraphics[width=0.32\textwidth]{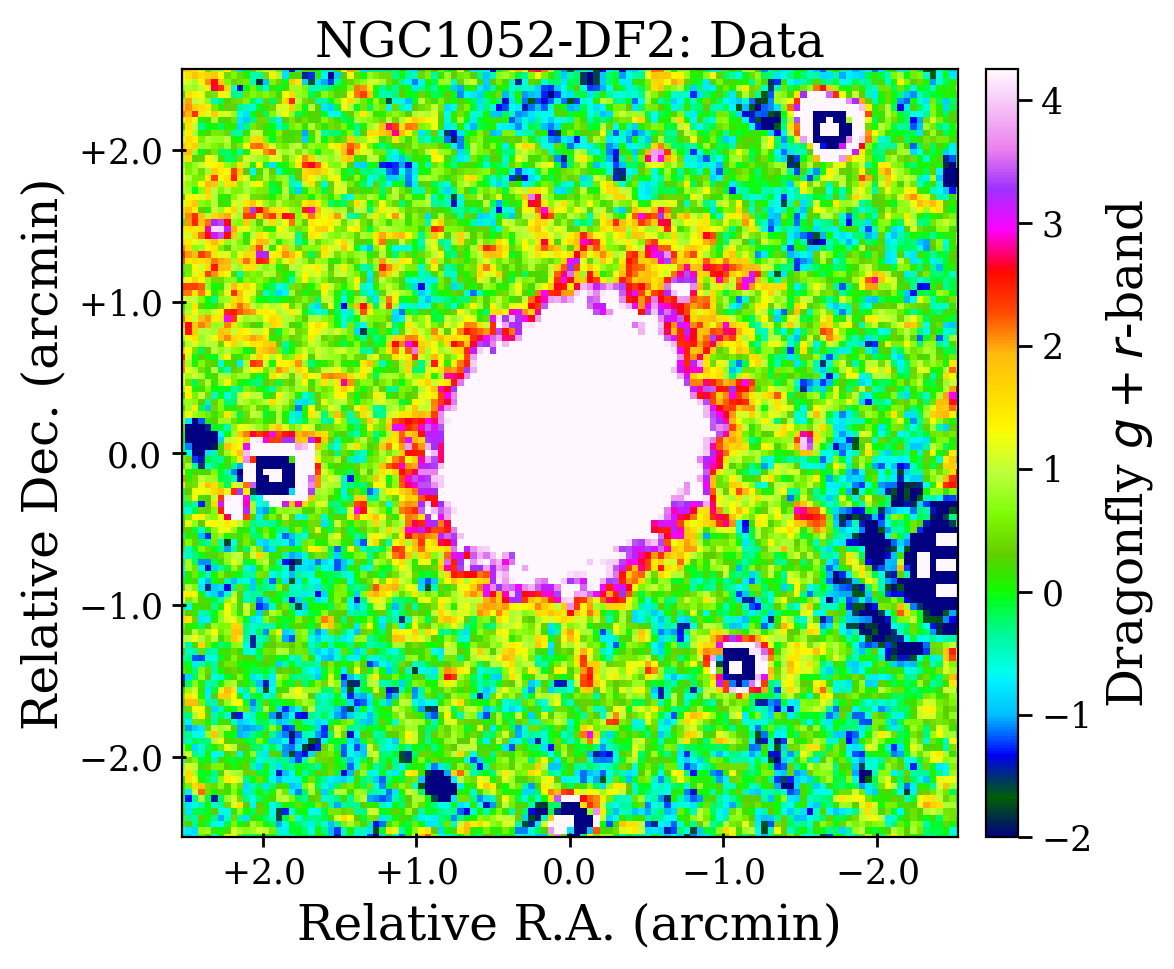} \includegraphics[width=0.32\textwidth]{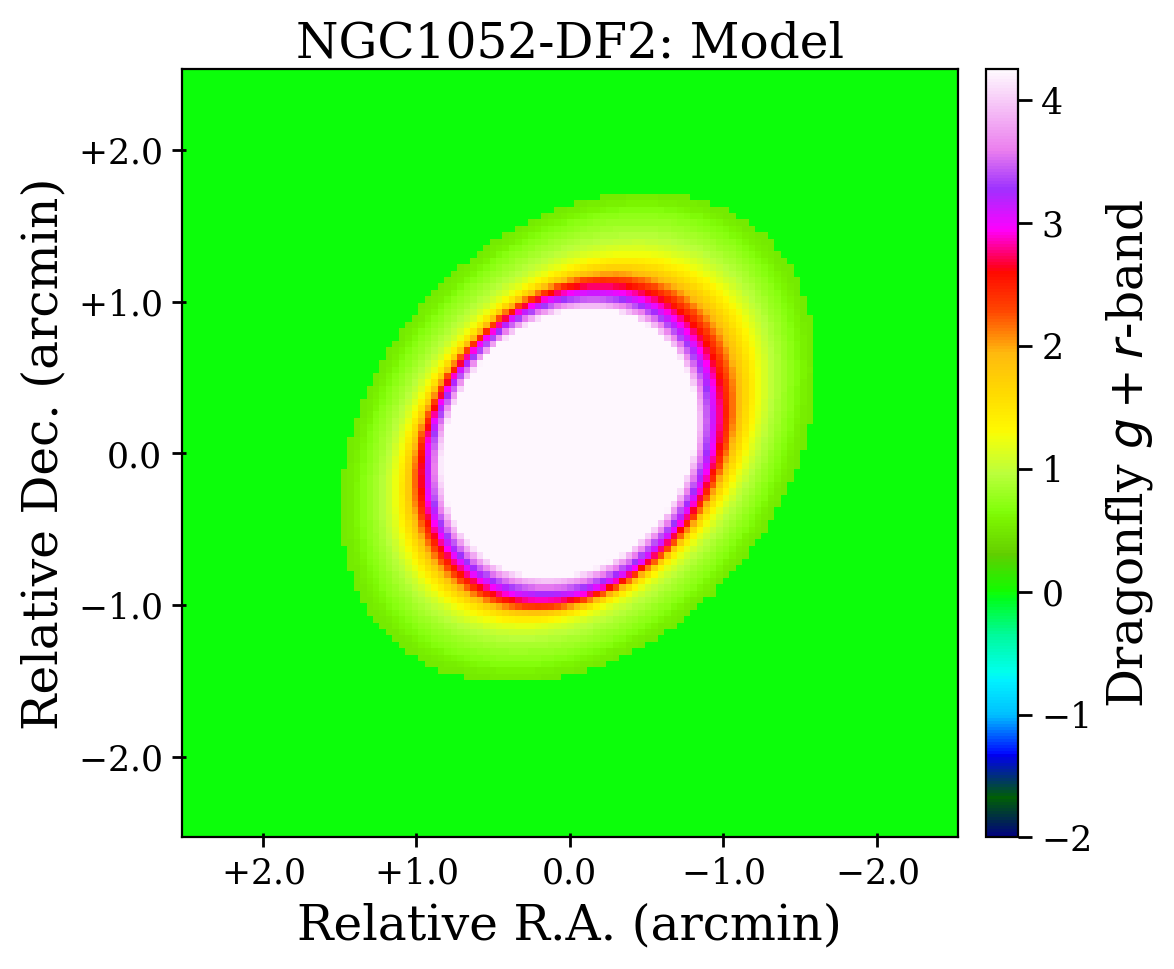} \includegraphics[width=0.32\textwidth]{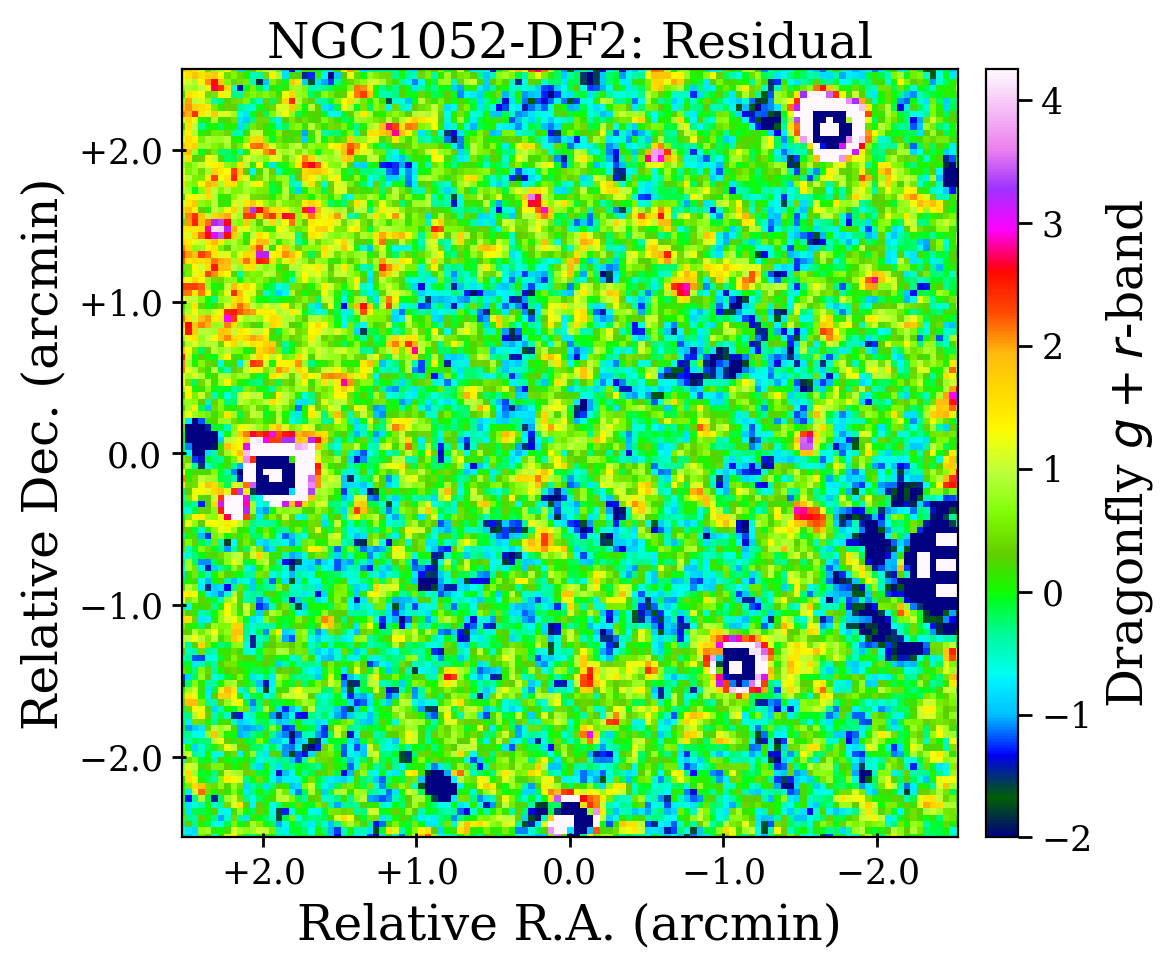} \\    
    \caption{Dragonfly data (\textit{left column}), fit models (\textit{middle column}), and fitting residuals (\textit{right column}) for NGC1052-DF5 (\textit{first row}), NGC1035 (\textit{second row}), NGC1052-DF4 (\textit{third row}), and NGC1052-DF2 (\textit{fourth row}). The colorbar is scaled for optimal analysis of the galaxies' outskirts and the residual images. \label{Fig:LowResFits}}
\end{figure}

\begin{figure}
    \centering
    \includegraphics[width=0.32\textwidth]{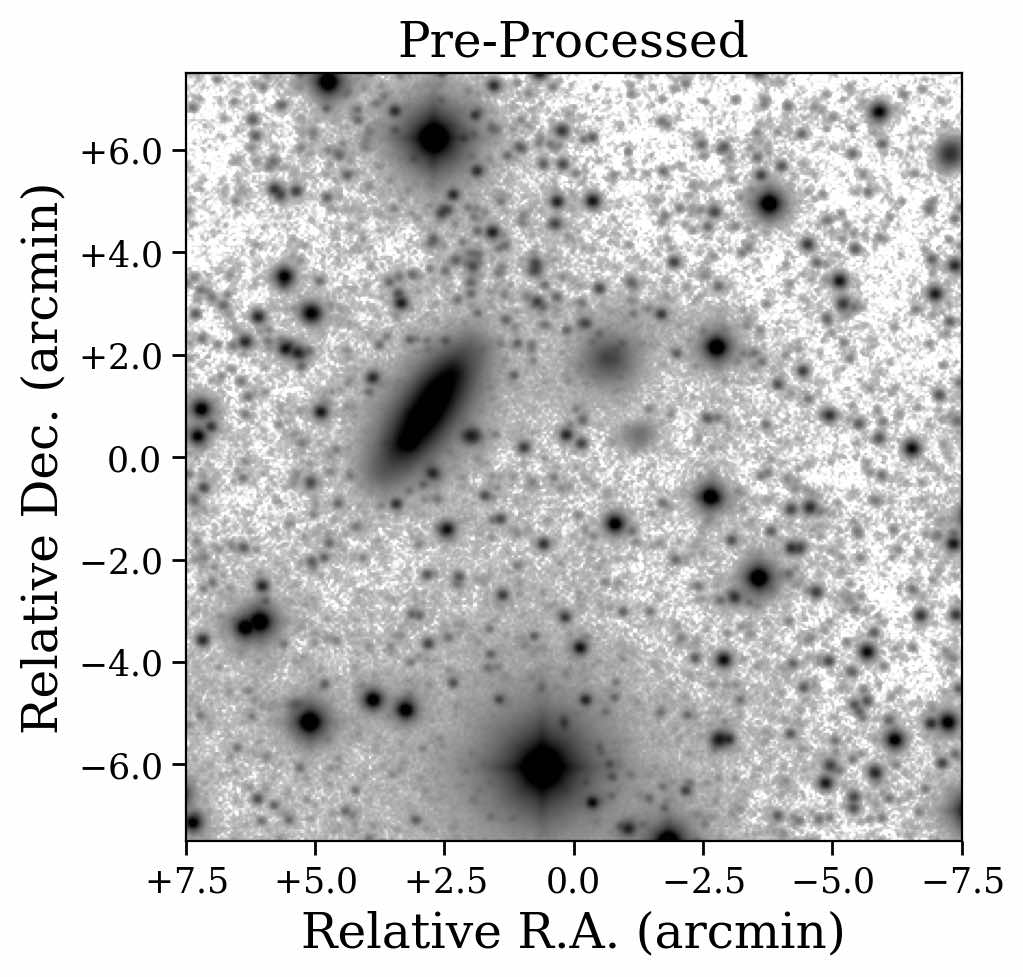}   \includegraphics[width=0.32\textwidth]{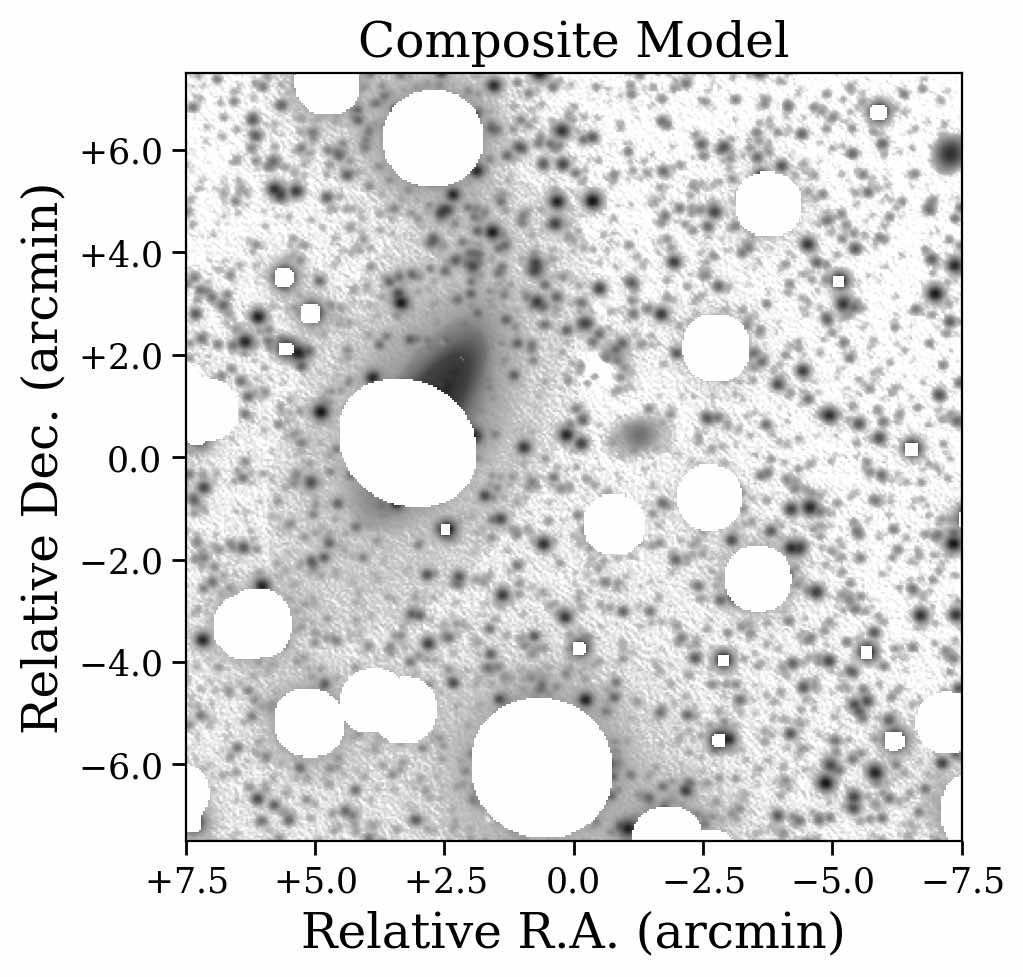}   \includegraphics[width=0.32\textwidth]{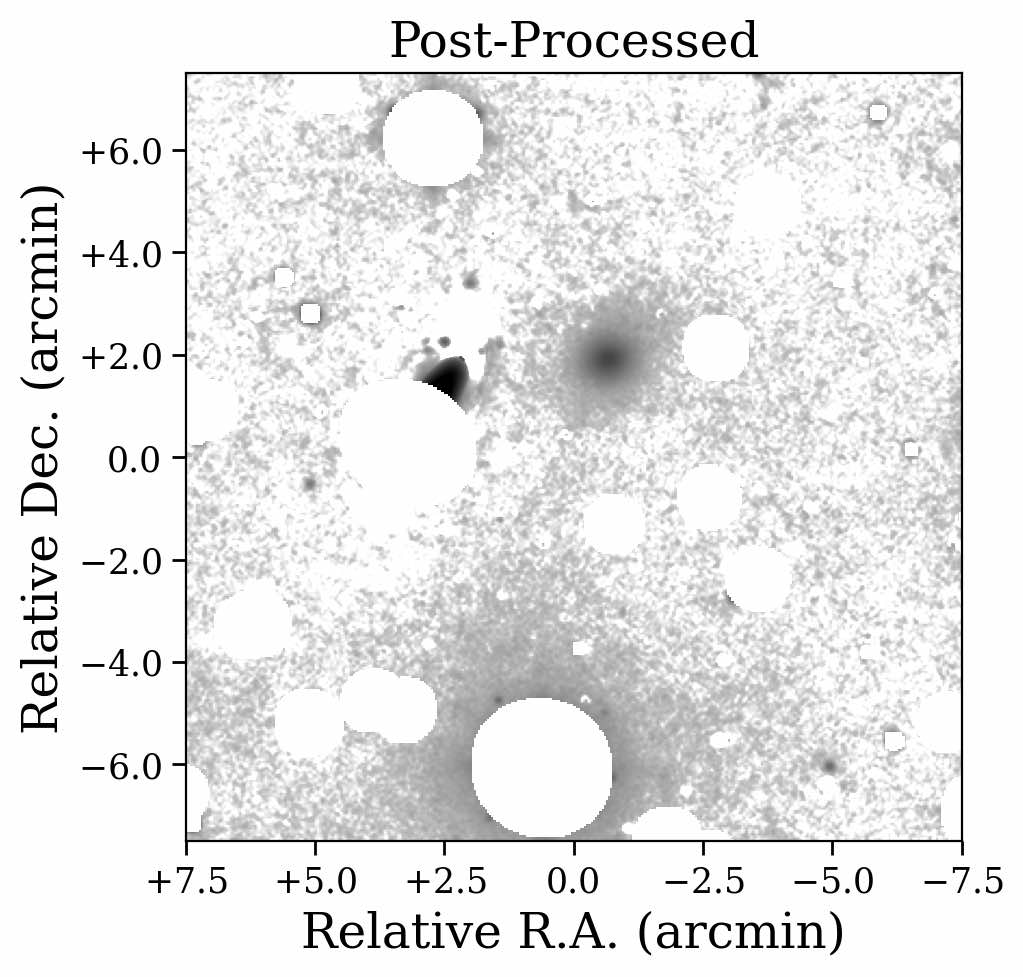}   \\   
    \caption{Dragonfly data before MRF (\textit{left panel}), the model subtracted by MRF along with models of NGC1052-DF5 and NGC1035 (\textit{middle panel}), and the final post-processed image analyzed in this work (\textit{right panel}). \label{Fig:CompModel}}
\end{figure}

In Sections~\ref{Sec:Inspection} and \ref{Sec:Fit} we outline the method we utilized to generate galaxy models. In Fig.~\ref{Fig:LowResFits}, we give the final model and residuals generated from the described techniques. In Fig.~\ref{Fig:CompModel} we give the complete subtracted model including that generated by the MRF procedure.

\section{Error Estimate} \label{Sec:Error}
\begin{figure}[hb]
    \centering
    \includegraphics[width=0.4\textwidth]{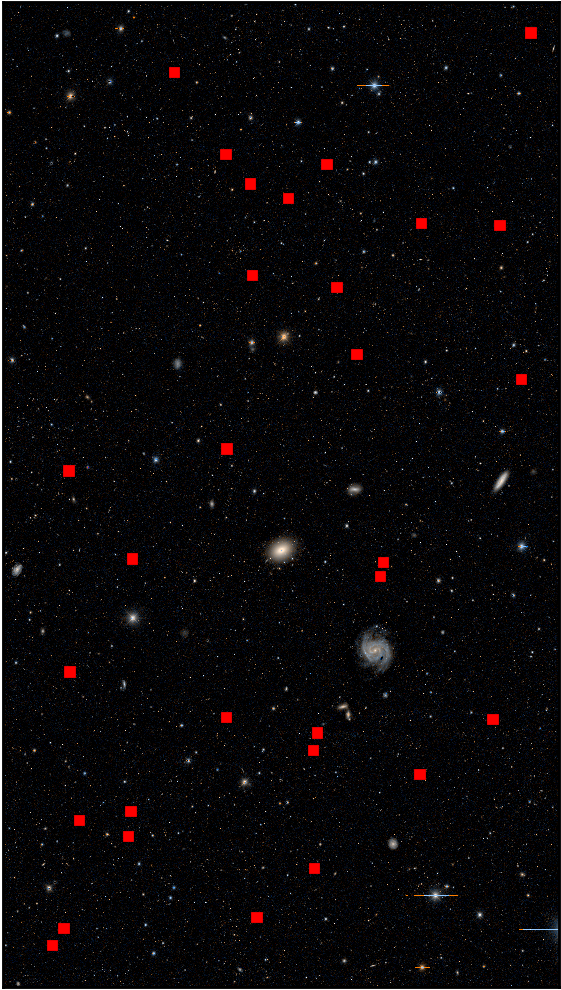}
    \caption{A visual depiction of Table~\ref{Table:ErrRegions} using DECaLS images.\label{Fig:Locs}}
\end{figure}

In order to estimate the uncertainty of our isophotal analysis, we added our models of NGC1052-DF2 and NGC1052-DF4 to 30 different locations near NGC1052 and repeated our fitting procedure. These locations are listed in Table~\ref{Table:ErrRegions} and plotted in Fig.~\ref{Fig:Locs}.

\begin{deluxetable}{ccc|cccc|ccc}
\tablecaption{Locations of Repeated Fitting for Error Analysis\label{Table:ErrRegions}}
\tablewidth{0pt}
\tablehead{
\colhead{R.A.} & \colhead{Dec.} & \quad & \quad & \colhead{R.A.} & \colhead{Dec.} & \quad & \quad & \colhead{R.A.} & \colhead{Dec.} \\
\colhead{(deg)} & \colhead{(deg)} & \quad & \quad & \colhead{(deg)} & \colhead{(deg)} & \quad & \quad & \colhead{(deg)} & \colhead{(deg)}
}
\startdata
\hline
39.8162006 & -7.3358901 & \quad & \quad & 40.1860284 & -7.5700095 & \quad & \quad & 40.3688542 & -8.5515120 \\
39.8321589 & -7.9515584 & \quad & \quad & 40.2025026 & -8.5785179 & \quad & \quad & 40.4626738 & -7.4066194 \\
39.8716085 & -7.6782209 & \quad & \quad & 40.2078246 & -8.8191706 & \quad & \quad & 40.5392747 & -8.2695813 \\
39.8834997 & -8.5543915 & \quad & \quad & 40.2103322 & -8.6098382 & \quad & \quad & 40.5418874 & -8.7182538 \\
40.0147628 & -7.6744495 & \quad & \quad & 40.2560314 & -7.6300094 & \quad & \quad & 40.5471793 & -8.7628797 \\
40.0163777 & -8.6526040 & \quad & \quad & 40.3130732 & -8.9059635 & \quad & \quad & 40.6362913 & -8.7345189 \\
40.0825248 & -8.2766276 & \quad & \quad & 40.3205831 & -7.7666937 & \quad & \quad & 40.6526426 & -8.4704841 \\
40.0885769 & -8.3006303 & \quad & \quad & 40.3242276 & -7.6040095 & \quad & \quad & 40.6543169 & -8.1134915 \\
40.1314397 & -7.9069662 & \quad & \quad & 40.3674423 & -8.0746604 & \quad & \quad & 40.6642989 & -8.9257553 \\
40.1675125 & -7.7875162 & \quad & \quad & 40.3688038 & -7.5520090 & \quad & \quad & 40.6851804 & -8.9555278 \\
\enddata
\end{deluxetable}

\clearpage

\section{Scattered Light} \label{Sec:PSF}
\begin{figure}
    \centering
    \includegraphics[width=0.75\textwidth]{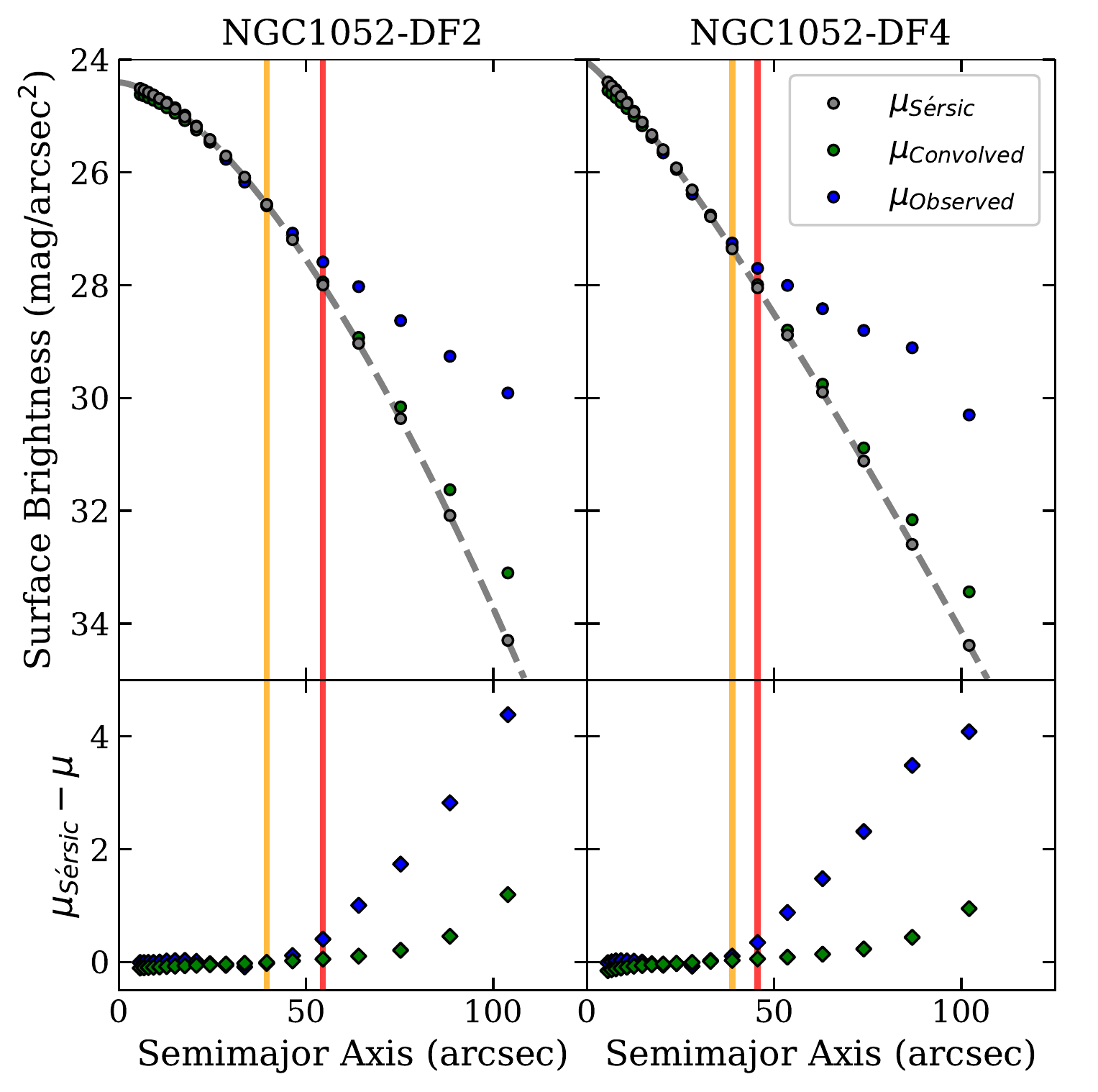}
    \caption{The effect of Dragonfly's PSF on the outskirts of NGC1052-DF2 (\textit{left}) and NGC1052-DF4 (\textit{right}). Surface brightness profiles (\textit{top}) are given for 2D S\'ersic models of each galaxy (\textit{grey circles}) as well as the same models convolved with Dragonfly's PSF (\textit{green circles}). We compare to the observed profile (\textit{blue circles}) and give the differences (\textit{bottom}) between the S\'ersic model and both the PSF-convolved model (\textit{green diamonds}) and observed profile (\textit{blue diamonds}), showing that PSF up-bending is not responsible for the observed break. $r_{\rm distort}$ and $r_{\rm break}$ are indicated by orange and red lines. \label{Fig:PSF}}
\end{figure}
In this work we consider elongation, position angle twists, and surface brightness profile breaks as evidence of tidal interactions. Scattered light from the center of the galaxies may also contribute to an up-bending which may mimic a tidal break, though this effect is expected to be small given Dragonfly's well controlled PSF wings \citep{2014PASP..126...55A,2022ApJ...925..219L}. To ensure the tidal break is not the result of Dragonfly's PSF, in Fig.~\ref{Fig:PSF} we convolve 2D S\'ersic models of each galaxy without the break with Dragonfly's wide-angle PSF (as utilized in Section~\ref{Sec:MRF}), and compare to the observed surface brightness profile. We find that the effect of the PSF's up-bending on the galaxies' outskirts is negligible compared to the observed break and lies below the surface brightness limit of the image.

\clearpage

\section{Photometric Data} \label{Sec:Profiles}
\begin{figure}
    \centering
    \includegraphics[width=0.75\textwidth]{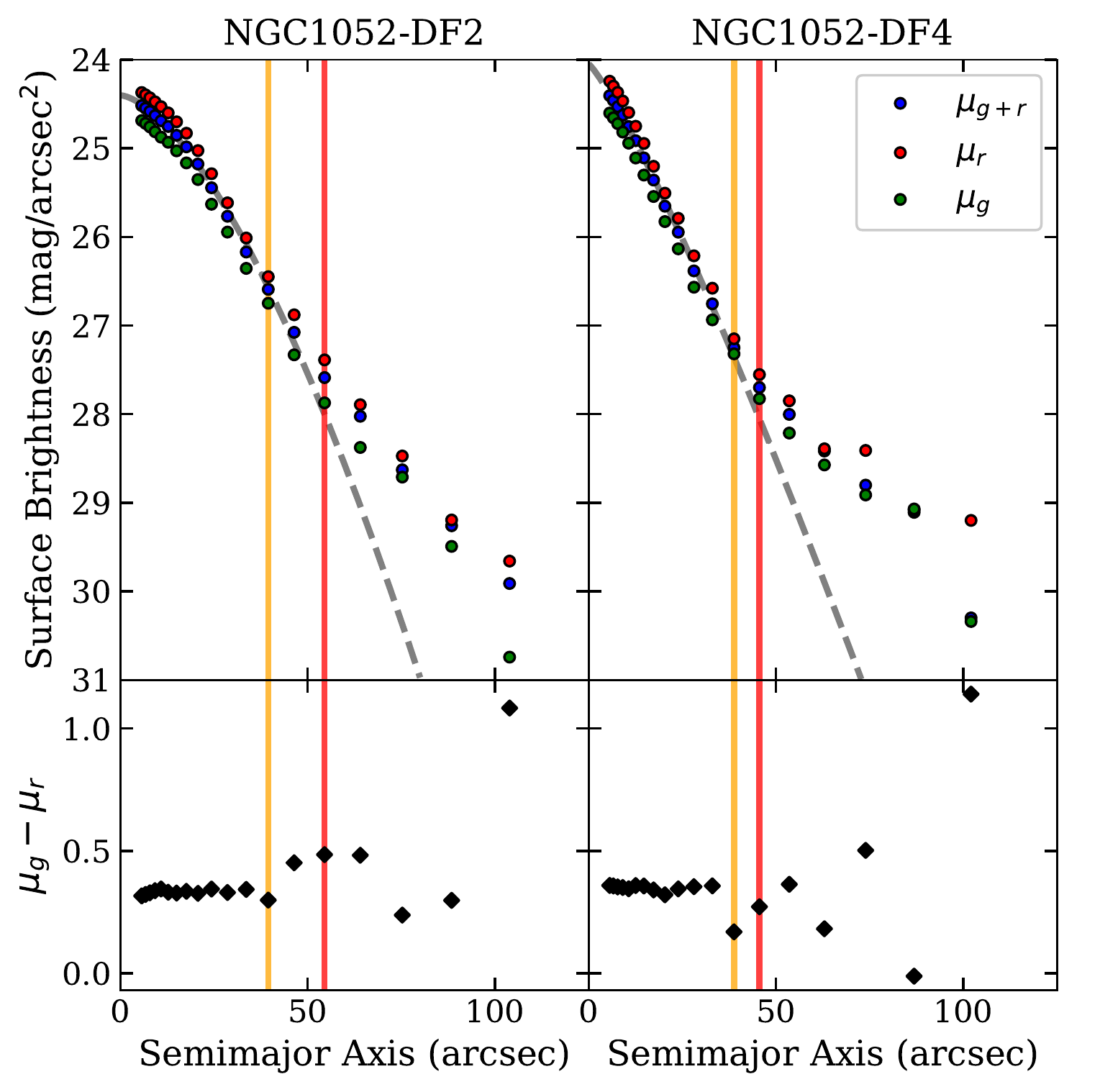}
    \caption{Surface brightness profiles in the $g$ and $r$ bands (\textit{top}) and color profiles (\textit{bottom}) for NGC1052-DF2 (\textit{left}) and NGC1052-DF4 (\textit{right}). S\'ersic fits, $r_{\rm distort}$, and $r_{\rm break}$ are indicated by dashed grey, solid orange, and solid red lines. \label{Fig:Colors}}
\end{figure}
This work largely focused on data averaged over the $g$ and $r$ photometric bands. For the use of interested readers, in Fig.~\ref{Fig:Colors} we give the surface brightness profiles in each band as well as the $g-r$ color profile. These were computed using the same isophote fitting technique as described in the main text.

%--------------------------------------------------------------
% References
%--------------------------------------------------------------

\bibliographystyle{aasjournal}
\bibliography{bibliography}
\end{CJK*}
\end{document}